\documentclass{ar-1col}
\usepackage[comma]{natbib}
\usepackage[mathscr]{euscript}
\usepackage{mathtools}
\usepackage{color}

\usepackage{color, colortbl}
\usepackage{multirow}
\usepackage{amssymb}
\usepackage{amsmath}
\usepackage{media9}
\usepackage{stmaryrd}

\jname{Xxxx. Xxx. Xxx. Xxx.}
\jvol{AA}
\jyear{2020}
\doi{10.1146/((please add article doi))}

\def\la{\mathrel{\mathchoice {\vcenter{\offinterlineskip\halign{\hfil$\displaystyle##$\hfil\cr<\cr\sim\cr}}}
{\vcenter{\offinterlineskip\halign{\hfil$\textstyle##$\hfil\cr<\cr\sim\cr}}}
{\vcenter{\offinterlineskip\halign{\hfil$\scriptstyle##$\hfil\cr
<\cr\sim\cr}}}
{\vcenter{\offinterlineskip\halign{\hfil$\scriptscriptstyle##$\hfil\cr><cr\sim\cr}}}}}

\def\ga{\mathrel{\mathchoice {\vcenter{\offinterlineskip\halign{\hfil$\displaystyle##$\hfil\cr>\cr\sim\cr}}}
{\vcenter{\offinterlineskip\halign{\hfil$\textstyle##$\hfil\cr>\cr\sim\cr}}}
{\vcenter{\offinterlineskip\halign{\hfil$\scriptstyle##$\hfil\cr
<\cr\sim\cr}}}
{\vcenter{\offinterlineskip\halign{\hfil$\scriptscriptstyle##$\hfil\cr><cr\sim\cr}}}}}

\newcount\longrefs
\def\aap{\ifnum\longrefs=1 {Astron.\ Astrophys.\ }\else 
                           {A\hbox{\rm \&}A}\fi}
\def\aapl{\ifnum\longrefs=1 {Astron.\ Astrophys.\ Lett.\ }\else 
                           {A\hbox{\rm \&}A}\fi}
\def\aapr{\ifnum\longrefs=1 {Astron.\ Astrophys.\ Rev.\ }\else 
                            {A\hbox{\rm \&}AR}\fi}
\def\aaps{\ifnum\longrefs=1 {Astron.\ Astrophys.\ Suppl.\ }\else 
                            {A\hbox{\rm \&}AS}\fi}
\def\aj{\ifnum\longrefs=1 {Astron.\ J.\ }\else 
                          {AJ}\fi} 
\def\ao{\ifnum\longrefs=1 {Applied Optics}\else 
                           {Appl.\ Opt.}\fi} 
\def\aspcs{\ifnum\longrefs=1 {Astron.\ Soc.\ Pacific Conf. Series}\else 
                           {ASP Conf.\ Ser.}\fi} 
\def\apj{\ifnum\longrefs=1 {Astrophys.\ J.\ }\else 
                           {ApJ}\fi} 
\def\apjl{\ifnum\longrefs=1 {Astrophys.\ J.\ Lett.\ }\else 
                            {ApJ}\fi} 
\def\aplett{\ifnum\longrefs=1 {Astrophys.\ J.\ Lett.\ }\else 
                            {ApJ}\fi} 
\def\apjs{\ifnum\longrefs=1 {Astrophys.\ J.\ Suppl.\ }\else 
                            {ApJS}\fi}
\def\apss{\ifnum\longrefs=1 {Astrophys.\ and Space Science}\else 
                            {Ap\hbox{\rm \&}SS}\fi}
\def\araa{\ifnum\longrefs=1 {Ann.\ Rev.\ Astron.\ Astrophys.\ }\else 
                            {ARA\hbox{\rm \&}A}\fi}
\def\azh{\ifnum\longrefs=1 {Astronomicheskii Zhurnal}\else 
                            {Astron.\ Zhur.}\fi}
\def\baas{\ifnum\longrefs=1 {Bull.\ Am.\ Astron.\ Soc.\ }\else 
                            {BAAS}\fi}
\def\bain{\ifnum\longrefs=1 {Bull.\ Astronom.\ Institutes Netherlands}\else
                            {Bull.\ Astr.\ Inst.\ Neth.}\fi}
\def\gca{\ifnum\longrefs=1 {Geochim.\ Cosmochim.\ Acta}\else 
                           {Geochim.\ Cosmochim.\ Acta}\fi}
\def\grl{\ifnum\longrefs=1 {Geophys.\ Res.\ Lett.\ }\else 
                           {Geoph.\ Res.\ Lett.\ }\fi}
\def\iaucirc{\ifnum\longrefs=1 {IAU Circulars}\else 
                          {IAU Circ.}\fi}
\def\icarus{\ifnum\longrefs=1 {Icarus}\else 
                          {Icarus}\fi}
\def\ip{\ifnum\longrefs=1 {in press}\else 
                          {in press}\fi}
\def\jchemp{\ifnum\longrefs=1 {J.\ Chem.\ Phys.\ }\else 
                           {J.\ Chem.\ Phys.\ }\fi}  
\def\jcp{\ifnum\longrefs=1 {J.\ Chem.\ Phys.\ }\else 
                           {J.\ Chem.\ Phys.\ }\fi}  
\def\jgr{\ifnum\longrefs=1 {J.\ Geophys.\ Res.\ }\else 
                           {J.\ Geophys.\ Res.\ }\fi}  
\def\jmolspec{\ifnum\longrefs=1 {J.\ Mol.\ Spectrosc.\ }\else 
                           {J.\ Mol.\ Spectrosc.\ }\fi}  
\def\jqsrt{\ifnum\longrefs=1 {J.\ Quant.\ Spectrosc.\ Radiat.\ Transfer}\else 
                           {J.\ Quant.\ Spectrosc.\ Radiat.\ Transfer}\fi}  
\def\jrasc{\ifnum\longrefs=1 {J.\ Royal Astron.\ Soc.\ Canada}\else 
                           {JRAS Can.\ }\fi}  
\def\mnras{\ifnum\longrefs=1 {Mon.\ Not.\ Roy.\ Astron.\ Soc.\ }\else 
                             {MNRAS}\fi} 
\def\nat{\ifnum\longrefs=1 {Nature}\else 
                           {Nat}\fi}
\def\pasj{\ifnum\longrefs=1 {Pub.\ Astron.\ Soc.\ Japan}\else 
                            {PASJ}\fi} 
\def\pasp{\ifnum\longrefs=1 {Pub.\ Astron.\ Soc.\ Pacific}\else 
                            {PASP}\fi} 
\def\physscr{\ifnum\longrefs=1 {Physica Scripta}\else 
                            {Phys.\ Scrip.\ }\fi} 
\def\planss{\ifnum\longrefs=1 {Planetary \& Space Science}\else 
                            {Plan. \& Space Sci.\ }\fi} 
\def\procspie{\ifnum\longrefs=1 {Proc.\ SPIE}\else 
                            {Proc.\ SPIE}\fi} 
\def\qjras{\ifnum\longrefs=1 {Quarterly J.\ Royal Astron.\ Soc.\ }\else 
                            {QJRAS}\fi} 
\def\sa{\ifnum\longrefs=1 {Soviet Astron.\ }\else 
                               {Sov.\ Astron.\ }\fi}
\def\skytel{\ifnum\longrefs=1 {Sky \& Telescope}\else 
                            {Sky \& Tel.\ }\fi} 
\def\solphys{\ifnum\longrefs=1 {Solar Phys.\ }\else 
                               {Solar Phys.\ }\fi}
\def\ssr{\ifnum\longrefs=1 {Space Science Rev.\ }\else 
                               {Space\ Sci.\ Rev.\ }\fi}

\def\dutch{\def\refname{Referenties}\def\abstractname{Samenvatting}%
  \def\bibname{Bibliografie}\def\chaptername{Hoofdstuk}%
  \def\appendixname{Bijlage}\def\contentsname{Inhoudsopgave}%
  \def\listfigurename{Lijst van figuren}\def\listtablename{Lijst van tabellen}%
  \def\indexname{Index}\def\figurename{Figuur}\def\tablename{Tabel}%
  \def\partname{Deel}\def\enclname{Bijlage(n)}\def\ccname{Ter attentie van}%
  \def\headtoname{Aan}\def\headpagename{Pagina}%
  \def\today{\number\day\space\ifcase\month\or januari\or februari\or maart\or%
     april\or mei\or juni\or juli\or augustus\or september\or oktober\or%
     november\or december\fi \space\number\year}%
  \typeout{
              >>>>> use hlatex209 for Dutch hyphenation <<<<< 
         }}
\newcounter{onefig} \newcounter{fignumber}
\newcount\nocaptions \newcount\nofigures \newcount\figwidth
\newcount\viewgraphs
  \def\paper{}  \def\figlabel{} 
\long\def\nextfig#1{\setcounter{figure}{\value{fignumber}}
  \addtocounter{fignumber}{1}
  \ifnum \viewgraphs=1 \newpage \pagestyle{empty} \fi 
  \ifnum\value{onefig}=0 #1 \fi                 
  \ifnum\value{onefig}=\value{fignumber} #1 \fi}
\def\figwidths#1#2{\ifnum \nocaptions=1 #2mm \else #1mm \fi}  
\def\paper#1{}  
\long\def\plotfig#1#2{\ifnum \nofigures=1 \else #2 \fi}
\long\def\captiontext#1{\ifnum \nofigures=1 \raggedright \fi 
   \ifnum \nocaptions=1 \paper
     \ifnum \viewgraphs=0 
       \newline  \mbox{}\hrulefill\mbox{} \newline 
       \newline label:~\{\figlabel\} 
     \fi 
     \else \ifnum \nofigures=0 \fi 
   #1 \fi}

\newcount\panelwidth \newcount\panelheight 
\newcount\bxmin \newcount\bymin \newcount\bxmax \newcount\bymax
\newcount\tbxmin \newcount\tbymin
\newcount\tpanelwidth \newcount\tpanelheight \newcount\tpdif
\def\panelsize #1,#2;{\panelwidth=#1 \panelheight=#2}  
\def\setbb #1,#2;#3,#4;#5,#6;{
  \tbxmin=#1 \tbymin=#2    
  \bxmin=#3 \bymin=#4      
  \bxmax=#5 \bymax=#6}     
\def\barepanel #1{%
  \ifnum\panelheight=0 
    \tpdif=\bymax \advance\tpdif by -\bymin
    \multiply \tpdif by \panelwidth
    \tpanelheight=\tpdif
    \tpdif=\bxmax \advance\tpdif by -\bxmin
    \divide \tpanelheight by \tpdif
  \else \tpanelheight=\panelheight \fi
  \epsfig{file=#1,%
     bbllx=\bxmin bp,bblly=\bymin bp,bburx=\bxmax bp,bbury=\bymax bp,clip=,%
     width=\panelwidth mm,height=\tpanelheight mm}}
\def\labelypanel #1{
  \ifnum\panelheight=0 
    \tpdif=\bymax \advance\tpdif by -\bymin
    \multiply \tpdif by \panelwidth
    \tpanelheight=\tpdif
    \tpdif=\bxmax \advance\tpdif by -\bxmin
    \divide \tpanelheight by \tpdif
  \else \tpanelheight=\panelheight \fi
  \tpdif=\bxmax \advance\tpdif by -\tbxmin
  \tpanelwidth=\panelwidth \multiply \tpanelwidth by \tpdif
  \tpdif=\bxmax \advance\tpdif by -\bxmin
  \divide \tpanelwidth by \tpdif
  \epsfig{file=#1,%
    bbllx=\tbxmin bp,bblly=\bymin bp,bburx=\bxmax bp,bbury=\bymax bp,%
    clip=,width=\tpanelwidth mm,height=\tpanelheight mm}}
\def\labelxpanel #1{%
  \ifnum\panelheight=0 
    \tpdif=\bymax \advance\tpdif by -\bymin
    \multiply \tpdif by \panelwidth
    \tpanelheight=\tpdif
    \tpdif=\bxmax \advance\tpdif by -\bxmin
    \divide \tpanelheight by \tpdif
  \else \tpanelheight=\panelheight \fi
  \tpdif=\bymax \advance\tpdif by -\tbymin
  \multiply \tpanelheight by \tpdif
  \tpdif=\bymax \advance\tpdif by -\bymin
  \divide \tpanelheight by \tpdif
  \epsfig{file=#1,%
    bbllx=\bxmin bp,bblly=\tbymin bp,bburx=\bxmax bp,bbury=\bymax bp,%
    clip=,width=\panelwidth mm,height=\tpanelheight mm}}
\def\labelxypanel #1{%
  \ifnum\panelheight=0 
    \tpdif=\bymax \advance\tpdif by -\bymin
    \multiply \tpdif by \panelwidth
    \tpanelheight=\tpdif
    \tpdif=\bxmax \advance\tpdif by -\bxmin
    \divide \tpanelheight by \tpdif
  \else \tpanelheight=\panelheight \fi
  \tpdif=\bxmax \advance\tpdif by -\tbxmin
  \tpanelwidth=\panelwidth \multiply \tpanelwidth by \tpdif
  \tpdif=\bxmax \advance\tpdif by -\bxmin
  \divide \tpanelwidth by \tpdif 
  \tpdif=\bymax \advance\tpdif by -\tbymin 
  \multiply \tpanelheight by \tpdif
  \tpdif=\bymax \advance\tpdif by -\bymin
  \divide \tpanelheight by \tpdif
  \epsfig{file=#1,%
    bbllx=\tbxmin bp,bblly=\tbymin bp,bburx=\bxmax bp,bbury=\bymax bp,%
    clip=,width=\tpanelwidth mm,height=\tpanelheight mm}}

\def\CC{\par \vspace*{-2ex} \footnotesize \baselineskip=8pt \begin{verbatim}}

\long\def\startignore #1\stopignore{}   

\def\setlistparams{         
  \topsep=0.7ex                 
  \itemsep=0.7ex                
  \leftmargini=3ex}             
\setlistparams                  

\newcounter{alistindex}       

\newcounter{romenumnr}

\newlength{\minipagewidth}

\newsavebox{\boxcontent}
\newcommand{\ovalhead}[1]{
  \unitlength=1cm
  \sbox{\boxcontent}{\mbox{~~{#1}~~}}
  \begin{center}
    \ifdim\wd\boxcontent>6ex 
    \ifdim\wd\boxcontent<8cm 
    \begin{picture}(8,3) \thicklines     
      \put(4.0,0.8){\oval(8,1.6)} 
      \put(0.0,0.7){\parbox{8cm}{
         \begin{center} \usebox{\boxcontent} \end{center}}}
    \end{picture}
    \else \ifdim\wd\boxcontent<12cm 
    \begin{picture}(12,3) \thicklines     
        \put(6.0,0.8){\oval(12,1.6)} 
        \put(0.0,0.7){\parbox{12cm}{
           \begin{center} \usebox{\boxcontent} \end{center}}}
    \end{picture}
    \else
    \begin{picture}(16,3) \thicklines     
        \put(8.0,0.8){\oval(16,1.6)} 
        \put(0.0,0.7){\parbox{16cm}{
           \begin{center} \usebox{\boxcontent} \end{center}}}
    \end{picture}
    \fi \fi \fi
  \end{center}} 

\newcounter{headnr}            
\newcounter{subheadnr}[headnr]
\newcounter{subsubheadnr}[subheadnr]
\def\head #1\par{
  \stepcounter{headnr}                          
  \vspace{2ex} \noindent                        
  {\bf \theheadnr~~~~#1}\\[1ex] \noindent}      
\def\subhead #1\par{  
  \stepcounter{subheadnr}
  \vspace{1.3ex} \noindent
  {\bf \theheadnr.\arabic{subheadnr}~~~#1}\\[0.3ex] \noindent}
\def\subsubhead #1\par{
  \stepcounter{subsubheadnr}
  \vspace{1.0ex} \noindent
  {\bf \theheadnr.\arabic{subheadnr}.\arabic{subsubheadnr}~~~#1}\\ \noindent}

\font\dropfont= cmr12 scaled \magstep5
\def\dropcap#1#2{{\noindent
    \setbox0\hbox{\dropfont #1}\setbox1\hbox{#2}\setbox2\hbox{(}%
    \count0=\ht0\advance\count0 by\dp0\count1\baselineskip
    \advance\count0 by-\ht1\advance\count0by\ht2
    \dimen1=.5ex\advance\count0by\dimen1\divide\count0 by\count1
    \advance\count0 by1\dimen0\wd0
    \advance\dimen0 by.25em\dimen1=\ht0\advance\dimen1 by-\ht1
    \global\hangindent\dimen0\global\hangafter-\count0
    \hskip-\dimen0\setbox0\hbox to\dimen0{\raise-\dimen1\box0\hss}%
    \dp0=0in\ht0=0in\box0}#2}

\def\level #1 #2#3#4{$#1 \: ^{#2} \mbox{#3} ^{#4}$}   

\def\deg{\hbox{$^\circ$}}       

\def\arcsec{\hbox{$^{\prime\prime}$}}

\def\Teff{\hbox{$\rm{T}_{\rm eff}$}}            

\def\Msun{\hbox{M$_{\odot}$}}               
\def\Lsun{\hbox{L$_{\odot}$}}               

\def\Mstar{\hbox{M$_{\star}$}}               
\def\Rstar{\hbox{R$_{\star}$}}               
\def\Lstar{\hbox{L$_{\star}$}}               
\def\Tstar{\hbox{T$_{\star}$}}               
\def\Teff{\hbox{T$_{\rm{eff}}$}}             

\def\Mdot{\hbox{$\dot{M}$}}                     
\def\mathstacksym#1#2#3#4#5{\def#1{\mathrel{\hbox to 0pt{\lower 
    #5\hbox{#3}\hss} \raise #4\hbox{#2}}}}

\mathstacksym\lta{$<$}{$\sim$}{1.5pt}{3.5pt} 
\mathstacksym\gta{$>$}{$\sim$}{1.5pt}{3.5pt} 
\mathstacksym\lrarrow{$\leftarrow$}{$\rightarrow$}{2pt}{1pt} 
\mathstacksym\lessgreat{$>$}{$<$}{3pt}{3pt} 

\definecolor{captioncolor}{cmyk}{1.0,0.4,0,0}
 \newcolumntype{a}{>{\columncolor{shadecolor}}c}

\begin{document}

\newcommand{\blue}[1]{\textcolor{blue}{#1}}
\newcommand{\red}[1]{\textcolor{red}{#1}}

\newcommand{\captioncolor}{\textcolor{fignumcolor}}

\newcommand*\sq{\mathbin{\vcenter{\hbox{\rule{1ex}{1ex}}}}}

\markboth{Decin}{Daedalean Life of Cool Ageing Stars}

\title{Evolution and Mass Loss of Cool Ageing Stars: a Daedalean Story}

\author{Leen Decin$^{1,2}$ 
\affil{$^1$ Institute of Astronomy, KU\,Leuven, Celestijnenlaan 200D,
B-3001 Leuven, Belgium; email: Leen.Decin@kuleuven.be}
\affil{$^2$ University of Leeds, School of Chemistry, Leeds LS2 9JT, United Kingdom}}

\begin{abstract}

The chemical enrichment of the Universe; the mass spectrum of planetary nebulae, white dwarfs and gravitational wave progenitors;
the frequency distribution of Type I and II supernovae; the fate of exoplanets $\ldots$
a multitude of phenomena which is highly regulated 
by the amounts of mass that stars expel through a powerful wind.
For more than half a century, these winds of cool ageing stars have been interpreted within the common interpretive framework of 1-dimensional (1D) models. 
I here discuss how that framework now appears to be highly problematic.  
\begin{itemize}
	\item[$\sq$]	Current 1D mass-loss rate formulae  differ by orders of magnitude, \\ rendering contemporary stellar evolution predictions highly uncertain. 
\end{itemize}
These stellar winds  harbour 3D complexities which bridge 23 orders of magnitude in scale, ranging from the nanometer up to thousands of astronomical units. 
We need to embrace and understand these 3D spatial realities if we aim to quantify mass loss and assess its effect on stellar evolution. We therefore need to gauge
\begin{itemize}
	\item[$\sq$]	the 3D life of molecules and solid-state aggregates: the gas-phase \\ clusters that form the first dust seeds are not yet identified. This limits  \\ our ability to predict mass-loss rates using a self-consistent  approach. 
	\item[$\sq$]	the emergence of 3D clumps: they contribute in a non-negligible \\ way to the mass loss, although they seem of limited importance \\ for the wind-driving mechanism.
	\item[$\sq$]	the 3D lasting impact of a (hidden) companion:  unrecognised binary\\ interaction has biased previous mass-loss rate estimates towards values \\that are too large. 
\end{itemize}
Only then will it be possible to drastically improve our predictive power of the evolutionary  path in 4D (classical) spacetime of any star.
\end{abstract}
\begin{keywords}
evolved stars,
stellar winds,
stellar evolution,
clumps,
binaries,
astrochemistry
\end{keywords}
\maketitle

\tableofcontents

\section{Deterministic and conceptual perspective} \label{Sec:Perspective}

Stars are born and die. Much of humankind's attention and imagination is directed towards the star-forming process, the phase during which new planets are formed around young stars that harbour the tantalising potential for new life forms to arise.
Traditionally, the end-phases of stellar evolution have received much less attention. 
Supernovae, neutron stars and black holes avoid the perception of being `unglamorous old stars', 
but the research field focussing on the late stages of stellar evolution of low and intermediate mass stars 
is far less blessed by public excitement.

\begin{marginnote}[10.5truecm]
	\entry{Low-and intermediate mass stars}{stars that have an initial mass between $\sim$0.8\,--\,8\,\Msun\ and end their life as white dwarfs if single}
	\entry{High-mass stars}{stars with initial mass $M \ga 8\,$M$_\odot$ that undergo core collapse at the end of their life
		to form a neutron star or a black hole}
	\entry{Stellar wind}{flow of gas (and dust) particles ejected from the upper atmosphere of a star}
	\entry{Interstellar medium}{matter and radiation that exist between the stars in a galaxy}
\end{marginnote}
Cool ageing stars possess, however, some characteristics that turn them into key objects for  
both \textit{deterministic} and \textit{conceptual} questions in the broad field of astrophysics. 
The \textit{deterministic} facet is linked to the chemical enrichment of our Universe. 
These stars are nuclear power plants that create new atoms inside their hot dense cores, including carbon, the basic building block of life here on Earth. Through their winds, they contribute $\sim$85\% of gas and $\sim$35\% of dust to the total enrichment of the interstellar medium  \citep[ISM;][]{Tielens2005pcim.book.....T}, and are the dominant suppliers of pristine building blocks of interstellar material. 
The \textit{deterministic} aspect seeks to answer the \textit{how} question --- \textit{how} and \textit{how} much do cool ageing stars contribute to the galactic chemical enrichment? 
Given some initial conditions at birth, in what way can evolved stars determine the evolution of a galaxy?
Key questions include: which atoms are created and in what quantity? 
What are the mechanisms that transport the newly created elements from the core to the outer atmosphere? 
Which conditions can lead to the formation of molecules and solid-state dust species? 
Under which circumstances can a stellar wind form, and what are their resulting wind velocities and mass-loss rates?

The \textit{conceptual} aspect addresses the more general and overarching \textit{why} question --- \textit{why} do cool ageing stars contribute to the galactic chemical enrichment? 
This question is posed in the sense of identifying the general physical and chemical laws, and their interaction. 
These laws are not only applicable to cool ageing stars, but to all of astrophysics. 
We are convinced of course that the laws of physics and chemistry are universal laws, but we must also admit that our knowledge of chemical processes is still very Earth-centric. 
As I will discuss below, cool ageing stars are unique laboratories which offer the exquisite possibility of teaching us about extraterrestrial, and hence universal chemistry. 
The \textit{how} and \textit{why} question inform one  another and are interrelated. Ideally, we want to optimize our insight from both the \textit{deterministic} and \textit{conceptual}  perspective so as to obtain as complete a picture as possible of the late stages of stellar evolution.

Understanding the crucial role of cool ageing stars, the Asymptotic Giant Branch (AGB) stars and their more massive counterparts the red supergiant (RSG) stars,  at the level of the \textit{conceptual} framework requires further explanation. 
Although not always recognized, these stars deserve this critical status exactly because they are thought to be `simple'. 
Since the first identification of high luminosity stars by Maury in \citeyear{Maury1897AnHar..28....1M}, their subsequent classification as giant stars by Hertzsprung, somewhere between \citeyear{Hertzsprung1905WisZP...3..442H} and \citeyear{Hertzsprung1911POPot..22A...1H}, and the first solid evidence of matter escaping from a red giant star by \citeauthor{Deutsch1956ApJ...123..210D} in 1956, their atmospheric and wind structure were thought to have an overall spherical symmetry, and hence are described by one-dimensional (1D) equations. 
A large variety of chemical reactions occur in their wind, including unimolecular, 2- and 3-body reactions, cluster growth and grain formation. To date, more than 100 different molecules, and their isotopologues, and $\sim$15 different solid-state species have  been detected. 
The simple thermodynamical structure and chemically rich environment makes these stars ideal candidates for disentangling the physical and chemical processes, and unravelling the general laws governing not only these stars and their winds, but also those in other chemically rich astrophysical environments, including high-mass star-forming regions, young stellar objects, protoplanetary disks, exoplanets, novae, supernovae, and interstellar shocks.
\begin{marginnote}[5truecm]
\entry{AGB stars}{cool stars, with luminosity between $\sim$2300\,--57\,500\,\Lsun, that represent one of the last stages of evolution of low- to intermediate mass stars and exhibit strong stellar winds}
\entry {RSG stars}{cool massive stars, with luminosity of about 2\,000\,--\,300\,000 \Lsun, that evolve from stars with initial mass between $\sim$10\,--\,30\,\Msun\ and develop powerful stellar winds}
\entry {Isotopologues}{molecules which differ only in their isotopic composition}
\entry{Daedalus}{craftsman and artist in the Greek mythology, father of Icarus, having built the paradigmatic Labyrinth for King Minos of Crete, and seen as symbol of wisdom, knowledge, and power}
\entry{Gordian knot}{metaphor for an intractable problem}
\end{marginnote}

However, recent findings complicate this picture. 
As I will discuss, the \textit{conceptual} argument still remains valid, but its underlying reasoning gets sharpened, while essential aspects deeply grounded in the \textit{deterministic} question will encounter a reformulation. 
Ground-breaking observations, theoretical insights, numerical simulations, and laboratory experiments bear ample evidence that our notion of spherical stars and winds was oversimplified. 
The cool ageing stars, that are the focus of this review, have an incredibly fascinating life and harbour 3D complexities that bridge 23 orders of magnitude in scale, from the nanometer up to thousands of astronomical units. 
The 3D life of molecules and solid-state aggregates, the emergence of 3D clumps, and the lasting 3D impact of a (hidden) companion 
offer a challenge against which our physical knowledge and chemical understanding need to be reevaluated. 
Just as Daedalus, these stars are a symbol of wisdom, knowledge, and power; fortunately the challenge that they pose is not a Gordian knot,
but can be taken up successfully through an intensive collaboration with and access to modern observatories, state-of-the-art theoretical models, laboratory experiments, and  high performance computation (HPC) facilities. 
Only then, will it be possible to better quantify the \textit{deterministic} aspects of cool ageing stars in the cosmological context, and to have a more coherent picture of these stars for the \textit{conceptual} framework.

\section{Setting the stage: the 1D world}\label{Sec:Setting}

Quantifying the contribution of the cool evolved AGB and RSG stars to the chemical enrichment of our Universe implies that we need to know the \textit{absolute} rate by which these stars eject matter into the ISM --- the mass-loss rate  as a function of time, \Mdot$(t)$\ --- and the \textit{relative} fraction of atoms, molecules, and solid dust species in their stellar wind as functions of radial distance $r$ and time $t$. 
It was not until the mid seventies, that estimates of the mass-loss rate were within reach and Goldreich and Scoville presented 
the first detailed physical/chemical model of a circumstellar envelope (CSE) after the initial detections of circumstellar molecular line emission 
a few years earlier \citep{Goldreich1976ApJ...205..144G}. 
Some brief historical reminders of this enlightening period are presented in Section~\ref{Sec:history}. 
Beginning with  their seminal work, I introduce the reader to some basic theoretical ingredients of stellar wind physics in Section~\ref{Sec:Setting_theory}. 
In that section I mainly focus on the one crucial parameter of stellar evolution, which I foresee will experience major improvements in the next couple of years: the mass-loss rate during the end-phases of stellar evolution. 
Granted,  our understanding of the \textit{relative} contributions of atoms, molecules, and dust grains will see some major breakthroughs, 
but for the  reasons I will describe below, I think that a quantitative understanding of these will take at least a decade.
\begin{marginnote}[]
\entry{Circumstellar envelope}{the region surrounding the star, encompassing the extended atmosphere, the stellar wind and the bow shock; see \textbf{Figure~\ref{Fig:drawing}}}
\end{marginnote}

\begin{figure}[htp]
\includegraphics[width=.95\textwidth]{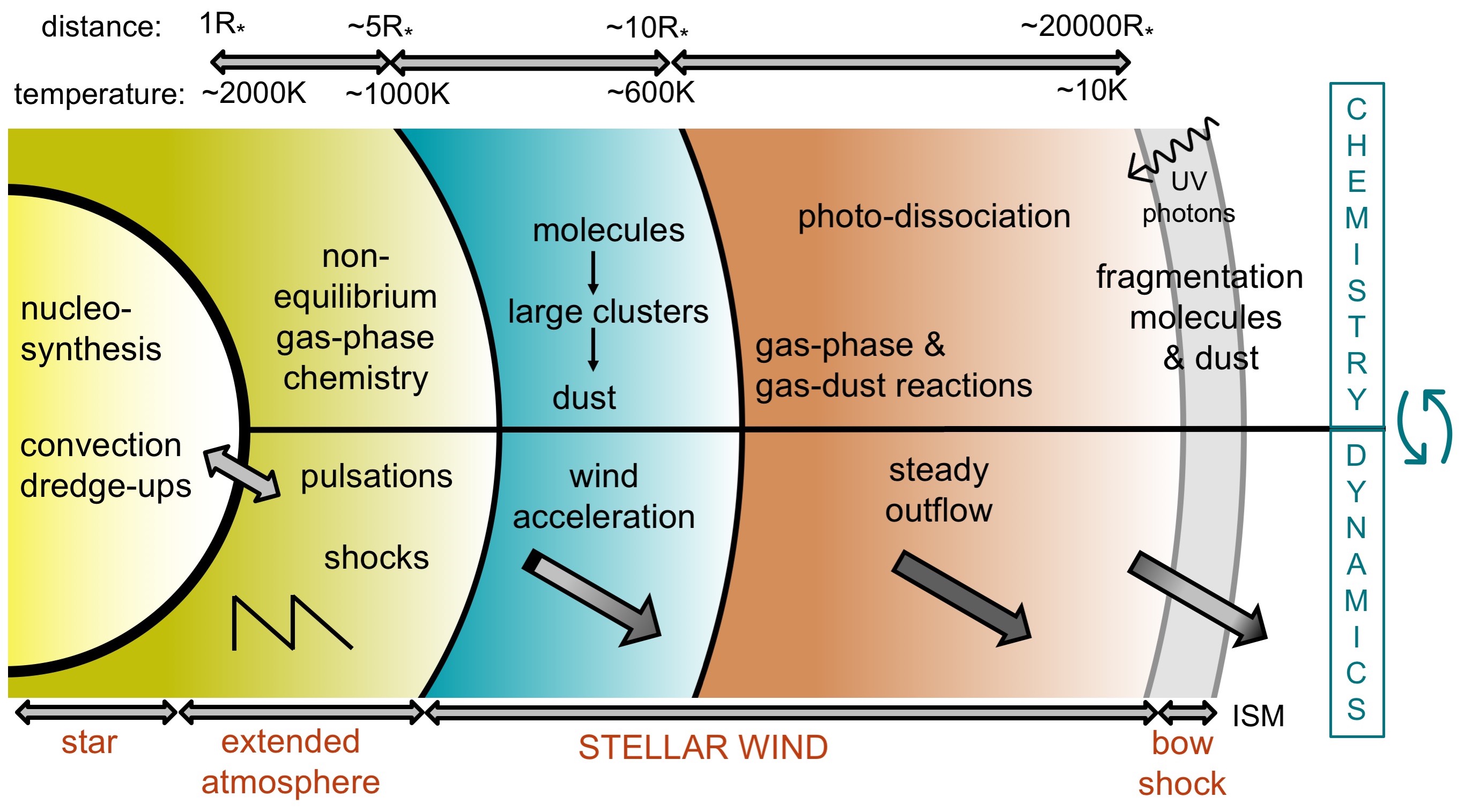}
\caption{Schematic  representation of an evolved AGB star and its circumstellar environment assuming spherical symmetry. The upper part illustrates the dominant chemical processes on length scales of nanometers to micrometers. The bottom part delineates the dynamical behaviour acting on scales of astronomical units (au). Physical and chemical processes are closely interrelated, including an intense interaction with the radiation field. 
Typical radial distances and temperatures are indicated at the top of the figure, where the stellar radius, \Rstar, is of the order of $\sim$1\,au.}
\label{Fig:drawing}
\end{figure}

\vspace*{-1ex}
\subsection{Bits of history} \label{Sec:history}

The first theoretical study of dust condensation in cool stars was performed by \citet{Wildt1933ZA......6..345W}. Using thermochemical calculations, Wildt considered the possibility that dust may contribute to the opacity in cool stellar atmospheres and showed that some very refractory solid compounds could be formed. 
\begin{marginnote}[]
	\entry{Refractory material}{material resistant to decomposition by heat, pressure, or chemical attack, having a high melting point and maintaining its structural properties at high temperature}
	\entry{Graphite}{crystalline form of the element carbon with the atoms arranged in a hexagonal structure}
\end{marginnote}
\hspace*{-4ex} But it took three decades until the far-reaching consequences of this work were realised. Motivated by the problem of identifying the origin of interstellar dust, \citet{Hoyle1962MNRAS.124..417H} suggested that graphite grains can condense in carbon-rich cool stars and can then be driven out by radiation pressure. (See the sidebar titled Carbon- and oxygen-rich cool stars.) Meanwhile, \citet{Deutsch1956ApJ...123..210D} provided the first evidence that matter is escaping from the RSG star $\alpha$~Her: matter is flowing beyond the orbit of its G star companion at a speed of $\sim$10\,km\,s$^{-1}$, well above the escape velocity at that distance. The estimated mass-loss rate was around $3\times10^{-8}$ solar masses per year (\Msun\,yr$^{-1}$). Parker described and solved the momentum equation (see Section~\ref{Sec:Setting_theory}) for the solar wind in 1958  and introduced the term `stellar wind' in 1960 \citep{Parker1958ApJ...128..664P, Parker1960ApJ...132..821P}. Meanwhile, the ejection of matter was also detected for other red giants, such as the RSG star Betelgeuse \citep{Weymann1962ApJ...136..844W}. 
The difficulty was explaining the apparently constant outflow of matter beyond 10 stellar radii with a speed of $\sim$10\,km\,s$^{-1}$, which is less than
the escape velocity from the star \citep{Weymann1962ApJ...136..476W}. \citet{Wickramasinghe1966ApJ...146..590W} were the first to propose that 
radiation pressure on grains can push the grains and the gas out of the stellar gravitational potential owing to momentum exchange between the
dust and the gas. 

\begin{textbox}[htp]\section{Carbon- and oxygen-rich cool stars}
	For stars in the early AGB or RSG phase, the chemical abundances reflect the chemical composition of the matter from which the star was formed. The galactic carbon-to-oxygen abundance ratio, C/O, is generally lower than 1, with C/O\,$\sim$\,0.56 in the solar vicinity and attaining lower values for lower metallicities \citep{Chiappini2003MNRAS.339...63C, Akerman2004A&A...414..931A}. 
	This implies that at the start of the AGB or RSG phase, stars are still oxygen-rich (O-rich, C/O$<$1) and are categorized as M-type stars. 
	During the AGB and RSG phase, carbon is fused in the stellar core owing to the triple-$\alpha$ process and is brought to the surface by convection. 
	For AGB stars with an initial mass between $\sim$1.5\,--\,4\,\Msun\ \citep{Straniero1997ApJ...478..332S}, the C/O ratio eventually  becomes larger than 1, leading to a carbon star. 
	Due to the exceptionally high C---O bond dissociation energy in CO, the less abundant of the two atoms (C or O) is completely 
	bound in CO and cannot partake in the formation of solids \citep{Gilman1969ApJ...155L.185G} and other molecules. 
	However, recent observations have challenged this idea: molecules such as CO$_2$, CS, and HCN are detected in O-rich winds, 
	and H$_2$O and SiO are detected in C-rich winds (see Section~\ref{Sec:molgrains_observations}).
\end{textbox}

Direct evidence of late-type stars with wind mass-loss rates well above a few $10^{-8}$\,\Msun\,yr$^{-1}$ came to light at the end of the 1960s, but before that several indirect arguments were put forward to prove that stars must lose a significant fraction of their mass during the final evolutionary phases. 
One such argument is due to \citet{Auer1965ApJ...142..182A} who found that the Hyades cluster contains about a dozen white dwarfs; each should have a mass below the Chandrasekhar limit of 1.4\,\Msun. 
But the Hyades cluster is a young group and stars with a mass of 2\,\Msun\ are still on the main sequence. 
Hence, the white dwarf progenitors must have lost at least 0.6\,\Msun\ during the post-main sequence phase, however they had not been observed  at that time.

\begin{marginnote}[3.6cm]
\entry{Late-type star}{terminology used to indicate stars that are cool, here of spectral types K and M}
\end{marginnote}
Ultimately, these mass-losing stars were detected in the late 1960s with the birth of infrared (IR) astronomy. Late M-type red giants were shown to often have an excess emission in the infrared, an effect that was attributed to circumstellar dust. In 1968/1969 \citet{Gillett1968ApJ...154..677G} and \citet{Woolf1969ApJ...155L.181W} identified  silicate grains in oxygen-rich AGB and RSG stars.  \citet{Gehrz1971ApJ...165..285G} derived the dust mass around a number of M-type stars, and using the expansion velocity from Deutsch,  mass-loss rates between $10^{-7}$\,--\,10$^{-5}$\,\Msun\,yr$^{-1}$ were obtained. 
The first circumstellar molecular rotational transition detected at radio wavelengths was the OH maser line at 1612\,MHz toward the RSG NML~Cyg \citep{Wilson1968Sci...161..778W}; the first thermally excited line was the CO v=0 J=1-0 transition detected a few years later toward the carbon star CW~Leo \citep{Solomon1971ApJ...163L..53S}. 
\begin{marginnote}[1.5cm]
	\entry{Maser}{Microwave Amplification by Stimulated Emission of Radiation, typically visible in the micrometer and radio wavelength domain. OH, the hydroxyl radical, is the first astronomical maser ever discovered \citep{Weaver1965Natur.208...29W}}
		\entry{Planetary nebula}{short ($\sim\!10^4$\,yr) evolutionary phase between the AGB and white dwarf phase; characterized by a hot central star that ionizes the gas ejected during the previous giant phase}
\end{marginnote}

\begin{textbox}[htp]\section{Superwind}
	The mass of the convective envelope decreases in time owing to both the stellar wind  and nuclear burning ($4\,^1$H\,$\rightarrow\,1\,^4$He), the latter effect resulting in an increase in core mass and hence luminosity \citep[the Paczy{\'n}ski-relation;][]{Paczynski1970AcA....20...47P}. 
	The difference in mass between the four fusing hydrogen nuclei and the newly created helium nucleus is converted to energy according to Einstein's equation $E\!=\!m c^2$ \citep{Einstein1905AnP...323..639E}. The energy production per gram by H-burning, $E_H$, is 6.45$\times$10$^{18}$\,erg/g. It follows that the nuclear burning rate, \Mdot$_c$, is given by \Mdot$_c$\,=\,\Lstar/$E_H$\,=\,1.02$\times$10$^{-11}$\,\Lstar/\Lsun\ (in \Msun\,yr$^{-1}$). 
	For wind mass-loss rates above  the nuclear burning rate, the associated timescale for stars to shed their envelope by a stellar wind is shorter than the nuclear burning timescale, such that mass loss determines the further evolution. Some authors denote this transition, occurring at a wind mass-loss rate of a few 10$^{-7}$\,\Msun\,yr$^{-1}$ (see \textbf{Figure \ref{Fig:Mdots}} in Section \ref{Sec:Mdots_theor_emp}), as the \textit{superwind} phase \citep{Lagadec2008MNRAS.390L..59L, Zijlstra2009ASPC..412...65Z}. However, we opt here to use the historical terminology of the word \textit{superwind}, as first expressed by \citet{Renzini1981ASSL...88..431R}, to indicate a mass-loss rate which greatly exceeds that prescribed by Reimers' law \citep[see Eq.~\eqref{Eq:Mdot_Reimers};][]{Reimers1975MSRSL...8..369R}. We recently argued that the maximum mass-loss rate during the \textit{superwind} phase is a few 10$^{-5}$\,\Msun\,yr$^{-1}$, and hence is around the single-scattering radiation pressure limit, indicating that the ratio of the wind momentum per second, \Mdot $v_\infty$, to the photospheric radiation momentum, $L_\star/c$, is around 1 \citep{Decin2019NatAs...3..408D}.
\end{textbox}

Using observations of similar RSG binary systems as Deutsch, \citet{Reimers1975MSRSL...8..369R} was the first to derive an empirical mass-loss rate relation of the form  
\begin{equation}
	\Mdot\,=\,4\times10^{-13} \eta L/g R\,,
	\label{Eq:Mdot_Reimers}
\end{equation}
 with \Mdot\ the mass-loss rate in units of \Msun\,yr$^{-1}$, $\eta$ a unitless parameter of the order of unity, and the stellar luminosity $L$, gravity $g$, and radius $R$ in solar units. For $\eta\sim1$, the AGB lifetime is of the order of one million years, and the maximum AGB mass-loss rate is a few $10^{-6}$\,\Msun\,yr$^{-1}$ \citep{Renzini1981ASSL...88..431R}. 
Renzini argued that a Reimers-like wind cannot explain the characteristics of planetary nebulae --- the descendants of the AGB stars --- and he suggested the existence of a \textit{superwind} developing at the high luminosity tip of the AGB phase and  with mass-loss rate of at least a few $10^{-5}$\,\Msun\,yr$^{-1}$ \citep{Renzini1981ASSL...88..431R}. 
(See the sidebar titled Superwind.) In the same year, \citet{Glass1981Natur.291..303G} established the first linear relation between the $K$-band magnitude and the logarithm of the period in regularly pulsating Mira-type AGB variables. (See the sidebar titled Variability and pulsation modes.) Pulsations are thought to be an essential ingredient for the wind driving in AGB stars: matter is levitated by shock waves induced by pulsations resulting in densities that are high enough at a few stellar radii for dust to condense and in sufficient momentum coupling between the gas and the grains. Observations indicate that the wind mass-loss rate ranges between $\sim\!10^{-8}$\,--\,$10^{-4}$\,\Msun\,yr$^{-1}$ and the expansion velocity between $\sim$5\,--\,30\,km\,s$^{-1}$.
The discussions that took place during that period had a considerable impact on the field of stellar evolution modelling: in all calculations before, say, 1980 the assumption was made that the mass of a star did not change either by mass loss or mass accretion. 
Because the mass is the prime parameter determining the evolution and lifetime of a star, any modification to the stellar mass over time has large repercussions on its evolutionary path. A proper understanding of stellar evolution can thus not be achieved without a detailed understanding of wind mass-loss rates, and hence wind physics.

\begin{textbox}[h!]\section{Variability and pulsation modes}
Variability in brightness is a common feature of AGB stars and is mainly caused by pulsations. 
The classification of pulsating AGB stars into Miras, Semiregulars (SR) and Irregulars was originally based solely on the appearance of light curves, without an understanding of the physical process at work. Mira variables have regular, large amplitude variations (variation in the visible light $\delta V\!>$2.5\,mag); semiregular variables are of smaller amplitude $\delta V\!<$2.5\,mag with some periodicity; and irregular variables show little periodicity although this is often due to a lack of detailed light curves. It turned out to be possible to trace the properties of variable stars through the period-luminosity ($P-L$) diagram, in which stars form distinct sequences depending on the pulsation mode responsible for their variability \citep{Wood1999IAUS..191..151W}; see Fig.~1 in \citet{McDonald2019MNRAS.484.4678M}. Pulsating stars are often multiperiodic, and normally only the period with the largest amplitude is used in the $P-L$ diagram.  Mira variables are generally located on sequence C, which is due to pulsations in the fundamental mode. Sequences B and C$^\prime$ are due to pulsation in the first overtone mode, and sequences A and A$^\prime$ to pulsation in the second and third overtone modes, respectively. Sequences D and E are, respectively, due to long secondary periods and binary stars \citep{Wood2015MNRAS.448.3829W}. The semiregular variables occupy sequence A and B and the lower half of sequence C.  

The energy transport, which determines the stability and growth rates of pulsation, is dominated by convection. 
Excitation of the pulsation modes in  linear, non-adiabatic 1D models occurs through the H and first He ionization zones (the $\kappa$-mechanism). 
Most of the layers below the top of the H ionization zone with a temperature of $\sim$8\,000\,K contribute to the determination of the
period of the fundamental mode, whereas all layers (including the surface layers) contribute in determining the higher overtones modes \citep{Fox1982ApJ...259..198F}.
As a star evolves on the AGB, it rises on the $P-L$ diagram and traverses the $P-L$ sequences from left to right: specific overtone modes gradually become stable, and the primary mode shifts towards lower radial orders \citep[][see \textbf{Figure~\ref{Fig:modes}}]{Trabucchi2019MNRAS.482..929T}. 
\end{textbox}

\begin{marginnote}[]
	\entry{Non-adiabatic}{occurring with loss or gain of heat or mass between the thermodynamic system and its surroundings}
\end{marginnote}

	\begin{figure}[htp]
	\includegraphics[width=\textwidth]{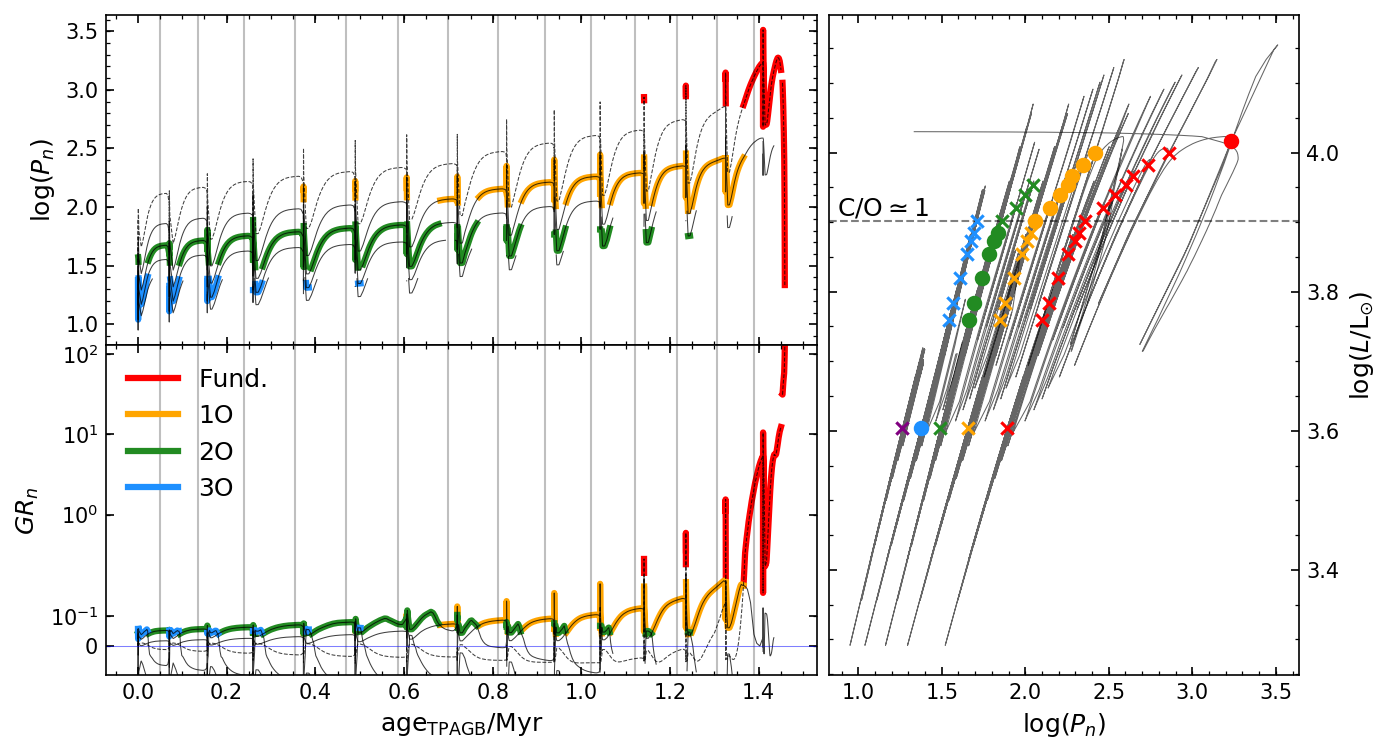}
	\caption{Pulsation periods, growth-rates, and period-luminosity diagram for a star evolving along the AGB phase.
	    A linear, non-adiabatic 1D pulsation model is applied to an evolutionary track for  a star with mass $M\!=\!2.6\,{\rm M}_{\odot}$ and metallicity $Z$\,=\,0.008 at the beginning of the AGB phase.
		Left-hand panels: pulsation periods (top; in units of days) and growth rates (bottom) as a function of time elapsed since the beginning of thermally pulsing AGB (TP-AGB) phase. 
		Radial pulsation modes are identified by their radial order $n$, with $n$\,=\,0 corresponding to the fundamental mode, $n$\,=\,1 to the first overtone (1O) mode, $n$\,=\,2 to the second overtone (2O) mode, and so on.
		For each mode $n$ with eigenfrequency $\omega_n\!=\!\omega_{R,n} + {\rm{i}} \omega_{I,n}$, the imaginary part of $\omega$ is the angular frequency of oscillation and
		the period of the $n$th mode is defined as $P_n = 2\pi/\omega_{I,n}$ with 
		the amplitude growth rate  calculated as $G\!R_n\!=\!\exp\left(2 \pi \frac{\omega_{R,n}}{\omega_{I,n}}\right) -1$.
		Dominant modes are highlighted by thick solid lines in colours. Modes other than the dominant are shown as thin solid lines, except for the fundamental mode which is shown as dashed thin lines to be more easily distinguishable. 
		Vertical lines mark the point of maximum luminosity of quiescent evolution at each thermal pulse cycle.  
		As time elapses, the dominant mode gradually shifts towards lower radial orders.
		Right-hand panel: theoretical period-luminosity diagram. Symbols correspond to quiescent evolutionary points, with the dominant mode represented by a filled circle.
		The quiescent evolutionary points bend towards longer periods as the luminosity increases, especially for the fundamental mode, owing to the effect of mass loss.
		This bending is emphasized by the models transitioning to C-rich --- indicated by the horizontal grey dashed line --- which causes an increase of radius with respect to O-rich models at the same luminosity.
		Image reproduced with permission from \citet{Trabucchi2019MNRAS.482..929T}.}
	\label{Fig:modes}
\end{figure}

\subsection{Stellar wind physics}\label{Sec:Setting_theory}

A central goal of stellar wind research is to derive a relation between the mass-loss rate and fundamental stellar parameters, such as the Reimers' relation (see Eq.~\eqref{Eq:Mdot_Reimers}). 
Here I wish to address a fundamental issue in the inductive method used to derive that relation --- i.e., the issue of \textit{forward} versus \textit{retrieval} approaches. 
As will become clear, the \textit{forward} method tends more toward the \textit{reductionist} approach in the sense that one tries to understand phenomena in terms of the interaction of the constituent parts, while the \textit{retrieval} method is more inclined toward the detection of (unexpected) emergent properties and hence can be argued to be more \textit{holistic} in its approach, i.e.,\ `the whole is more than the sum of its parts'. Both approaches are not mutually exclusive, rather they inform each other. 
Let me briefly describe these approaches, both in their benefits and their apparent shortcomings.

\begin{marginnote}[-4truecm]
	\entry{Radial pulsation}{occurs when a star oscillates around the equilibrium state by changing its radius symmetrically over the whole surface}
	\entry{Thermal pulse}{caused by a helium shell flash, occurring over periods of 10\,000 to 100\,000 years, lasting only a few hundred years}
\end{marginnote}

The \textit{forward} approach is more mathematically oriented in that one seeks to describe all explicit and implicit relations between the quantities involved in a mathematically and physically \textit{consistent} way. A \textit{self-consistent} approach is even more restrictive and implies that all explicit input functions are the result of solving the system of basic equations of the problem, without introducing ad hoc assumptions. 
Resorting to these \textit{theoretically} predicted mass-loss rates has the advantage of being able to study  the explicit dependence on 
individual input parameters. 
However, the predictive power of any theoretical model is dictated by the level of description of the physical and chemical processes, and their interaction.
 In general, any realistic model should account for the thermodynamics, the hydrodynamics, the radiative transfer, and the chemistry including gas-phase and solid-state species; see \textbf{Figure~\ref{Fig:Jels}}. 
Their combined action determines the local and global physical and chemical properties of the wind, and hence the mass-loss rate. 
Various interactions shown in \textbf{Figure~\ref{Fig:Jels}} are highly non-linear both with regard to the chemical and physical description, and the mathematical and numerical treatment. 
The \textit{forward} approach can be highly demanding for the central processing unit (CPU), and hence a bottleneck for analysing large samples of observational data.

\begin{figure}[htp]
\includegraphics[width=\textwidth]{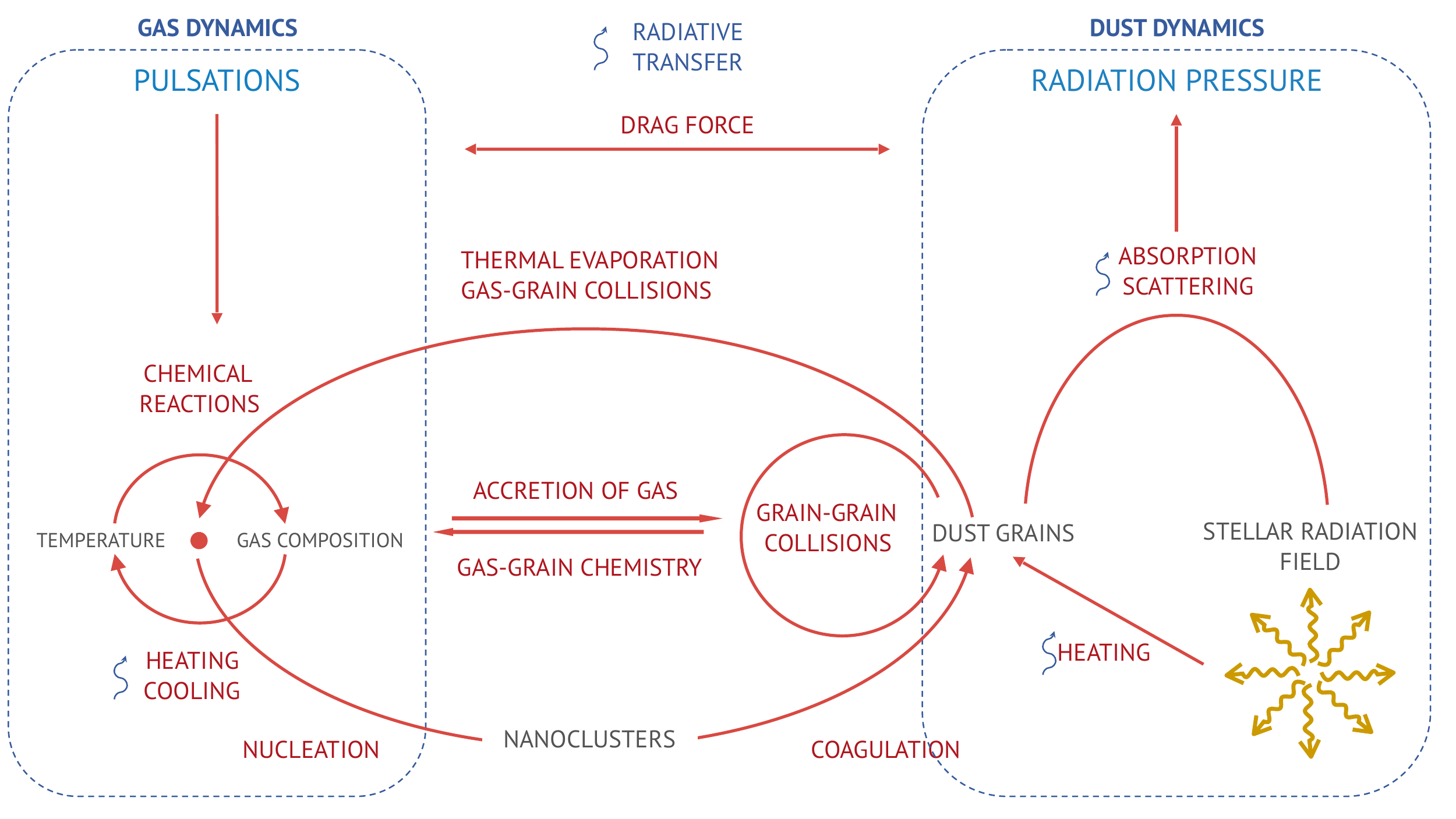}
\caption{Schematic overview of the physical and chemical processes occurring in AGB winds. The two main driving forces
are the pulsations of the star and the radiation pressure on dust grains. Both are self-consistently connected via the mechanisms
depicted in the figure. 
Note that processes annotated with a curly blue arrow require the inclusion of radiative transfer calculations. 
Indicated in navy blue are wind driving forces, in black are physical and chemical quantities, in
red are interactions, and in dark blue are different physics categories (dynamics and radiative transfer).  
This Figure is adapted from \citet{BoulangierPhD}.}
\label{Fig:Jels}
\end{figure}

To model observed quantities, one often resorts to \textit{retrieval} modelling. 
\textit{Retrieval} modelling implies that one prescribes externally parameters  that are basically internal parameters. 
These (often simplifying) parametrizations are informed by the outcome of detailed \textit{forward} modelling, or by an astrophysical hypothesis
informed by specific observations and their analyses. 
As I will demonstrate in Section~\ref{Sec:model_outcomes}, the solution to the momentum equation is often simplified by applying the $\beta$-velocity power law, 
and the solution to the energy equation by a temperature power law. 
The advent of new ground-based and space-borne missions has led to considerable progress in \textit{empirically} derived mass-loss rate relations 
based on the use of  \textit{retrieval} methods. However, it turns out that these relations --- which sometimes show dependencies on different fundamental parameters --- are not always mutually consistent; see Section~\ref{Sec:Mdots_theor_emp}. The reasons for this discrepancy can be traced back to difficulties in determining highly accurate fundamental stellar parameters of AGB and RSG stars and the close entanglement of various of these parameters (such as luminosity, mass, age, and pulsation period). In addition, the parametrizations inherent in the \textit{retrieval} method can yield a systematic bias in the derived mass-loss rates, systematic selection effects on observed samples might induce an unrecognised bias, and it is well established that correlation does not imply causality. I will deal with some of these threats shortly.

The \textit{reductionist} approach allows one to better demonstrate the various ingredients of stellar wind physics and chemistry. 
I therefore refer to that method in Section~\ref{Sec:standard_CSE}.
As illustrated in \textbf{Figure~\ref{Fig:drawing}}, the description of the CSE can be divided into three regions, and research groups tend to focus on the detailed description of one of them: (1)~the \textit{extended atmosphere} in which pulsation-induced shocks result in a chemistry that is not in thermodynamic equilibrium \citep[e.g.][]{Willacy1998A&A...330..676W, Cherchneff2006A&A...456.1001C, Gobrecht2016A&A...585A...6G, Hofner2018A&ARv..26....1H, Bladh2019A&A...626A.100B}, (2)~the \textit{wind formation zone} in which radiation pressure on newly formed dust grain leads to the onset of the stellar wind, hence initiating the mass loss \citep[e.g.][] {Dominik1993A&A...277..578D, Gail1999A&A...347..594G, Hofner2018A&ARv..26....1H, Bladh2019A&A...626A.100B}, and (3)~the \textit{steady outflow zone} in which the wind is freely expanding and is interacting with the surrounding interstellar radiation field resulting in photo-dissociation of molecules and further modification of the grain spectrum \citep[e.g.][]{Willacy1997A&A...324..237W, PatzerPhD, Glassgold1999IAUS..191..337G, Agundez2010ApJ...724L.133A, Li2014A&A...568A.111L, VandeSande2018A&A...616A.106V}. While time dependency is inherent in the description of region~1, the other two regions are often modelled using a stationary approach.

\subsubsection{Standard CSE model} \label{Sec:standard_CSE}

For our discussion, it is sufficient to  remind the reader of the general conservation laws describing the stellar wind structure under the assumption of (i)~stationarity, because it allows us to obtain an insight in the complex interplay between various physical and chemical processes, 
and (ii)~spherical symmetry, because this is suggested by various observations of extended circumstellar envelopes.
In this situation, the hydrodynamics as expressed in the equations of mass and momentum conservation can be written as \citep[e.g.][]{Goldreich1976ApJ...205..144G}
\begin{equation}
  \frac{{\rm{d}}M(r)}{{\rm{d}}t} = \Mdot(r) = 4 \pi r^2 \rho(r) v(r) \,,
  \label{Eq:Mdot}
\end{equation}
\begin{equation}
  v(r) \frac{{\rm{d}}v(r)}{{\rm{d}}r} = (\Gamma(r) -1) \frac{G M_{\star}}{r^2}\,,
\label{velstructure}
\end{equation}
where \Mdot$(r)$ refers to the mass-loss rate of the gas at a radial distance
$r$ from the star, $\rho(r)$ is the gas density, $v(r)$ the gas velocity,
$M_{\star}$ the stellar mass, $G$ the gravitational constant, and $\Gamma(r)$ the ratio of the radiation 
pressure force on the dust to the gravitational force which can be written as 
\citep{Decin2006A&A...456..549D}
\begin{equation}
  \Gamma(r) = \frac {3 v(r)}{16 \pi \rho_s c G M_{\star}
    \dot{M}(r)}\int\!\!\!{\int{\frac{Q_{\lambda}(a,r) L_{\lambda} \dot{M}_d(a,r)}{a
        [v(r)+v_{\rm{drift}}(a,r)]}}} \,{\rm{d}}\lambda\,{\rm{d}}a \,,
\label{Eq:Gamma}
\end{equation}
with $\rho_s$ the specific density of dust, $c$ the speed of light, $\dot{M}_d$  mass-loss rate of the dust, $v_{\rm{drift}}(a,r)$ the drift velocity of a grain of size $a$, $Q_{\lambda}(a,r)$ the dust extinction efficiency, and $L_{\lambda}$ the monochromatic stellar luminosity 
at wavelength $\lambda$. 
\begin{marginnote}[]
	\entry{Spherical symmetry}
	{\includegraphics[width=2.8cm]{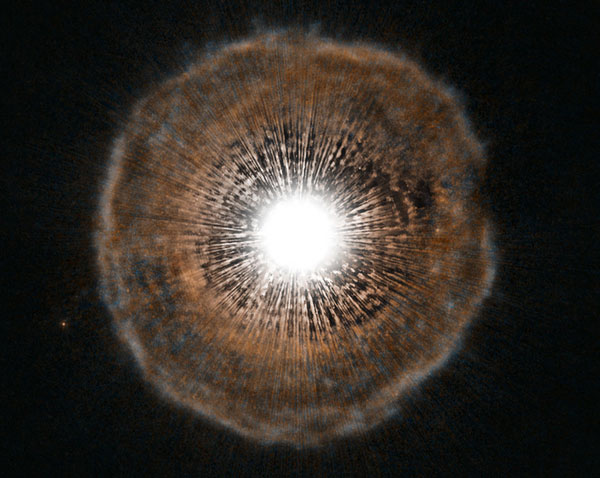}
		The spherical shell around the AGB star U Cam. Credit: ESA\,/\,Hubble\,/\,NASA /\,H.~Olofsson}
	\entry{Drift velocity}{difference between the dust and gas velocity}
\end{marginnote}

From the first law of thermodynamics expressing the conservation of energy, the perfect gas law, and the equation of mass conservation, the thermal structure of the gas is governed by the relation \citep{Goldreich1976ApJ...205..144G}
\begin{equation}
\frac{1}{T(r)} \frac{{\rm{d}}T(r)}{{\rm{d}}r} = - \frac{4}{3r} \left(1 + \frac{1}{2} \frac{{\rm{d}} \ln v(r)}{{\rm{d}} \ln r} \right) + \frac{2}{3} \frac{H(r)-C(r)}{k \,T(r) \, v(r) \, n_{\rm{H_2}}(r) \, (1 + f_{\rm{H}}(r))}
   \label{Eq:Tgas}
\end{equation}
with $H(r)$ and $C(r)$ the total heating and cooling rate per unit volume, respectively, $n_{\rm{H_2}}(r)$ the number density of H$_2$,
and $f_{\rm{H}}(r)$ the number fraction of atomic to molecular hydrogen. 
The first term on the right hand side of Eq.~(\ref{Eq:Tgas}) represents the cooling due to adiabatic expansion in case of constant mass loss. 
The second term, which represents the balance of the heating and collision-driven radiative cooling processes, needs a proper treatment of the radiative transfer and the radial abundance profile of molecules such as H$_2$, H$_2$O, CO and HCN and of the grains to calculate $C(r)$ and $H(r)$; 
see for example \citet{Decin2010A&A...516A..69D}.

The balance of radiative energy gain and radiative energy loss is used to calculate the temperature of an individual grain of dust, $T_d$, 
\begin{equation}
\int_0^\infty \sigma_{d,\nu}^{\rm{abs}} \left[J_\nu(r) - B_\nu(T_d(a,r))\right] {\rm{d}}\nu =0\,,
\label{Eq:Tdust}
\end{equation}
with $\sigma_{d,\nu}^{\rm{abs}}$ the absorption opacity of the dust grain at frequency $\nu$, 
$J_\nu(r)$ the local mean radiation intensity, and $B_\nu$ the Planck function. 
At temperatures higher than the condensation temperature, the grain will sublimate.

For the purpose of radiation hydrodynamics, the treatment of the radiative transfer is often simplified. A well-known method has been proposed by \citet{Mihalas1974ApJS...28..343M} and is based on the zero- and first-order moment transport equations
\begin{eqnarray}
\frac{{\rm{d}}H_\nu(\tau_\nu)}{{\rm{d}}\tau_\nu} & = & J_\nu(\tau_\nu) - S_\nu(\tau_\nu) \label{Eq:zerothmoment}\\
\frac{{\rm{d}}K_\nu(\tau_\nu)}{{\rm{d}}\tau_\nu} & = & H_\nu(\tau_\nu) \label{Eq:firstmoment}\,,
\end{eqnarray}
with $\tau_\nu$ the optical depth, $S_\nu$ the source function, $H_\nu$ the flux, and $K_\nu$ being related to the radiation pressure $p_\nu = (4\pi/c)K_\nu$. 
Eqs.~\eqref{Eq:zerothmoment}\,--\,\eqref{Eq:firstmoment} is a system of coupled integro-differential equations that can be mathematically closed by defining the 
Eddington factor
\begin{equation}
f_\nu(r) = \frac{K_\nu(r)}{J_\nu(r)}\,,
\end{equation}
with $f_\nu(r)$ approaching 1/3 for an isotropic radiation field.

The chemical evolution of the composition of a closed system is dictated by a set of chemical formation and destruction reactions. Mathematically, this is a set of coupled ordinary differential equations where the change in number density of the $i$th species is given by 
\begin{equation}
\frac{{\rm{d}}n_i}{{\rm{d}}t} = \sum_{j \in F_i} \left(k_j \prod_{r\in R_j} n_r \right) - \sum_{j \in D_i} \left( k_j \prod_{r \in R_j} n_r \right)\,.
\label{Eq:balance}
\end{equation}
The first term, within the summation, represents the rate of formation of the $i$th species by a single reaction $j$ of a set of formation reactions $F_i$. The second term is the analogue for a set of destruction reactions $D_i$. 
Each reaction $j$ has a set of reactants $R_j$, where $n_r$ is the number density of each reactant and $k_j$ the rate coefficient of this reaction.   For chemistry in thermodynamic equilibrium, Eq.~\eqref{Eq:balance}, involving both gaseous and dust species, reduces to the well-known law of mass action \citep{Gail2013pccd.book.....G}
\begin{equation}
\prod_{\shortstack{i \\ {\rm{all\ gases}}}} \left(\frac{p_i}{p_0}\right)^{\nu_i} \prod_{\shortstack{i \\ {\rm{all solids}}}} \left(a_i^c\right)^{\nu_i} = {\rm e}^{-\Delta G/RT}\,,
\label{Eq:mass_action}
\end{equation}
with $\nu_i$ the stoichiometric coefficients, the activity $a^c_i$ of species $i$ defined as $a^c_i=p_i/p$, $p$ the pressure, $p_i$ the partial pressure of species $i$, $p_0$ the standard pressure of 1 bar, $R$ the gas constant in units as used for the data of $\Delta G$,
and $G$ the Gibbs function so that 
\begin{equation}
\Delta G = \sum_i \nu_i G_i(p_0,T)\,,
\label{Eq:Gibbs}
\end{equation}
with $G_i(p,T)$ the partial free enthalpy of 1 mole of species $i$ at temperature $T$ and pressure $p$. Chemical equilibrium is established if the chemical reaction time scales are small compared with other competing time scales governing the considered concentrations, so that Eq.~\eqref{Eq:mass_action} is only dependent on temperature.
\begin{marginnote}[]
	\entry{Stoichiometric coefficient}{number written in front of each compound in a chemical reaction to balance the number of each element on both the reactant and product sides of the equation}
	\end{marginnote}

Together with suitable boundary and initial conditions, the resulting mathematical system $\mathcal{M}$ described in 
Eqs.~\eqref{Eq:Mdot}\,--\,\eqref{Eq:Gibbs} constitutes a complete and well-posed set of coupled equations given a set of independent fundamental stellar parameters. 
The solution provides a theoretical prediction of the physical and chemical quantities, including the mass-loss rate and the spectral appearance. 
By choosing the stellar mass, temperature, luminosity, and abundance composition \{$M_\star$, $T_\star$, $L_\star$, \{$\epsilon_X$\}\} as independent stellar parameters, we can express this formally as \citep{Gail2013pccd.book.....G}
\begin{equation}
\{M_\star, T_\star, L_\star, \{\epsilon_X\}\} \xRightarrow[]{\text{$\mathcal{M}$}} \Mdot\,.
\label{Eq:math_model_Mdot}
\end{equation}
Admittedly, the chemical abundances \{$\epsilon_X$\}\ are genuinely free parameters only for stars in the early AGB phase.
In principle the elemental abundances result from stellar evolution --- i.e., in particular nucleosynthesis and convection-induced dredge-up processes --- so 
the stellar mass and chemical abundances are not independent. 
However, the time scales involved with stellar evolution are much larger than those of the physical and chemical processes considered here. This implies that mass and chemical abundances only vary on secular time scales and can be considered as independent stellar parameters.

\subsubsection{Outcome of stationary 1D model predictions} \label{Sec:model_outcomes}

Given the assumption of stationarity, any model prediction applies for a genuinely \textit{dust-driven} wind; but see Section~\ref{Sec:limitations}. The radial structure for a low-mass carbon-rich star at the tip of the AGB phase is shown in \textbf{Figure~\ref{Fig:model}}. Efficient dust nucleation around 3 stellar radii enables a dust-driven wind. 
The gas pressure, and hence also the density $\rho$, decreases approximately exponentially near the photosphere, with $\rho \propto r^{-2}$ further out due to the condition of mass conservation. The gas expansion velocity, $v$, exceeds the local escape velocity, $v_{\rm{esc}}$, around 3 stellar radii and the wind becomes gravitationally unbound reaching a terminal wind velocity, $v_\infty$, of around 22\,km\,s$^{-1}$. In \textit{retrieval} modelling, this particular behaviour of the wind acceleration is often approximated 
 by the so-called $\beta$-type velocity law \citep{Lamers1999isw..book.....L} 
\begin{equation}
v(r) = v_0 + (v_\infty - v_0) \left( 1 - \frac{R_{\rm{dust}}}{r}\right)^\beta\,,
\label{Eq:velocity}
\end{equation}
\begin{marginnote}[5truecm]
	\entry{Dust condensation radius}{radial distance at which the first solid-state species are formed}
	\entry{Terminal wind velocity}{at large distance from the star, the velocity asymptotically approaches the terminal wind velocity}
\end{marginnote}
with $r$ the distance to the star and $v_0$ the velocity at the dust condensation radius $R_{\rm{dust}}$. 
Low values for $\beta$ describe a situation with a high wind acceleration. 
In the same vein the gas and dust temperature structure are often approximated by a power law
\begin{equation}
  T(r) = T_{\star} \left(\frac{R_{\star}}{r}\right)^{\zeta}
\end{equation}
with $\zeta \approx 0.5-0.6$.

\begin{figure}[htp]
\begin{center}
\includegraphics[width=\textwidth]{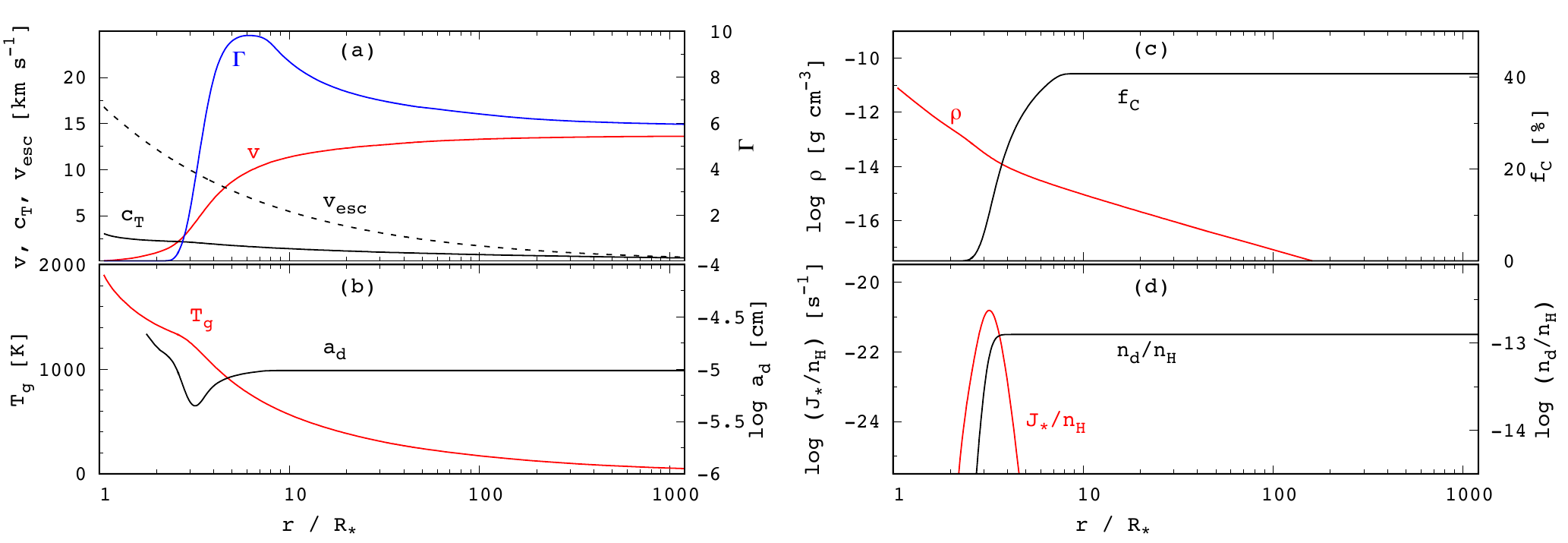}
\vspace*{-2ex}
\caption{Radial structure for a stationary dust-driven wind with independent fundamental parameters \Mstar\,=\,0.7\,\Msun, \Lstar\,=\,2.4$\times$10$^4$\,\Lsun, \Mdot\,=\,8$\times$10$^{-5}$\,\Msun\,yr$^{-1}$, \{$\epsilon_{X\not= C}$\}$_\odot$, $\epsilon_{\rm{C}}/\epsilon_{\rm{O}}$\,=\,1.4 and  assuming homogeneous nucleation with C$_1$ as basic monomer. (\textit{a})~gas velocity $v$, isothermal sound speed $c_T$, escape velocity $v_{\rm{esc}}$, and radiative acceleration in units of the gravitational acceleration $\Gamma$; (\textit{b})~radiative equilibrium gas temperature $T_g$ and dust grain size $a_d$; (\textit{c})~mass density $\rho$ and degree of condensation, that is the fraction $f_c$ of the amount of carbon condensate relative to the condensible carbon present in the gas phase; (\textit{d})~nucleation rate $J_\star$ and grain particle density $n_d$, both quantities normalized to the number density of hydrogen nuclei $n_{\rm{H}}$. The nucleation rate $J_\star$ corresponds to the formation rate of the gas-phase cluster that is the least abundant, and hence will dictate the dust formation rate; see Section~\ref{Sec:grains}. Figure courtesy H.-P.\ Gail.}
\label{Fig:model}
\end{center}
\end{figure}

Stationary wind models typically predict terminal wind velocities between 5\,--\,20\,km\,s$^{-1}$, and mass-loss rates between 
$\sim$5$\times$10$^{-8}$\,--\,3$\times$10$^{-4}$\,\Msun\,yr$^{-1}$ for carbon-rich winds, where the maximum mass-loss rate 
is a factor of a few lower for oxygen-rich winds \citep{DeBeck2010A&A...523A..18D, Gail2013pccd.book.....G, Decin2019NatAs...3..408D}. 
A higher luminosity and lower temperature induce higher mass-loss rates and expansion velocities due to the larger radiation pressure on the dust grains and the potential for more efficient dust nucleation and growth, respectively.
The mass-loss rate is very sensitive to the stellar mass via its effect on the gravity and hence $\Gamma(r)$: a reduction of the stellar mass 
by a factor of 2 leads to an increase of the mass-loss rate by a factor 3 to 100 \citep{Gail2013pccd.book.....G}. 
This effect might hence be a natural candidate to explain the \textit{superwind} scenario proposed by \citet{Renzini1981ASSL...88..431R}, 
since the AGB stellar mass will be reduced significantly by the preceding mass loss.

For RSG stars the role of grains close to the star remains unresolved, and radiation pressure on molecular lines, turbulent pressure, acoustic waves and Alfv\'en waves have been proposed as alternative mechanism \citep{Josselin2007A&A...469..671J, Bennett2010ASPC..425..181B, Scicluna2015A&A...584L..10S, Montarges2019MNRAS.485.2417M, Kee2020}. In general, these alternative processes might also \textit{support} the AGB stellar wind, although their role in \textit{driving} the wind is very much debated \citep{Wood1990ASPC...11..355W, Gustafsson2003agbs.conf..149G}.
\begin{marginnote}[]
	\entry{Turbulent pressure}{pressure caused by small-scale motions of stochastic nature; $P_{\rm{turb}}\!=\!\zeta \rho v_{\rm{turb}}^2$ with $v_{\rm{turb}}$ the turbulent velocity and $\zeta$ equal to 1 in the case of isotropic turbulence}
	\entry{Acoustic wave}{mechanical and longitudinal wave resulting from 3D fluctuations in the pressure field}
	\entry{Alfv\'en wave}{a transverse electromagnetic wave propagating along the magnetic field lines of a plasma and resulting from an interaction of the magnetic field and the electric currents within it \citep{Alfven1942Natur.150..405A} }
	\end{marginnote}

\subsubsection{Limitations of the stationary approach}\label{Sec:limitations}
The standard 1D CSE description (Section~\ref{Sec:standard_CSE}) is time independent, and hence does not treat the extended atmosphere in which pulsations induce shock waves that levitate the gas to distances where dust can form. 
It is generally accepted that pulsations are a key ingredient of AGB mass loss 
\citep[][See the sidebar titled Variability and mass-loss rate]{Bowen1988ApJ...329..299B, McDonald2016ApJ...823L..38M, McDonald2019MNRAS.484.4678M}, 
however the nature of the pulsations and their impact on the density scale height continues to be a source of debate.
The stellar interior  where  these  variations  originate is an  optically  thick  region dominated by convection that has proven difficult to model. 
Linear, non-adiabatic 1D pulsation models have been successful in predicting the overtone modes expected to occur 
in early-type AGB semiregular variables, but poor agreement is found for the fundamental mode Mira-type pulsators \citep{Trabucchi2017ApJ...847..139T, Trabucchi2019MNRAS.482..929T}. 


A global 3D radiation-hydrodynamics approach of the convection, and related pulsations, has recently been explored by \citet{Freytag2017A&A...600A.137F}. Due to computational constraints, the models only reach up to $\sim$2\,\Rstar.  Irregular structures with convection cells dominate in the interior part and propagating shocks in the outer atmosphere. The models develop radial and non-radial pulsations, but with a different frequency between the inner and outer part of the model. For models in which the radial fundamental mode dominates, the pulsation periods range between $\sim$300\,--\,630 days, in good agreement with observations for Mira-type variables. The exact mechanism for the mode excitation is, however, not yet fully understood, and possibilities such as stochastic excitation by convection, excitation by the $\kappa$-mechanism, and acoustic noise are explored by the authors. 
\begin{marginnote}[]
	\entry{Non-radial pulsation}{some parts of the stellar surface are moving inwards, while others move outwards at the same time}
\end{marginnote}

\begin{textbox}[htp]\section{Variability and mass-loss rates}
For Mira-type variable stars with a luminosity above $\sim$2\,000\,\Lsun\ and pulsation period $P$ between $\sim$300\,--\,800 days, a linear relation exists between the period and the logarithm of the mass-loss rate ($\sim$10$^{-7}$\,\Msun\,yr$^{-1}$\,$<$\,\Mdot\,$<$\,3$\times$10$^{-5}$\,\Msun\,yr$^{-1}$), suggesting that the corresponding increase in luminosity causes the radiation pressure on dust to be more effective. Semiregular variables with $P\!\la\!200$ days cover essentially the same mass-loss rate regime as the Mira variables with period between 200\,--\,400 days, while a maximum mass-loss rate of a few $10^{-5}$\,\Msun\,yr$^{-1}$ seems to be reached for $P\!\ga\!800$ days \citep{Vassiliadis1993ApJ...413..641V, DeBeck2010A&A...523A..18D}. Between $\sim$60 and $\sim$300 days, an approximately constant mass-loss rate of $\sim$3.7$\times$10$^{-7}$\,\Msun\,yr$^{-1}$  is found, while for $P\!<$\,60 days the mass-loss rate is a factor of $\sim$10 smaller. 
This rapid increase of mass-loss rate and dust production when the star first reaches a pulsation period of $\sim$60 days coincides approximately 
with the point when the star transitions to the first overtone pulsation mode, while the second rapid mass-loss-rate enhancement at 
$P\!\!\sim$300 days coincides with the transition to the fundamental pulsation mode. 
This indicates that stellar pulsations are the main trigger for the onset of the AGB mass loss and are significant in controlling the mass-loss rate \citep{McDonald2016ApJ...823L..38M, McDonald2019MNRAS.484.4678M}. 
\end{textbox}

Owing to their complex nature various researchers have treated the effects of pulsations in a 1D parametrized way following the \textit{piston} approach \citep{Bowen1988ApJ...329..299B, Gauger1990A&A...235..345G, Hofner1995A&A...297..815H}. 
These  wind  models typically have an  inner boundary  situated  just  below  the  photosphere of  the star where the radius and luminosity are assumed to have sinusoidal variations characterized by the pulsation period $P$ and velocity amplitude $\Delta v_p$. 
The pulsation period  can be  derived  from  the  period-luminosity ($P-L$) relation. The velocity amplitude, $\Delta v_p$, typically ranges between 2\,--\,4\,km\,s$^{-1}$, corresponding to shock amplitudes of $\sim$15\,--\,20\,km\,s$^{-1}$ in the inner atmosphere \citep{Bladh2019A&A...626A.100B}. This standard inner boundary condition is meant to describe the pulsation properties of Mira variables pulsating in the fundamental mode, but a similar simplified approach for semiregular variables has not yet been pursued. 
Formally, we can express the 1D \textit{piston} models as \citep{Gail2013pccd.book.....G}
\begin{equation}
\{M_\star, T_0, L_0, \{\epsilon_X\}, P, \Delta v_p\} \xRightarrow[]{\text{$\mathcal{M}$}} \Mdot\,,
\label{Eq:math_model_Mdot2}
\end{equation}
with $T_0$\,=\,\Tstar($t=0$) and $L_0$\,=\,\Lstar($t=0$). The predicted wind velocities and mass-loss rates show no significant differences compared to more complex models in which (the mean of) the dynamical properties predicted by the 3D radiation-hydrodynamic models are used as inner boundary condition for the dust-driven wind \citep{Liljegren2018A&A...619A..47L}.

Using the \textit{piston} approach, model grids for carbon- and oxygen-rich winds have been published by \citet{Arndt1997A&A...327..614A, Eriksson2014A&A...566A..95E} and \citet{Bladh2019A&A...626A.100B}, respectively. 
The grid of 48 models by \citet{Arndt1997A&A...327..614A} self-consistently calculates the dust nucleation by assuming chemical equilibrium (CE) and homogeneous nucleation with C$_1$ as the basic monomer, while the more extensive grids of \citet{Eriksson2014A&A...566A..95E} and \citet{Bladh2019A&A...626A.100B} assume the presence of dust \textit{seeds} that can act as further building blocks for grain growth. More on these two different approaches can be found in Section~\ref{Sec:molgrains}.
\citet{Arndt1997A&A...327..614A} have presented a linear multivariate regression analysis by means of a multidimensional maximum-likelihood method to derive an explicit mass-loss rate formula for the implicit mass-loss relation of pulsation-enhanced dust-driven winds, with best-fit formula being
\begin{marginnote}[]
	\entry{Seed particle}{tiny solid particle, predicted using nucleation theory or assumed to pre-exist and typically chosen to consist of 1\,000 monomers or to have a radius of 1\,nm}
\end{marginnote}
\begin{eqnarray}
\log \Mdot_{\rm{fit}} & = & -4.95 -2.8 \log(\Mstar[\Msun]) + 1.65  \log\left(\frac{L_0[\Lsun]}{10^4}\right) -9.45  \log\left(\frac{T_0[{\rm{K}}]}{2\,600}\right) \nonumber \\
& &  + 0.470  \log\left(\frac{\epsilon_{\rm{C}}/\epsilon_{\rm{O}}}{1.8}\right)  -0.146  \log\left(\frac{P[{\rm{days}}]}{650}\right) + 0.449  \log\left(\frac{\Delta v_p [{\rm{km\,s^{-1}}}]}{2}\right)\,,
\label{Eq:Mdot_Arndt1}
\end{eqnarray}
with \Mdot$_{\rm{fit}}$ in units of \Msun\,yr$^{-1}$. From the regression coefficients it is clear that \Mdot$_{\rm{fit}}$ is strongly influenced by $T_0$, \Mstar, and $L_0$ and is only weakly dependent on $\Delta v_p$, $\epsilon_{\rm{C}}/\epsilon_{\rm{O}}$, and $P$. This outcome renders the possibility of a reduced fit, with an equally high correlation coefficient,
\begin{equation}
\log \Mdot_{\rm{fit}}  =  -4.93 -2.88  \log(\Mstar[\Msun]) + 1.53  \log\left(\frac{L_0[\Lsun]}{10^4}\right)  -8.26 \log\left(\frac{T_0[{\rm{K}}]}{2\,600}\right)\,.
\label{Eq:Mdot_Arndt}
\end{equation}
The results published by \citet{Bladh2019A&A...626A.100B} allow for a similar linear multivariate regression analysis, yielding
\begin{equation}
\log \Mdot_{\rm{fit}}  =  -5.26 -3.82  \log(\Mstar[\Msun]) + 3.17 \log\left(\frac{\Lstar[\Lsun]}{10^4}\right)  -6.47  \log\left(\frac{\Teff[{\rm{K}}]}{2\,600}\right)\,.
\label{Eq:Mdot_Bladh}
\end{equation}
While Eqs.~\eqref{Eq:Mdot_Arndt1}\,--\,\eqref{Eq:Mdot_Arndt} refer to the temperature and luminosity at time $t=0$ where the piston position takes its mean value over the pulsation period and is moving outward with maximum speed, 
 Eq.~\eqref{Eq:Mdot_Bladh} uses the effective temperature and stellar luminosity of the hydrostatic dust-free model that was used as starting structure for the calculations.
Although any impact of the pulsation practically cancels in Eqs.~\eqref{Eq:Mdot_Arndt}\,--\,\eqref{Eq:Mdot_Bladh}, the pulsation quantities have an important implicit influence. The reason is that pulsations increase the density scale height allowing for an efficient condensation and growth of dust species. This outcome explains why \textit{empirically} derived mass-loss rate formula, such as the one proposed by Reimers, can be expressed in terms of fundamental stellar parameters  without notion of the pulsation characteristics; for Reimers' law being \Lstar, \Rstar, and \Mstar\ 
(see Eq.~\eqref{Eq:Mdot_Reimers} in Section~\ref{Sec:history}).

\subsubsection{Theoretical versus empirical mass-loss rate relations} \label{Sec:Mdots_theor_emp}

The similar dependence between the mass-loss rate and some of the fundamental stellar parameters identified by 
the \textit{forward} and the \textit{retrieval} approaches has nurtured the idea that \textit{empirically} derived mass-loss rate relations could 
provide an alternative approach for understanding the essence, if not the detail, of the process by which mass loss  occurs.  
The advent of new observing facilities resulted in tremendous progress in the field of observational astrophysics. Molecular lines and dust emission have been used to retrieve the mass-loss rate of the stars under study \citep[][and references therein]{Hofner2018A&ARv..26....1H}.
Circumstellar CO rotational and OH maser emission are the molecular diagnostics most often used to estimate the gas mass-loss rates \citep[e.g.][]{Baud1983A&A...127...73B, Schoier2002A&A...391..577S, Decin2006A&A...456..549D, Ramstedt2008A&A...487..645R, DeBeck2010A&A...523A..18D}. The benefit of analysing molecular lines is that one obtains the expansion velocity.
The disadvantages are that the observation of emission from thermally excited lines is typically limited to nearby stars within $\sim$2\,kpc from the Sun \citep[although the sensitivity of ALMA is now opening up the field to larger distances, such as the Large Magellanic Cloud;][]{Groenewegen2016A&A...596A..50G}, the unknown fractional abundance of the molecule, and the fact that the analysis often requires a non-local thermodynamic equilibrium (non-LTE) radiative transfer analysis which can be quite CPU intensive. 
The latter aspect implies that sample studies seldom exceed $\sim$50 stars \citep{Danilovich2015A&A...581A..60D}; 
whereas the calculation of dust spectral features --- and the related analysis of the spectral energy distribution (SED) --- are  readily applied to large samples, owing to the inherently simpler radiative transfer calculations. 
However, the identification of dust features is more ambiguous, and a reliable estimate of the mass-loss rate of the dust can only be achieved if several dust features with differing optical depths are combined. 
In addition, one needs to assume a dust expansion velocity and a gas-to-dust mass ratio to convert the derived dust densities into gas mass-loss rates \citep[e.g.,][]{Heras2005AA...439..171H, Verhoelst2009AA...498..127V, Groenewegen2009A&A...506.1277G}.

Supported by the incredible increase in computational power during the last two decades, a whole family of mass-loss rate relations has been derived. Without any attempt for completeness, I have summarized some of these \textit{theoretical}, \textit{empirical} and \textit{semi-empirical} relations in the \textbf{Supplemental Text}, where I focus on those relations which have an explicit dependence on two of the main fundamental stellar parameters:  the luminosity and the effective temperature. 
For an AGB star of stellar mass 2\,\Msun\ or with an effective temperature of 2\,800\,K, these mass-loss rate relations are shown in the upper panels of \textbf{Figure~\ref{Fig:Mdots}}, for a RSG star of mass 15\,\Msun\ or with an effective temperature of 3\,500\,K, the mass-loss prescriptions are shown in the bottom panels of \textbf{Figure~\ref{Fig:Mdots}}; more detailed information is provided in the \textbf{Supplemental Text}. 
The large diversity between prescriptions is remarkable, with differences up to 2 orders of magnitude or higher. 
To be fair, not all mass-loss rate formula are applicable to all classes of cool ageing stars, so sometimes we might be comparing apples with oranges. But even within the class of the `apples', it seems that we are having different cultivars in the same basket; let's call them the medium-sized \textit{Golden Russets} which make extraordinary cider and the large red-coloured \textit{Haralsons} which are 
an excellent choice for pies. 
Both are different genomic expressions of the \textit{malus pumila} and in the same vein the relations we witness in \textbf{Figure~\ref{Fig:Mdots}} are different expressions of an `emergent property': the mass-loss rate. 
This statement deserves further explanation.

\begin{figure*}[htp]
\begin{minipage}[p]{0.437\textwidth}{
\includegraphics[height=4.7truecm]{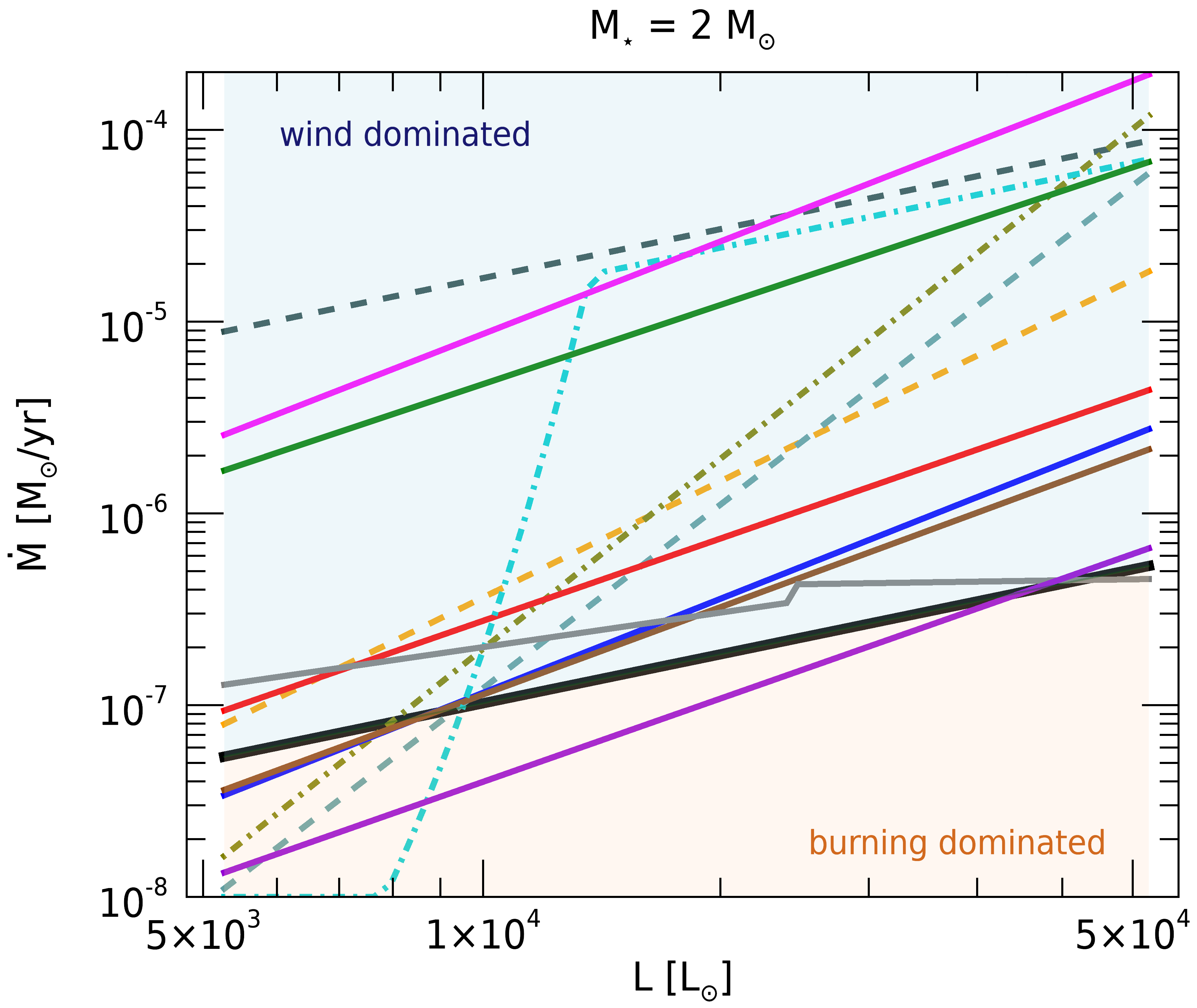}
}
\end{minipage}
\hfill
\begin{minipage}[p]{0.56\textwidth}{
\includegraphics[height=4.7truecm]{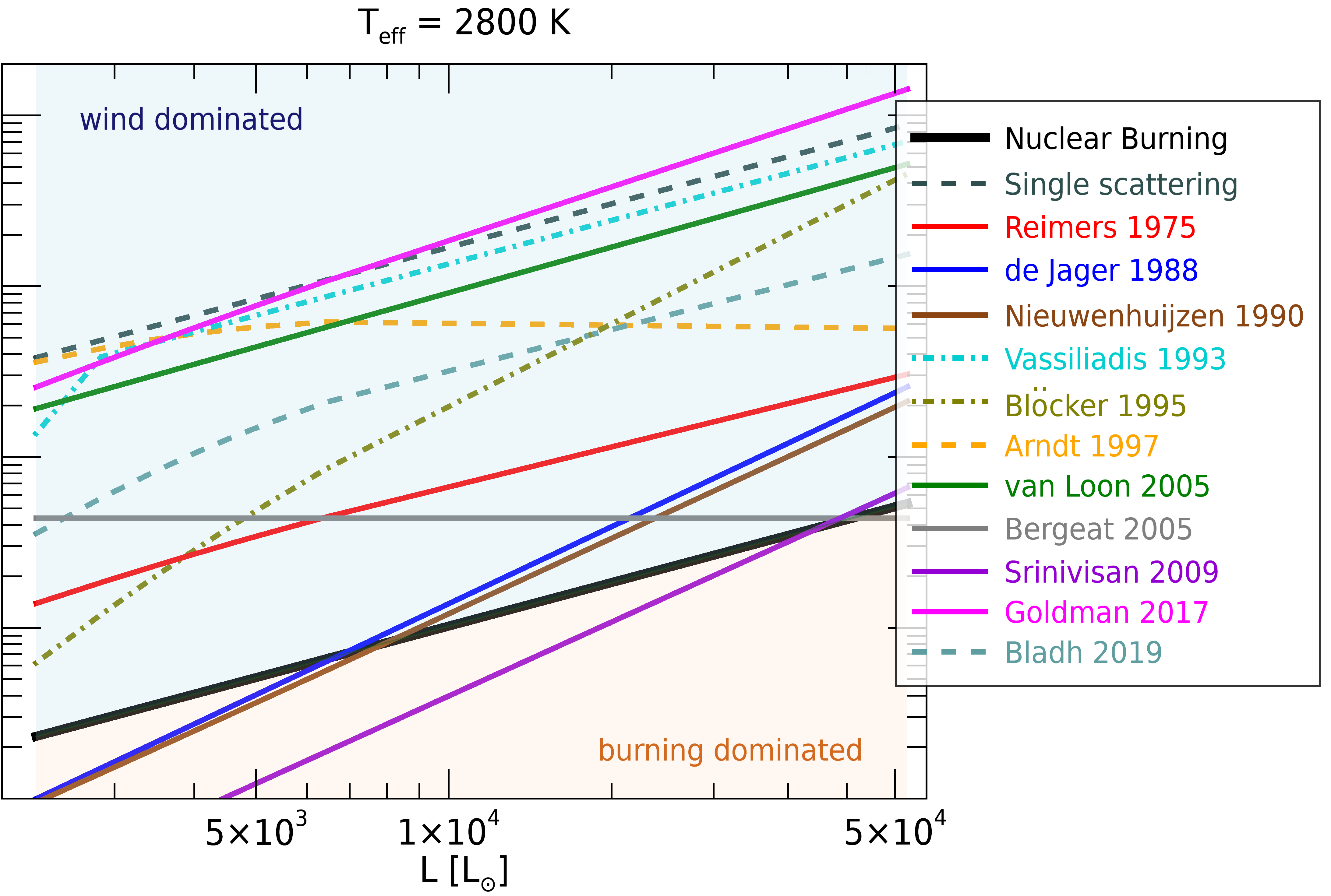}
}
\end{minipage}
\begin{minipage}[p]{0.437\textwidth}{
		\includegraphics[height=4.7truecm]{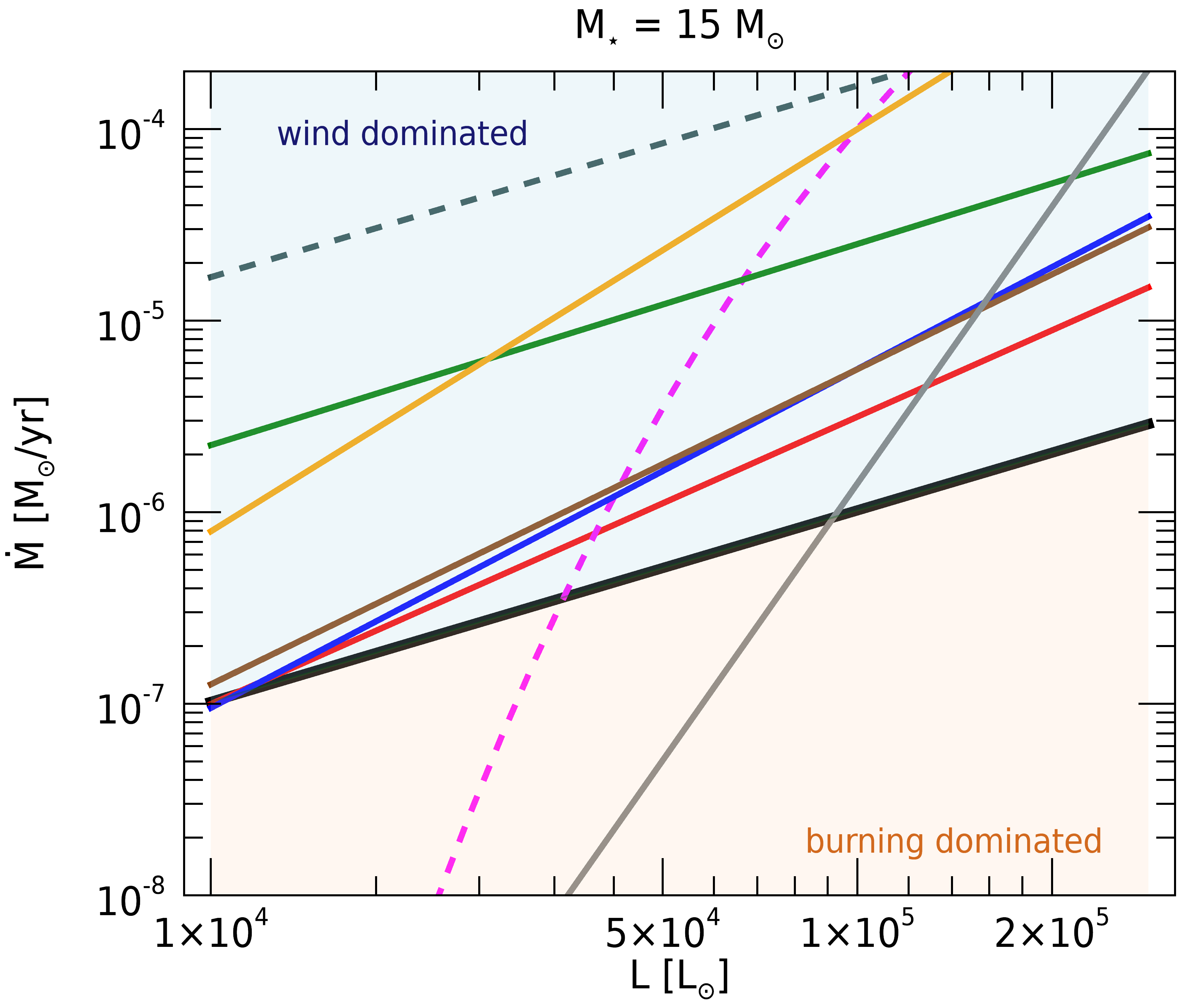}
	}
\end{minipage}
\hfill
\begin{minipage}[p]{0.56\textwidth}{
		\includegraphics[height=4.7truecm]{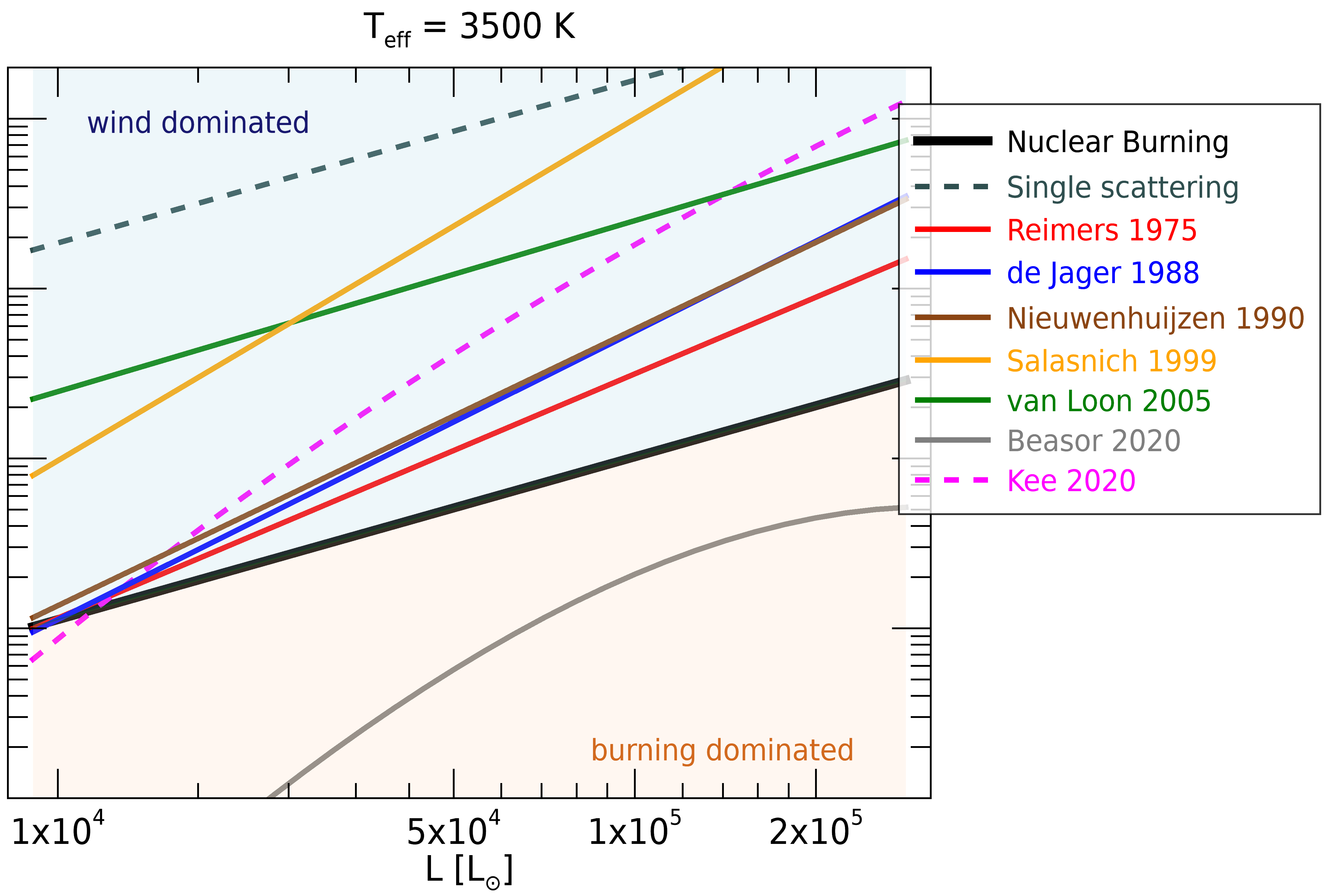}
	}
\end{minipage}
\caption{Mass-loss rate as function of luminosity.
The upper two panels show mass-loss rate prescriptions for AGB stars, at a fixed stellar mass of 2\,\Msun\ (\textit{left}) and fixed effective temperature of 2\,800\,K (\textit{right});
the bottom two panels apply to RSG stars, at a fixed stellar mass of 15\,\Msun\ (\textit{left}) and fixed effective temperature of 3\,500\,K (\textit{right}).
\textit{Empirical} mass-loss rate relations are displayed with a full line, \textit{semi-empirical} relations with a dash-dotted line, and \textit{theoretical} relations with a dotted line. 
The rate at which hydrogen is consumed by nuclear burning, \Mdot$_c$\,=\,1.02$\times$10$^{-11}$\Lstar, is shown as thick black line;
the single-scattering radiation pressure limit for an expansion velocity of 12\,km\,s$^{-1}$ is shown as dashed dark grey line.
Stellar mass loss rules the evolution of the AGB and RSG stars if the wind mass-loss rate exceeds the nuclear burning rate, as indicated by the light-blue region; the hydrogen-burning dominated region is indicated by the light-orange region.
The change in the slope for the \citet{Vassiliadis1993ApJ...413..641V} description is caused by the stellar mass exceeding the limit of 2.5\,\Msun\ (see Eqs.~\eqref{Eq:Mdot_VW93a}\,--\,\eqref{Eq:Mdot_VW93b}); the jump for the \citet{Bergeat2005A&A...429..235B} description in the left panel is due to the effective temperature getting below 2\,900\,K (Eqs.~\eqref{Eq:Mdot_B05b}\,--\,\eqref{Eq:Mdot_B05c}). More detailed information is provided in the \textbf{Supplemental Text}.}
\label{Fig:Mdots}
\end{figure*}

Indeed, while the various mass-loss rate proposals seem already disappointingly incompatible with huge differences, 
there is still another underlying conceptual complication.   
Given these differences, it is to be expected that the impact of a particular choice of mass-loss rate on stellar evolution calculations will be significant. This is illustrated in \textbf{Figure~\ref{Fig:evol_tracks}} where a set of simplified evolutionary tracks is shown for stars with an AGB mass at the first thermal pulse between 0.8\,--\,3\,\Msun\ (see the \textbf{Supplemental Text} for more information). 
For the \citet{Blocker1995A&A...297..727B} and \citet{Vassiliadis1993ApJ...413..641V} mass-loss rate relations which have a strong dependence on luminosity, the stars evolve on a nearly horizontal track --- where the mass remains approximately constant --- until the star reaches the locus where ${\rm{d}} \log \Mstar/{\rm{d}}t\!=\!{\rm{d}} \log \Lstar/{\rm{d}}t$ \citep[referred to as the `cliff' by][see filled red circles in \textbf{Figure~\ref{Fig:evol_tracks}}]{Willson2000ARA&A..38..573W}. 
For the set of models shown in \textbf{Figure~\ref{Fig:evol_tracks}}, the mass-loss rate at the `cliff' is around 0.5\,--\,1.7$\times$10$^{-6}$\,\Msun\,yr$^{-1}$. Passing beyond that `cliff' implies that the evolution is further ruled by the wind mass-loss rate, which for the \citet{Blocker1995A&A...297..727B} and \citet{Vassiliadis1993ApJ...413..641V} relations implies an asymptotic downhill behaviour.  
This particular behaviour let \citet{Willson2000ARA&A..38..573W} identify a strong selection effect concerning stars for which the mass-loss rate is measurable. Stars not yet near the cliff will have low mass-loss rates that are difficult to detect or to measure, while stars beyond the cliff will be short-lived causing a scarcity in the detection rates. This implies a selection bias towards stars near the cliff. 
Thus, the \textit{empirical} mass-loss laws `tell us the parameters of the stars that are losing mass, and not the dependence of the mass-loss 
rates on the parameters of any individual star' \citep{Willson2000ARA&A..38..573W}. 
Although the 1-dex width in mass-loss rate around the `cliff' is not necessarily that  narrow --- all depend on the particular mass-loss rate behaviour over time ---  \citeauthor{Willson2000ARA&A..38..573W}'s conclusion remains valid: \textit{empirical} mass-loss rate formulae `tell us which stars are losing mass rather than how stars are losing  mass over time'.

\begin{figure}
\includegraphics[width=\textwidth]{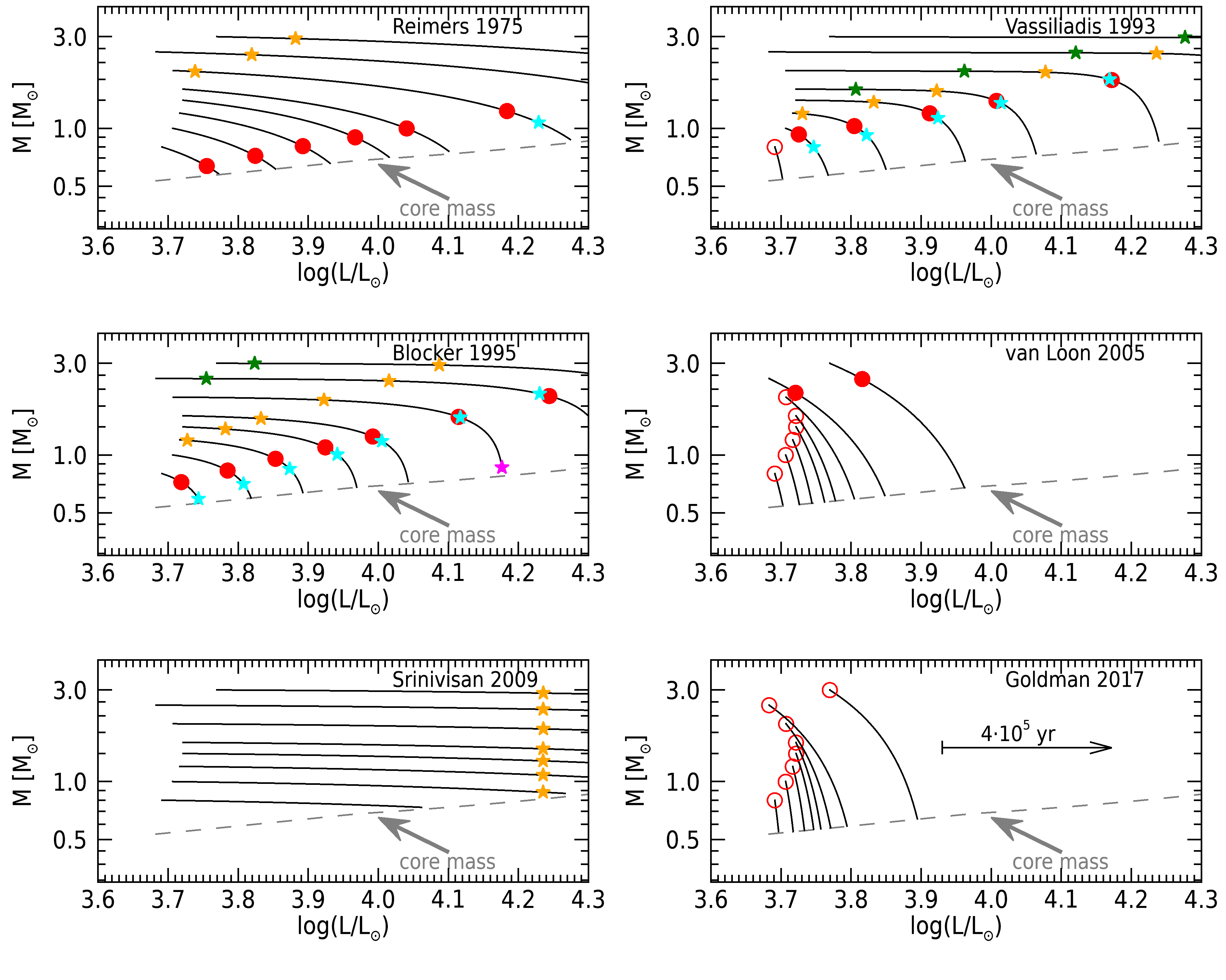}
\caption{Evolution in mass and luminosity for stars with a solar composition for different mass-loss rate prescriptions. 
The different tracks show the stellar evolution for AGB stars with a mass at the first thermal pulse of (0.8, 1.0, 1.2, 1.4, 1.6, 2.0, 2.5, 3)\,\Msun.
In each panel, a different mass-loss rate prescription has been used, as indicated in the top right corner. 
The core-mass-luminosity relation is displayed by the dashed gray line. 
The filled red circles indicate the locus where ${\rm{d}} \log \Mstar/{\rm{d}}t\!=\!{\rm{d}} \log \Lstar/{\rm{d}}t$; 
an open red circle implies ${\rm{d}} \log \Mstar/{\rm{d}}t$ is larger than ${\rm{d}} \log \Lstar/{\rm{d}}t$ at the start of the calculation.
The coloured stars specify the locus where \Mdot\ is equal to $10^{-8}$ (green), $10^{-7}$ (orange), $10^{-6}$ (cyan), $10^{-5}$ (magenta) \Msun\,yr$^{-1}$.
The rate of change of the star's abscissa in this plot is ${\rm{d}} \log \Lstar/{\rm{d}}t\!=\!0.605$\,Myr$^{-1}$, which is indicated by the black arrow in the lower right panel.}
\label{Fig:evol_tracks}
\end{figure}

This conclusion by \citeauthor{Willson2000ARA&A..38..573W} brings us back to the earlier discussion of the benefits (but also pitfalls) of the reductionist approach. At the risk of oversimplification, this approach is linked  to that termed `hierarchical reduction'; the idea that phenomena at one hierarchical level can be explained by using concepts from a lower hierarchical level. Thus we conventionally express astrophysical phenomena in terms of chemical principles, and chemical phenomena in physical terms, and physical phenomena in a mathematical language; or as the Nobel Prize physicist Steven Weinberg succinctly expressed `explanatory arrows always point downward' \citep{Weinberg1994}.
\textit{Empirical} mass-loss rate prescriptions do not follow the arrow downward, but are holistic expressions of an emergent property, 
and hence take a `top-down' approach in their attempt to unravel stellar evolution. 
Indeed, quite confusingly, `top-down' does not imply `point downward' in Weinberg's words. 
The risk of an `emergent property' is that it raises the expectation that the behaviour is understood, but that is not necessarily true. 
In this particular case, I  would not recommend  using \textit{empirical} mass-loss rate relations in stellar evolution models, 
since the seemingly logical relation might be causally wrong. 
In contrast the \textit{forward theoretical} approach is `bottom-up' (and points downward), and there are a number of factors that argue persuasively that
the bottom-up approach will change the landscape of mass-loss rates  considerably in the next few years. 
These `winds of change' come from different cardinal points, each of them inherently linked to the 3D reality of a stellar wind. 
They are steered by recent progress in quantum chemistry and astrophysical observations, and can now build up momentum thanks to the latest developments in supercomputing capabilities.

I do not want to leave this section with the reader having the feeling that these \textit{empirical} laws are deceptive or useless. 
On the contrary! If systematic biases in the retrieval approach can be avoided, the observations tell us the \textit{real} rates of mass loss.
Only after detailed theoretical descriptions are found that reproduce the retrieved rates and relations, may the models be used to 
extrapolate to populations not presently available for study, such as low-metallicity populations in the early Universe. 
For only then, we will have greater faith in our predictions of the maximum luminosity achieved by AGB stars; 
the mass spectrum of planetary nebulae and white dwarfs; 
the frequency of type~I and type~II supernovae, and possibly the masses of their progenitors; and 
the fate of stellar and planetary companions residing close to the mass-losing red giant primary star.  
But even if \textit{theoretical} and \textit{empirical} mass-loss rate relations agree, we must be extremely vigilant against 
any confirmation bias in our theoretical efforts. 
\begin{marginnote}[]
	\entry{Type I or Type II supernova}{a supernova is classified as type~II if the spectrum displays the hydrogen Balmer lines, otherwise it is Type~I}
	\entry{Type Ia supernova}{when a white dwarf is triggered into a runaway nuclear fusion, caused by the accretion of matter from a binary companion or a stellar merger}	
	\entry{Type Ib/c and Type II supernova}{caused by the gravitational collapse of the core of a massive star, resulting in a black hole or neutron star}
	\end{marginnote}

\subsection{Informative measures challenging the 1D world}\label{Sec:Setting_challenges}

Our earlier discussion has identified the explicit dependence of the mass-loss rate on fundamental parameters as \textit{the} nut that needs to be cracked. 
From the discussion of \textbf{Figure~\ref{Fig:Mdots}} it is clear that the \textit{theoretical} and \textit{empirical} mass-loss rate prescriptions appear to be similar, but they are in fact intrinsically different. 
There are challenges ahead of us for there to be a simple gateway between both types of relations, 
but recent progress in observations and quantum chemistry leads us to believe that the gap between them is not cavernous, and a bridge is gradually but assuredly coming into view. 
The central pillar of that bridge is the incorporation of a 3D view in all our measures of mass loss, either in the forward or the retrieval approach. 
As a first step, we need to question whether the retrieved mass-loss rates and relations are reliable. 
Here recent progress in observational techniques indicate that there is an elephant in the room --- well actually two elephants --- a smaller and a bigger one, which are caught in the act owing to the incredible capacities of novel high spatial resolution observing capabilities. 
On the one hand, there are flow instabilities induced by convection that result in the formation of granulation cells on the surface of the giant stars, and of small-scale density structures in the stellar wind, with sizes of between $\sim$1\,--\,50\,au (see \textbf{Figure~\ref{Fig:3D}}). 
 On the other hand --- and encompassing much larger geometrical scales --- there is mounting (indirect) evidence that most evolved cool giant stars with measurable mass-loss rates are surrounded by at least one stellar or planetary companion.  The companion will perturb the structure of the stellar wind (see \textbf{Figure~\ref{Fig:3D}}), and will under some circumstances 
induce an increase in mass-loss rate. 
As I will discuss in Section~\ref{Sec:partner_impact}, neglecting the 3D structural complexities in retrieval approaches can lead to mass-loss rates that are incorrect
by an order-of-magnitude, which --- as we have seen  in \textbf{Figure~\ref{Fig:evol_tracks}} --- has a huge impact on the outcome of stellar evolution models.

\begin{figure}[htp]
\centering
\includegraphics[width=.92\textwidth]{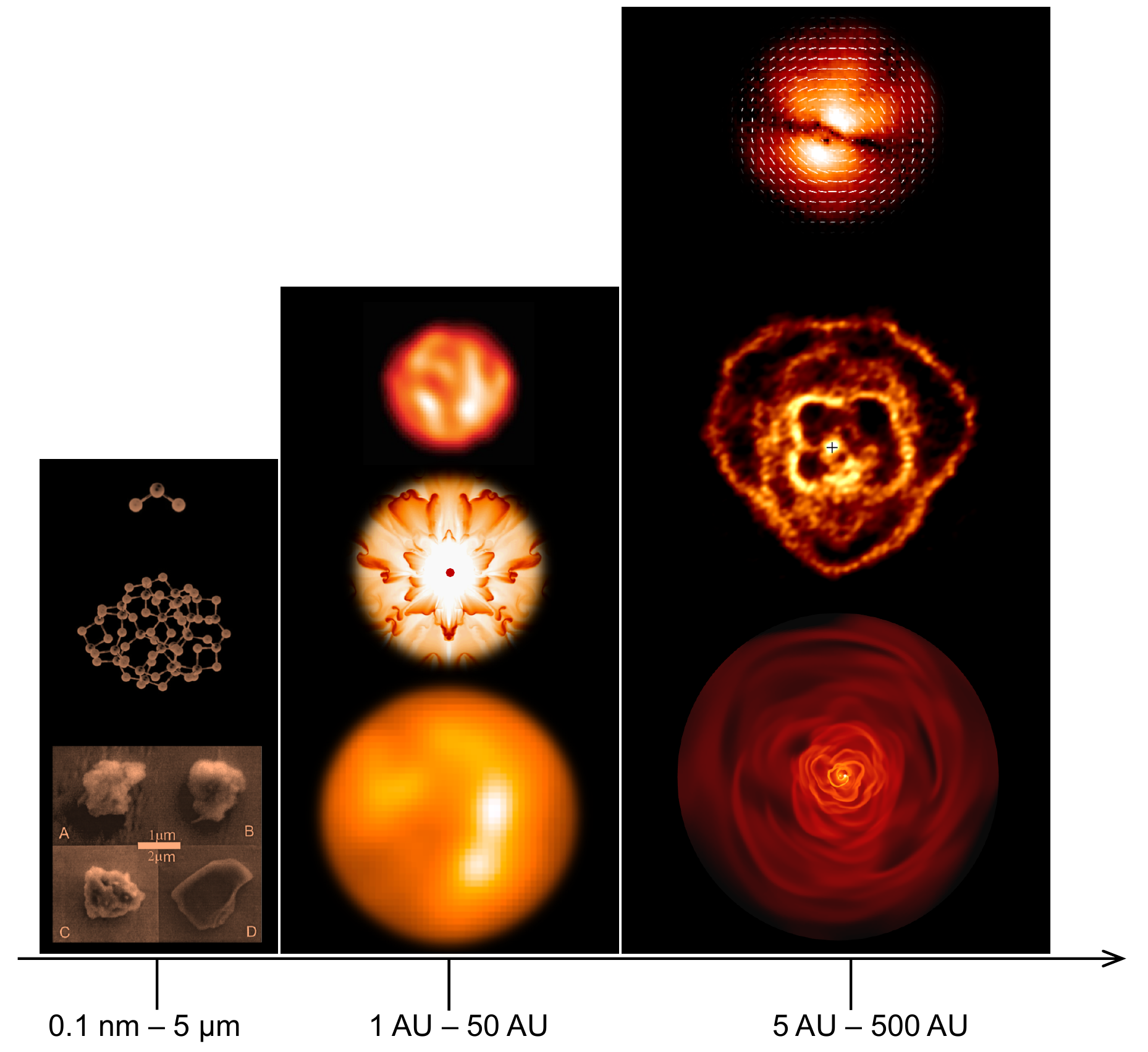}
\caption{3D structures of relevance for stellar wind research. 
Compilation based on observations, theoretical calculations, modelling, and laboratory experiments.
First column:
 the 3D life of molecules and dust grains: \textit{Top} -- TiO$_2$ proposed as candidate for the first dust condensate in O-rich winds \citep{Jeong2003A&A...407..191J, LamielGarcia2017}; 
\textit{Middle }-- global minima candidate for (TiO$_2$)$_n$ with $n=24$ obtained within a bottom-up approach by using a global optimisation algorithm searching the complex multidimensional potential energy surface \citep{LamielGarcia2017};
\textit{Bottom} -- presolar corundum grain with AGB origin, image obtained using a scanning electron microscope  by \citet{Choi1998Sci...282.1284C}.
Second column: the emergence of 3D clumps.
\textit{Top} -- due to convection in the stellar atmosphere large granulation cells form on the surface of the O-rich AGB star $\pi^1$~Gru and are now seen thanks to the PIONIER instrument mounted at the {\sl Very Large Telescope Interferometer} \citep[VLTI;][]{Paladini2018Natur.553..310P}; 
\textit{Middle} -- flow instabilities arising in a 2D model for a dust-driven wind of a carbon star \citep{Woitke2006A&A...452..537W}, the red dot represents the AGB star;
\textit{Bottom} -- 2.166\,$\mu$m image of the carbon star CW~Leo obtained with NACO, mounted at the {\sl Very Large Telescope} (VLT), showing the presence of small-scale structures (clumps) embedded in the stellar wind \citep{Menut2007MNRAS.376L...6M}.
Third column: the 3D lasting impact of a companion.
\textit{Top} -- detection of an equatorial dust lane in the wind of the carbon star CW~Leo using scattered light data obtained by ExPo mounted on the {\sl William Herschel Telescope} \citep{Jeffers2014A&A...572A...3J};
\textit{Middle} -- a rose-like spiral structure detected in the stellar wind of the O-rich AGB star R~Hya using ALMA \citep{DecinScience};
\textit{Bottom} --  3D hydrodynamical simulation for a binary system containing a mass-losing AGB star \citep{ElMellah2020arXiv200104482E}.
}
\label{Fig:3D}
\end{figure}

This discussion on the emergence of 3D clumps and the impact of a (hidden) companion provides insight into the \textit{deterministic} perspective of cool ageing stars and their contribution to the chemical enrichment of the Universe. 
However, there is an additional approach to the \textit{how} question that could potentially  provide useful information for the \textit{why} question as well. 
By probing what we believe are the early beginnings of a stellar wind, we might be able to grasp the essence of mass loss. To do that, we have to strip the outer layers of complexity, the large 3D structures with scales greater than 1\,au, to uncover what lies beneath. We need to reverse the time axis and dive deeply into the smallest 3D scales that matter. 
Doing so, we reach the nanometer-sized domain where dust nucleation is happening (see \textbf{Figure~\ref{Fig:3D}}). 
Arriving there, we will discover that, here as well, the bottom-up approach has recently been very instructive in providing new insights in the intricacies of this phase transition, but it requires us to embrace the complexities of the 3D structures of molecules, large gas-phase clusters and dust grains. 
Once we arrive at that scale, we may begin to understand \textit{why} cool ageing stars contribute to the galactic chemical enrichment 
and \textit{what} are the driving forces for the complex physicochemical processes governing the Universe. However, it also will become clear that untangling the tangled web of Schr\"odinger's equations -- describing the transition from molecules to dust --- still lies in the distant future. 

\section{The emergence of 3D clumps}\label{Sec:clumps}

The first unambiguous observation of structure on the stellar surface of a star other than the Sun  was reported in 1990, when \citeauthor{Buscher1990MNRAS.245P...7B} presented the detection of a bright surface feature on the surface of the red supergiant Betelgeuse. Betelgeuse has one of the largest apparent angular sizes in the night sky 
\citep[with a measured angular diameter of $\sim$44.2\,mas in the infrared,][]{Dyck1992AJ....104.1982D}, and is an  ideal candidate for spatially resolving its surface, not only in the optical, but  also in the UV and sub-millimeter wavelength domain \citep{Gilliland1996ApJ...463L..29G, OGorman2017A&A...602L..10O}. 
 \citet{Gilliland1996ApJ...463L..29G} interpreted the observed feature  as resulting from magnetic activity, atmospheric convection, or global pulsations and shock structures that heat the chromosphere. 
Recently, another profound realisation in observational astronomy occurred through the first image of large granulation cells on the surface of a (much smaller) AGB star, $\pi^1$~Gru; our Sun might show some resemblance to $\pi^1$~Gru once it becomes a red giant star in about 7.7 billion years. 
While the surface of the Sun is covered with about two million convective cells whose typical size is around 2\,000\,km across,  
$\pi^1$~Gru was shown to have only a few granules with a characteristic horizontal size of about 1.2$\times$10$^{8}$\,km, or $\sim$27\% of the stellar diameter \citep{Paladini2018Natur.553..310P}. 
These observations are consistent with the historical prediction by \citet{Schwarzschild1975ApJ...195..137S} that the surface of cool giants might be covered by a relatively small number of giant convection shells. 
Schwarzschild came to his hypothetical conclusion on the basis of the working hypothesis that the pressure or density scale-heights determine the size of the convective elements. 
Moreover he addressed the question of whether mass ejection could be triggered by photospheric convection. 
He argued in favour of that suggestion on the basis of observations which provided substantial evidence for non-spherical circumstellar dust clouds in the neighbourhood of red giant and supergiant stars, in which the polarized light signal shows variability on the same time scale (of $\sim$200\,days) as the irregular brightness variations caused by the giant convective cells. 
The cooler regions of the large-scale surface convective elements might enhance the production of dust grains resulting in an uneven distribution of dust and hence of the polarization signal.
With this section revolving around the \textit{deterministic} question and hence aspects of mass loss, 
we seek to answer whether, in principle, Schwarzschild's argument  is settled, and if and how these non-spherical circumstellar dust clouds 
can be used to trace the mass-loss mechanism.

\subsection{Weather map from cool ageing stars: dry with variable cloud cover}\label{Sec:clumps_observations}

The conjecture of Schwarzschild that there is a potential causal link between the dynamical, inhomogeneous stellar atmosphere driven by  convective flows and the formation of clumpy clouds in the CSE can be checked against observational evidence. In the early seventies there was only indirect evidence of non-spherical dust clouds in the close vicinity of cool ageing stars \citep{Shawl1972PhDT........10S, Schwarzschild1975ApJ...195..137S}, 
but the new revolution in observational techniques allowed for high-spatial resolution (reconstructed) images  
and provided direct observational evidence that the CSEs harbour small and large-scale inhomogeneities 
(for a discussion of large correlated density structures, see Section~\ref{Sec:partner}). 
The CSE of the carbon-rich AGB star CW~Leo (at a distance of $\sim\!150$\,pc) has long been known to be quite complex and continually evolving (see \textbf{Figure~\ref{Fig:3D}}). 
As the IR brightest  AGB star experiencing high mass loss,  it has been exhaustively studied by many  observational techniques. 
Fine structure on sub-arcsec scale was detected within $\sim$30 stellar radii from the central star \citep{Weigelt1998A&A...333L..51W, Haniff1998A&A...334L...5H}. 
The fragmentation of the shell in distinct clumps was suggested to be caused by inhomogeneous mass loss potentially induced by large-scale surface convection cells. 
However,  the stellar surface and inner CSE of this star are obscured by the optically thick envelope of carbon dust, 
hindering the identification of the observed clumpy features with processes occurring at the stellar surface, in the shock-dominated 
inner CSE, and in the dust-formation region. 
Even for carbon stars with a more modest mass loss, IR images can be difficult to interpret, as was shown recently by the example 
of R~Scl \citep{Wittkowski2017A&A...601A...3W}. 

The envelopes of O-rich giants tend to be more transparent in the visual and infrared wavelength regime than C-rich giants (see also Section~\ref{Sec:molgrains}), thereby providing better access to the surface and innermost CSE layers and hence allowing for a more detailed evaluation of Schwarzschild's conjecture.
 Observations of scattered light  of nearby  O-rich AGB and RSG stars   establish there are dust clouds
as close as $\sim$0.5 $R_{\star}$ to the star, 
whose size is $\la$20\,\Rstar\ and which change in morphology on time scales of weeks to months 
\citep{Khouri2016A&A...591A..70K, Ohnaka2017A&A...597A..20O, Adam2019A&A...628A.132A, Kaminski2019A&A...627A.114K, Cannon2020}. 
The dynamical time scale implied by changes in the circumstellar morphology are in good agreement with the characteristic time scales for 
convection; and the observed spatial scales (both the size of the clouds and the distance from the stellar surface) compare well with recent 
3D radiation hydrodynamic models simulating the outer convective envelope and dust-forming region 
\citep[][see also Section~\ref{Sec:clumps_theory} for some side notes on this claim]{Hofner2019A&A...623A.158H}. 
Thanks to these results, we have increased confidence that the Schwarzschild's conjecture actually might reflect reality. 
However, the recent two-dimensional mapping of the velocity field over the surface and inner CSE of the nearby red supergiant Antares 
throws a spanner in the works, because the maps reveal vigorous upwelling and downdrafting motions of several huge clumps of gas 
with velocities ranging from about $-$20 to $+$20\,km\,s$^{-1}$ in the inner CSE up to $\sim$1.7\,\Rstar\ \citep{Ohnaka2017Natur.548..310O}. 
Convection alone cannot explain the observed turbulent motions and atmospheric extension, suggesting that a process which has not yet been
identified is operating in the extended atmosphere \citep{Arroyo2015A&A...575A..50A, Ohnaka2017Natur.548..310O}. 
Admittedly, this result does not rule out Schwarzschild's conjecture, but it reveals that the apparent analogy and the correlation 
between the surface granulation pattern and the gas and clumps in the extended atmosphere and inner CSE does not in itself signify 
a causal relation and conceivably other processes might be important.

\subsection{How to model a turbulent life}\label{Sec:clumps_theory}

Whether we consider either a small Sun or a giant star, modelling the turbulent dynamical process of convection and its interaction with 
other physical, chemical and radiative processes, is a major challenge. 
This is not only due to the complexity of the problem, but is also due to the huge CPU demand dictated by the numerical resolution  required for a proper sampling of the time-dependent small- and large-scale fluctuations in density, temperature, velocity, and brightness. Detailed radiation hydrodynamics (RHD) simulations can help to understand qualitatively these processes and to model quantitatively the dynamical layers in and around these stars \citep{Freytag2019IAUS..343....9F}. Modelling convection and surface granulation in sun-like stars is only possible using \textit{local} 3D RHD simulations owing to the huge disparity in spatial and time scales: for a \textit{global} simulation of a sun-like star one would need a spatial resolution of at least a fraction of the photospheric pressure scale height of $\sim$150\,km
to cover the Sun (diameter of about 1\,400\,000\,km), and a time resolution of about a second for at least several rotation periods of about a month or, even better, several magnetic cycles, each spanning 22 years \citep{Freytag2019IAUS..343....9F}. 
In contrast, \textit{global} 3D RHD models covering the entire convective surface of cool giants are within the realm of possibility, because of Schwarzschild's prediction that only a few giant convection cells cover their surfaces. The first global 3D RHD simulations for a red supergiant were presented in 2002 \citep{Freytag2002AN....323..213F}, the domain of the smaller AGB stars was reached in 2017 \citep{Freytag2017A&A...600A.137F}. These models encompass part of the atmosphere with the outer boundary situated at $\sim$2\,\Rstar. 
The model dynamics are governed by the interaction of long-lasting giant convection cells, short-lived surface granules, and radial fundamental-mode pulsations (see also Section~\ref{Sec:limitations}). 
The models did not yet include dust formation and therefore no wind driving. Recently, a global model for an oxygen-rich AGB star was presented that incorporates both the outer convective atmosphere and the dust-forming region up to $\sim$2.8\,\Rstar\ \citep{Hofner2019A&A...623A.158H}. The current models do not yet describe wind acceleration and the kinetic treatment of grain growth does not account for nucleation, but assumes the presence of seed particles. I will expand on the challenge of dust nucleation shortly (Section~\ref{Sec:molgrains}), but for the time being it is sufficient to realize that we are still speculating as to the specific steps of seed formation. In the models grain growth is hypothesized to take place under the most optimum circumstances (sticking coefficient assumed to be one; see also Section~\ref{Sec:dust_nucl_short}).

Although the 3D RHD models including dust growth still have some limitations, already they offer great insight into the essence of mass loss at least for stars on the upper AGB and in the RSG phase where stars have the lowest surface gravities, and hence largest pressure scale heights and largest granules relative to the stellar radius. The large-scale convective flows and pulsations generate atmospheric shock waves: the shorter-wavelength disturbances cause a complex small-scale network of shocks in the innermost layers, while the fundamental-mode pulsation causes a more or less spherical shock front that is able to travel further away from the stellar surface. In the dense wake of the shock, gas is temporarily lifted to distances where dust formation may occur \citep{Freytag2017A&A...600A.137F}. Remarkably and importantly, the temperature shows a rather smooth, almost spherical pattern, in contrast to the gas densities, which are strongly affected by the local 3D dynamics and show a pronounced 3D clumpy morphology \citep{Hofner2019A&A...623A.158H}. Only when the temperature falls below a critical value does the gas become supersaturated.  Under these circumstances the molecules are  more prone to leave the gas than to rejoin it, so they become deposited on the surface of the solid particles and  grain growth is triggered. Since the temperature acts as a threshold for the onset of grain growth, while the gas densities determine the grain growth efficiency, the models show an almost ring-like dust number density distribution conditioned by the local temperature. 
For a 2D slice through the center of the grid, the dust layers appear narrow in the radial direction due to the rapidly decreasing densities with increasing distance. 
Even then, smaller patches of dust clouds appear where both the density and radius of the grains vary in response to the local gas density 
and velocity. 
This behaviour is not only characteristic for O-rich environments, but 2D carbon-rich wind models also show a similar pattern 
\citep[][see \textbf{Figure~\ref{Fig:3D}}]{Woitke2006A&A...452..537W}.
\begin{marginnote}[]
	\entry{Supersaturation}{a gas-solid or gas-liquid chemical system which is in a non-equilibrium state such that there are too many gaseous molecules for the present temperature-pressure conditions}
	\end{marginnote}

Given the fact that the models of \citet{Freytag2002AN....323..213F, Freytag2017A&A...600A.137F} and \citet{Hofner2019A&A...623A.158H} do not include the radiation pressure on dust grains, they do not directly address the speculation by Schwarzschild that mass ejection could be triggered by photospheric convection. 
However, given the results just described I argue here that clumps generated by convective motions are not the cause or the essence of mass loss, rather they are the consequence of the local 3D atmospheric conditions. 
This argument is based on i)~examination of the isotherms, which are set by non-local radiative processes, being nearly perfect spheres 
and ii)~the consideration that not only the onset of the dust growth, but also, even more elementally,  dust nucleation
is dictated by the local temperature 
\citep[see Fig.~8 in][and also Section~\ref{Sec:theoretical_outcomes_nucleation}]{Woitke2006A&A...452..537W}.
Convection-induced 3D cloudy structures are of second order for wind driving in the sense that grain growth can be locally of higher or lower efficiency due to a change in density,
but as long as the gas is not supersaturated no force can be generated which overcomes the stellar gravitational attraction.
This argument rests on the common idea of a dust-driven outflow, which is more appropriate for AGB than RSG winds. 
For RSG stars with their very low surface gravities the situation might be the reverse. 
A recent study shows that inferred atmospheric turbulent velocities yield turbulent pressure high enough to initiate mass loss 
even in the absence of circumstellar dust \citep{Kee2020}. 
In this regard, and for these fluffy RSG stars, Schwarzschild's conjecture still holds --- but is not yet proven --- and photospheric convection seems to be a viable mechanism to trigger mass ejection. 

\subsection{Impact on derived properties}\label{Sec:clumps_impact}

Clumps show up not only in the inner CSE, but they are also present at much larger distances from the star  --- i.e.,  up to the bow shock, where the stellar wind collides with the ISM \citep{Cox2012A&A...537A..35C, Decin2016A&A...592A..76D, Decin2018A&A...615A..28D, Montarges2019MNRAS.485.2417M, Kaminski2019A&A...627A.114K}. 
At present, it is still not understood which mechanism prevents clumps that are formed in the inner CSE, from dissipating during their travel through the huge CSE. 
Potentially, clumps can cool very efficiently inducing a reduction of the internal pressure.
But even without understanding the process, we are in principle able to quantify the impact of 3D clumps on mass-loss rate retrieval efforts by carefully measuring the amount of mass in the clumps and the surrounding smooth envelope. 
However, the need for highly performant 3D radiative transfer analysis often steers scientists towards simplistic analytic estimates.
A small mistake in the optical depth, however, might have a  substantial impact on the mass estimates because the optical depth enters in the exponent of the intensity estimate ($I_\nu \propto S_\nu \exp(-\tau_\nu)$). 
This leaves us with only a few examples, all RSG stars, to guide us in this exercise. 
These analyses indicate that clumps contribute  from a few up to $\sim$25\% of the total mass loss \citep{Ohnaka2014A&A...568A..17O, Montarges2019MNRAS.485.2417M, Cannon2020}, the only potential exception is the extreme RSG VY~CMa for which the derived dust mass in the clumps (of 0.47\,\Msun) seems unrealistically high to be compatible with a current stellar mass of 17\,\Msun\ 
\citep{Kaminski2019A&A...627A.114K}. 
With the exception of VY~CMa, this suggests that the mass loss in ejected clumps contributes non-negligibly to the total mass loss, but also that the clumps do not represent the main mass-loss mechanism.
This conclusion reinforces the final statement in the preceding section: convective-induced turbulent pressure might invoke mass ejection for RSG stars, but in all its generalities that mechanism only depends on the turbulent velocities and has no significant explicit dependence on any 3D clump characteristic.

Given this (tentative) outcome and the substantial  struggle for deriving reliable mass-loss rate relations (as described in Section~\ref{Sec:Mdots_theor_emp}), it seems fair to conclude that the story of 3D clumps only bears limited relevance for both the \textit{deterministic} and \textit{conceptual} question. However, this is too short-sighted for several reasons. 
First, forward theoretical models predict a considerably different molecular abundance pattern whether clumps are included or excluded  \citep{Agundez2010ApJ...724L.133A, VandeSande2018A&A...616A.106V}, where notable examples include the formation of water 
(H$_2$O) molecules in a carbon-rich wind and hydrogen cyanide (HCN) in O-rich envelopes at sufficiently high abundances to be detected
with very sensitive telescopes. 
These predictions are not only consistent with contemporary observations \citep[e.g.,][]{Lombaert2016A&A...588A.124L, VandeSande2018A&A...609A..63V}, but they also convey the message that our estimate of the chemical enrichment of the ISM by cool ageing stars is at best preliminary. 

Second, stars do not stand still. Instead stars travel through the ISM, where a striking testimony is  the appearance of bow shocks that are so beautifully imaged with the Herschel Space Observatory \citep{Cox2012A&A...537A..35C}. 
The difference in velocity and density between both media invoke the growth of Rayleigh-Taylor and Kelvin-Helmholtz instabilities, and kinetic temperatures are estimated to reach up to 10\,000\,K,  depending on the shock velocity $v_S$ \citep{vanMarle2011ApJ...734L..26V, Decin2012A&A...548A.113D}. 
This might foster the dramatic idea that the collision with the ISM can destroy any relic of the chemical processes occurring in stellar winds, because molecules can break up (dissociate) and grains can be destroyed. 
However, this is only true for the most violent collisions and even then the larger grains are shown to survive \citep{vanMarle2011ApJ...734L..26V}, leaving us with a grain size distribution modified toward larger grains. For low-velocity collisions ($v_S\!\le\!5$\,km\,s$^{-1}$), most of the kinetic energy dissipates via magnetic compression and through the rotational emission of molecules such as CO and H$_2$O or the fine structure lines of C$^+$ and O \citep{Godard2019A&A...622A.100G}. 
As the velocity increases, H$_2$ becomes the dominant coolant over a wide range of shock velocities and gas densities \citep{Lesaffre2013A&A...550A.106L, Flower2015A&A...578A..63F}. 
The general outcome is a higher fraction of ionized atoms, excited molecules, and sputtered grains. 
 Returning to the persistent clumps and their importance for the \textit{deterministic} aspect, 
if material is embedded within higher density clumps it will be less affected
by the collisions which occur in the bow shock,
by the harsh interstellar UV field, and by the possible entwinement of radiative and mechanical energies. 
In other words, clumps not only help us diagnose the genuine wind chemistry, they are also a key ingredient for quantifying the galactic chemical evolution and are, as such, of fundamental importance for the \textit{deterministic} question.

\vspace*{-2ex}
\section{The 3D lasting impact of a partner}\label{Sec:partner}

So while the smaller elephant (a.k.a.\ the 3D clumps) seems part of the furniture with limited repercussion on mass-loss rate estimates --- but with a profound effect on chemical processes --- there is still that other, bigger elephant, which came into view owing to the detection of large correlated density structures in stellar winds, and which exposes the scientific quandary that modern astrophysics has been contending with in recent years. 
It has been more often than not that unexplained phenomena in observational astrophysics were either `justified' or conversely `neglected' by
using the phrasal idiom `$\ldots$ binarity which is not within the scope of this paper (or conference talk)'; and magnetic fields suffer the same fate. 
Our Sun with its eight planets has no (known) stellar companion. 
So  akin to social psychology, where it is well known that humans tend to hire job candidates on the basis of similarities to themselves (the `similarity attraction bias'), 
a general thesis that formed the basis for stellar evolution  models took shape, namely that solar-like stars live their lives alone 
(and  the planets are inconsequential for the late stages of evolution). 
In retrospect that thesis now appears to be problematic, 
especially if we are pondering  the fate of solar-like stars experiencing substantial mass loss on the AGB. 
I will venture the idea that a planetary or stellar companion impacts the wind morphology of
almost all AGB and RSG stars with a detectable mass-loss rate (\Mdot\,$\ga$\,10$^{-7}$\,\Msun\,yr$^{-1}$).
Moreover, in a fraction of stars the companion induces a change in expansion velocity and mass-loss rate. 
This implies that most empirically derived mass-loss rates are obtained from samples containing a large fraction of stars that experience binary interaction with a (sub-)stellar companion. 
Therefore, our knowledge of the mass-loss rate is biased by the impact that companions can have  on the strength of the mass loss and on the observed diagnostics from which mass-loss rate values are retrieved. It goes without saying that this viewpoint touches directly on the \textit{deterministic} question, and --- unlike the small elephant --- this time we will not be bogged down in the minutiae of detail; a partner changes your life once and for all. 

\subsection{The key to finding the invisible partner}\label{Sec:partner_observations}

\begin{marginnote}[15truecm]
	\entry{Post-AGB star}{when the envelope mass of the AGB star constitutes less than $\sim$1\% of the stellar mass, the star becomes a post-AGB star; a phase which only takes a few thousand years before transiting to the PN phase}
\end{marginnote}

Binary companions orbiting mass-losing AGB stars somehow managed to escape the scientific picture.
Every so often the presence of companion was invoked to explain the metamorphosis from an overall spherical AGB wind to the aspherical morphologies seen in their descendants, the post-AGB stars and the planetary nebulae (PNe).
Indeed, already the first photographic atlas of PNe, published in \citeyear{Curtis1918PLicO..13....9C}, showed an astonishingly wide range of PNe morphologies and led \citeauthor{Curtis1918PLicO..13....9C} to question whether `it is possible to postulate any general form or forms, which shall be mechanically plausible, and to which the planetaries, or a considerable proportion of the planetaries, will more or less closely conform?'. Since then, the quest for the primary mechanical cause shaping PNe has been open. Whereas $\sim$80\% of AGB stars have a wind with overall spherical symmetry \citep{Neri1998A&AS..130....1N}, less than 20\,\% of PNe possesses a circular symmetry \citep{Parker2006MNRAS.373...79P, Sahai2011AJ....141..134S}. Naming only a few prominent examples imaged by, for example, the {\sl Hubble Space Telescope} (HST): the inner structure of the Eskimo Nebula, a bipolar double-shell PN, shows a complex `rose-like' structure  \citep{ODell2002AJ....123.3329O}, while the Helix Nebula is a bipolar PN with an `eye'-like morphology  \citep{ODell2004AJ....128.2339O,Su2007ApJ...657L..41S}. Biconical shapes are seen in various post-AGB stars and PNe, including the post-AGB star IRAS\,17150$-$3224 which also has a highly equatorially enhanced shell \citep{Ueta2007AJ....133.1345U} and the Owl Nebula (NGC~3587), a PN which has a barrel-like structure in its inner shell caused by bipolar cavities \citep{Guerrero2003AJ....125.3213G}. The Red Rectangle is an O-rich post-AGB binary system which has a Keplerian (rotating) disk and an outflow, the latter mainly being formed of gas leaving the disk \citep{Bujarrabal2016A&A...593A..92B}; in its outer halo regularly spaced arcs embedded in a bipolar outflow are detected \citep{Cohen2004AJ....127.2362C}.

For a long time, the development of these nonspherical structures during the post-AGB and PNe phase has been a matter of debate.
Several contending theories attempted to explain this morphological metamorphosis, including rapid stellar rotation and strong magnetic fields in single-star models  \citep{Garcia1999ApJ...517..767G}, and binary models with a particular focus on short-period binary systems formed after a common-envelope phase \citep[orbital period $P_{\rm{orb}}\la10$ days;][]{Soker1998ApJ...496..833S, Miszalski2009A&A...496..813M}. 
Even for the first two mechanisms, the presence of a binary companion is sometimes called upon to sustain the mechanism \citep{Sokder1992PASP..104..923S}.
In principle, the material ejected during the AGB phases does not have enough angular momentum to form Keplerian disks, which should only appear around binary
stellar systems, as these systems have the necessary angular momentum stored in their orbital movement. 
Axial structures in the form of collimated fast winds have then been proposed to be associated with such rotating disks, from which material would fall onto the star or a
companion during early post-AGB phases, powering very fast and collimated stellar jets  \citep[][]{Bujarrabal2001A&A...377..868B, Soker2001ApJ...558..157S, Ballick2002ARA&A..40..439B}. 
Most of these proposed PN shaping processes act on the AGB star itself and each of them is thought to operate over a short time, either during the final few hundred years of the AGB phase or during the early post-AGB phase.
The short lifetime of the post-AGB and PN phases, the strong observational bias toward detecting short-period binary post-AGB stars and PNe, and the high mass-loss rates at the end of the AGB phase leading to an obscuration of the  inner workings made the identification of the shaping mechanism and its time of occurrence observationally challenging.


\medskip

Aiming to understand some of the observable characteristics of axisymmetric or bipolar post-AGB stars and PNe, \citet{Mastrodemos1999ApJ...523..357M} were the first to perform pioneering hydrodynamical simulations of dusty red giant winds in detached binary systems ($P_{\rm{orb}}\!\ga\!1$\,yr), in which the effects of the companion are manifested on the wind of the red giant rather than the star itself. 
(See the sidebar titled  Binary interaction.) In these simulations, the primary star was a mass-losing red giant star and the companion's mass ranged between 0.25\,--\,2\,\Msun\ with orbital separation between 3\,--\,24\,\Rstar.
They derived a range of envelope geometries encompassing bipolar, elliptical and quasi-spherical geometries characterized by a continuously decreasing density contrast between the equatorial plane and the poles. The last category manifested a novel type of hydrodynamic wind solution in the form of  a spiral shock caused by the reflex motion of the mass-losing AGB star. Depending on the system's parameters and the inclination angle of the system, the two-dimensional projection of these spirals in the plane of the sky appears as an Archimedes spiral or as a series of rings in the wind. 
The year 1999 was also when the first-ever multiple, incomplete, concentric shells were detected in the envelope of the carbon star CW~Leo (\Mdot$\sim\!1.5\!\times\!10^{-5}$\,\Msun\,yr$^{-1}$) using the HST \citep{Mauron1999A&A...349..203M}. Probably unaware of the results by \citet{Mastrodemos1999ApJ...523..357M}, \citeauthor{Mauron1999A&A...349..203M} attributed the shells to mass-loss modulations with a time scale of $\sim$200\,--\,800\,yr, caused by an undefined episodic process intrinsic to the star. 
The impact of a binary companion was tentatively suggested, but was deemed insufficient to explain the structure of the shells farther out. 
Searching for the cause of the episodic process, \citet{Simis2001A&A...371..205S} suggested that an intricate non-linear interplay between gas-grain drift, grain nucleation, radiation pressure, and envelope hydrodynamics can result in gas and dust density shells that occur at irregular intervals of a few hundred years for a single-star model. However, the predicted shell density variations were much larger than the observed ones \citep{Decin2011A&A...534A...1D}.

It took an additional seven years before the same authors, \citeauthor{Mauron2006A&A...452..257M}, reported in 2006 the detection of the first conspicuous Archimedes spiral pattern in the wind of a carbon-rich AGB star (AFGL~3068, \Mdot\,=\,4.2$\times$10$^{-5}$\,\Msun\,yr$^{-1}$). 
This discovery was, and still is, highly significant and led to the unequivocal conclusion that binary interaction shapes the stellar wind during the AGB phase, and as such plays a role in carving out planetary nebulae. However, despite presenting this and analogous results (see below) to the scientific community, there still remains a fraction of skeptical astronomers  because the companion remains invisible to current instrumentation. 
As a renowned German astrophysicist told me after my lecture on 7~February 2020,  `You know, Leen, it all looks so fantastic, the observations are so fascinating, the current state-of-the-art models seem to do a pretty good job for interpreting the data, but in the end shouldn't we only believe what we actually can see?'. 
I remember that fraction of a second during which the old saying of Edgar Allan Poe (1845) jumped into my mind `Believe nothing you hear, and only one half that you see', but finally found myself arguing that much of astrophysics remains in the realm of conjectures, even \citeauthor{Einstein1915SPAW.......844E}'s theory of general relativity (1915) or, to bring it closer to my own field of expertise, even the fact that we believe that AGB stars are the ancestors of planetary nebulae. 
In a few exceptions, such as $o$~Ceti or W~Aql, the binary companion is actually seen \citep{Karovska1993ApJ...402..311K, Ramstedt2011A&A...531A.148R}, but the high luminosity outshines the faint companion for most sources.
 The {\sl Gaia} parallax' measurements can potentially give us some light at the end of the tunnel if corrections can be made for the turbulent AGB and RSG atmospheres \citep{Kervella2019A&A...623A.116K}.

 \begin{textbox}[htp]\section{Binary interaction}
While this is not the place to present a detailed physico-chemical description of binary interaction, the key point is easy to state: the gravitational potential of the companion alters the equation of motion and if the companion is UV-active, which is likely if it has lost its envelope, it can impact the chemistry of the surrounding envelope. 
The concept of Roche-lobe overflow (RLOF) has proven powerful in the description of binary evolution. The critical equipotential surface in the Roche potential, passing through the inner Lagrangian point $L_1$, defines two Roche lobes surrounding each star. The volume averaged radius of the Roche lobe can be approximated to an accuracy of  better than 1\% following \citeauthor{Eggleton1983ApJ...268..368E}'s formula (1983):
\begin{equation}
\frac{R_{L1}}{a}  = \frac{0.49\, q^{2/3}}{0.6 \,q^{2/3} + \ln(1+q^{1/3})}\,,
\nonumber
\label{Eq:Eggleton}
\end{equation}
where $q$ is the mass ratio $M_1/M_2$, and $a$ the orbital separation. Mass-transfer interaction from the primary to the secondary can be classified into four types in increasing order of interaction: Bondi–Hoyle–Lyttleton (BHL) accretion \citep{Hoyle1939PCPS...35..405H, Bondi1944MNRAS.104..273B}, wind-Roche lobe overflow  \citep[WRLOF;][]{Podsiadlowski2007BaltA..16...26P}, Roche lobe overflow (RLOF) when the Roche lobe surface is connected \citep{Paczynski1971ARA&A...9..183P}, and common envelope evolution  \citep[CEE;][]{Ivanova2013A&ARv..21...59I}. 
It is believed that for  wide binary systems ($a\!\ge\!2$\,au) the mass transfer does not occur through RLOF, but via WRLOF in which case wind material from the mass-losing AGB star fills the giant's Roche lobe and is transferred to the companion through a compressed channel which generally does not pass through $L_1$. 
\end{textbox}

\medskip
The opening of the {\sl Atacama Large Millimeter/submillimeter Array} (ALMA) in 2012 heralded a leap forward in the understanding of wind morphologies. 
Observational evidence was steadily accumulating that AGB and RSG winds exhibit large correlated density structures --- including arcs, shells, bipolar structures, tori, rotating disks, and spirals --- embedded in a smooth, radially outflowing wind \citep[see \textbf{Figure~\ref{Fig:3D}};][]{Maercker2012Natur.490..232M,  Ramstedt2014A&A...570L..14R, Kim2015ApJ...814...61K,  Decin2015A&A...574A...5D, Kervella2016A&A...596A..92K, Wong2016A&A...590A.127W, Guelin2018A&A...610A...4G, Homan2018A&A...616A..34H, Homan2018A&A...614A.113H, Bujarrabal2018A&A...616L...3B, Ramstedt2018A&A...616A..61R, Decin2019NatAs...3..408D, Randall2020A&A...636A.123R}. 
At the same time, we saw a revival of hydrodynamical simulations for binary systems in which the primary is a mass-losing AGB star. 
Depending on the parameters of the system --- such as binary separation, mass ratio, eccentricity, rotation and pulsations of the AGB star, mass-loss rate, wind velocity, $\ldots$ --- a wide variety of morphologies is predicted, including: 
(1) a spiral structure caused by the orbital motion of the mass-losing AGB star around the common center-of-mass, or by the accretion wake of the companion owing to a Bondi-Hoyle-Lyttleton (BHL) flow (see \textbf{Figure~\ref{Fig:binary}}), that can be bifurcated for a  non-circular orbit; 
(2) a circumbinary disk; 
(3) an accretion disk around the secondary; 
(4) a bipolar outflow, that can display a ripple-like structures if the AGB pulsations are included;
(5) an equatorial density enhancement (EDE), with a regular (Keplerian) or complex velocity vector field; 
(6) and even `spider' or `rose-like' structures
\citep[see \textbf{Figure~\ref{Fig:3D}};][]{Kim2012ApJ...759...59K,Chen2017MNRAS.468.4465C,Liu2017ApJ...846..117L,
Saladino2018A&A...618A..50S,Saladino2019A&A...629A.103S,Kim2019ApJS..243...35K,ElMellah2020arXiv200104482E,Chen2020ApJ...892..110C}.
In general, smaller values of the orbital separation or wind velocity and larger companion masses induce stronger interaction. 
These outcomes can be formally expressed in a similar way as Eq.~\eqref{Eq:math_model_Mdot} or Eq.~\eqref{Eq:math_model_Mdot2}, although it is obviously seen that the number of independent parameters is far larger.  Therefore, the amount of studies in which the model sensitivities can be tested for a large grid of parameters is still very restricted by the huge demand on CPU time and memory required.

\begin{figure}[htp]
\centering
\includemedia[width=.95\textwidth,activate=onclick,passcontext,transparent,addresource=videos/DecinVideo8.mp4,flashvars={source=videos/DecinVideo8.mp4}]{\includegraphics[width=.95\textwidth]{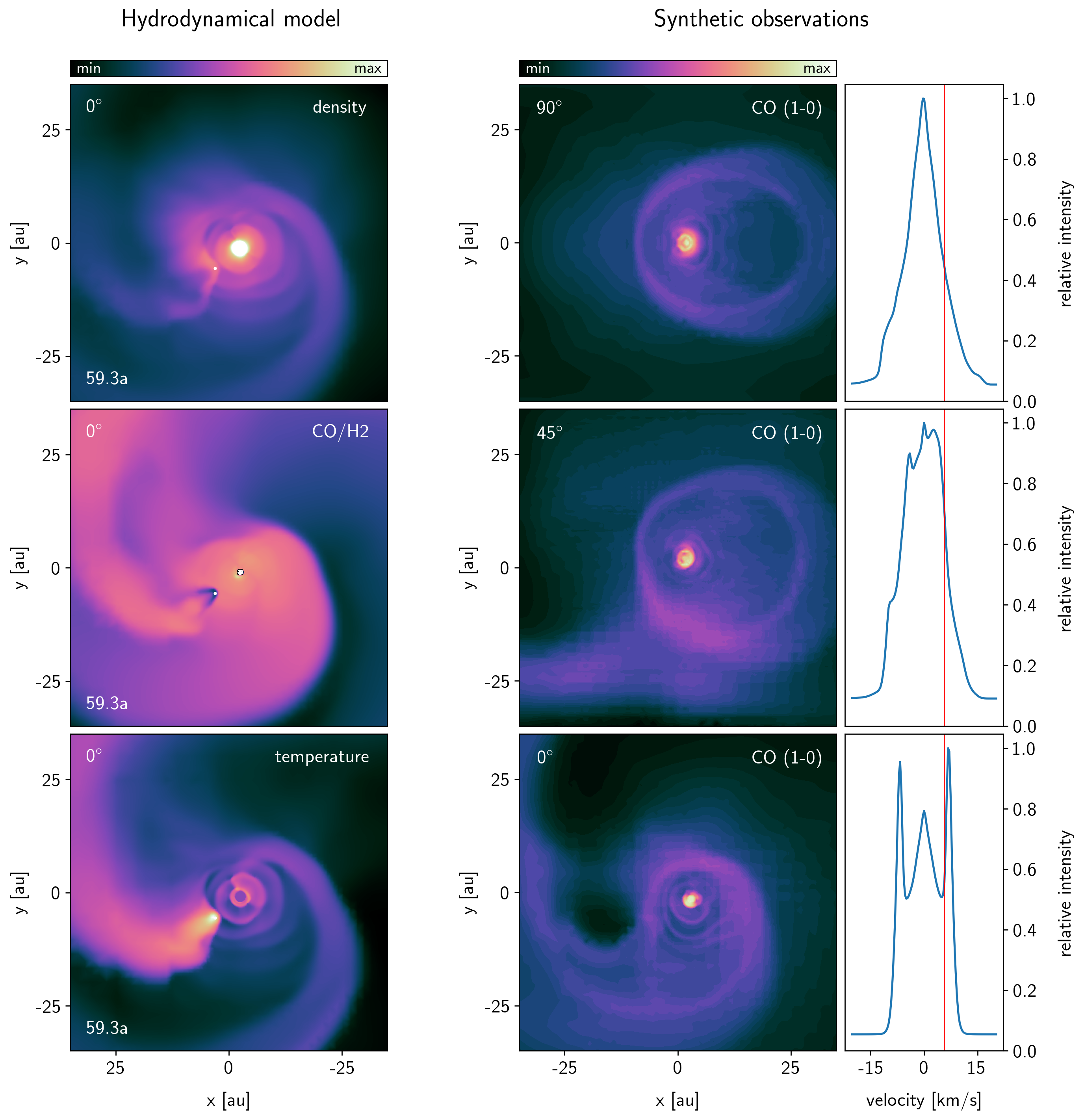}}{VPlayer.swf}
\caption{3D hydro-chemical simulation for a binary system containing a mass-losing AGB star. The left panels are a video showing the total density (top panel),  CO/H$_2$ number density (middle panel), and temperature (bottom panel) for a binary system model of which the AGB star has a mass of 1\,\Msun, effective temperature of 2\,900\,K, a radius of  0.9\,au, and a pulsation period of 1 year. The companion has a mass of 0.5\,\Msun\ and resides at a circular orbit with separation of 10\,au. The simulation time runs for 59.3\,yr.
Owing to dust formation occurring in the region where the temperature is lower than 1\,500 K, a wind is initiated with mass-loss rate of 4.7$\times$10$^{-6}$\,\Msun\,yr$^{-1}$. The formation of two types of spiral structures can readily be seen, one structure being caused by the gravity wake near the companion, the other one owing to the reflex motion of the AGB star. Both spiral structures merge at larger distances from the AGB star. The small ripples in the close vicinity of the AGB star are relics of the pulsation pattern. The same setup for the AGB star not having a companion yields a mass-loss rate of 7.6$\times$10$^{-7}$\,\Msun\,yr$^{-1}$ (Bolte et al.\ \textit{in prep.}). For a simulation time of 59.3\,yr, the hydrodynamical quantities are then displayed for viewing angles of the system ranging between 0\deg (edge-on view) and 90\deg (face-on view).
The right panels show the corresponding CO v=0 J=1-0 emission map and line profile at $t\!=\!59.3$\,yr for three different viewing angles (at 90\deg -- top panel, at 45\deg -- middle panel, and at 0\deg -- bottom panel)
in the observer's frame calculated  using the {\sc Magritte} 3D radiative transfer solver \citep{DeCeuster2020MNRAS.492.1812D, DeCeuster2021}. The video slices through the velocity channel map between $-$20 to $+$20\,km\,s$^{-1}$. Figure courtesy F.\ De Ceuster and J.\ Bolte.  }
\label{Fig:binary}
\end{figure}

The resemblance between the observed and theoretical morphologies supports the claim that binary interaction is a key architect   of AGB and RSG wind shaping. 
However, it also should be acknowledged that the observations are often more complex than any model prediction, owing to simplification in the models and to the fact that other mechanisms, such as magnetic fields, might contribute to an initial wind anisotropy. 
Generally, deducing the model functions from the behaviour of a complex system is an inverse problem that is difficult to solve, but I contend that this approach will see a significant growth in the coming decade thanks to increasing CPU power and, in particular, the use of artificial intelligence; see for example the paper by  \citet{DeMijolla2019A&A...630A.117D}. 

\medskip

\begin{figure}[htp]
	\centering	\includegraphics[width=.8\textwidth]{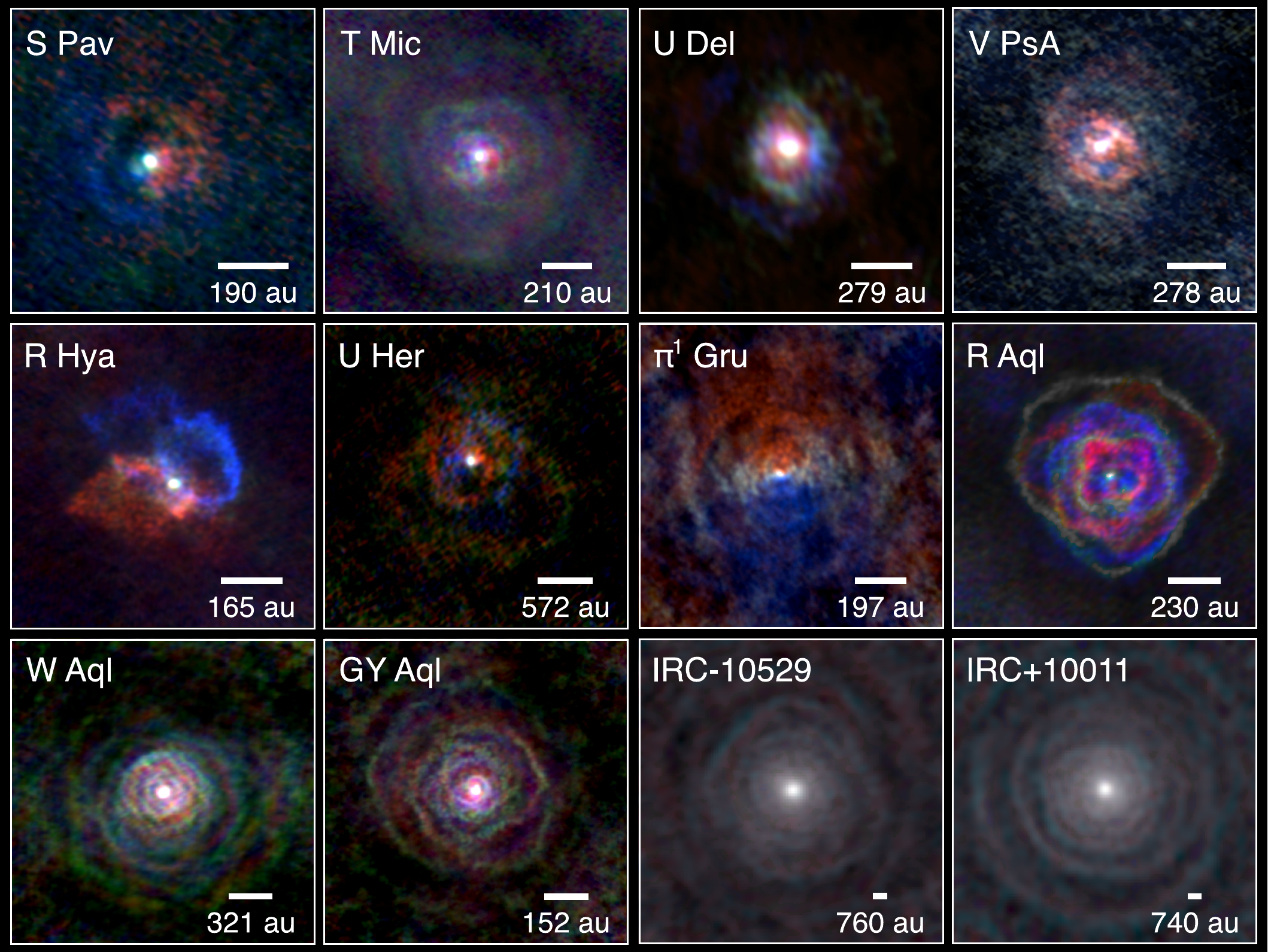}
	\caption{Gallery of AGB winds. Emission maps of 12 AGB stars are shown, derived from the {\sc atomium}
		$^{12}$CO v=0 J = 2-1 data. For each star, emission that is redshifted with respect to the local standard of rest
		velocity is shown in red, blueshifted emission is in blue, and rest velocity is in white. The scale bars have an
		angular extent of 1\arcsec. 
		Sources are ordered by increasing mass-loss rate, from left to right, and from top to bottom.
		Figure based on Figure~1 of \citet{DecinScience}.}
	\label{Fig:ATOMIUM}
\end{figure}

Not only will a systematic approach elucidate the intricacies of binary interactions based on theoretical predictions, the same truism holds for observations. In 2018 the ALMA {\sc atomium}\footnote{{\sc atomium}: ALMA Tracing the Origins of Molecules In dUst-forming oxygen-rich M-type stars; https://fys.kuleuven.be/ster/research-projects/aerosol/atomium/atomium.}  Large Programme was granted substantial time on the telescope and a well-selected sample of oxygen-rich AGB and RSG stars was observed in a systematic and unbiased way with the specific aim of understanding the thermodynamical, chemical, and morphological properties of their stellar winds \citep{DecinScience, Gottlieb2020}. The {\sc atomium}  data revealed that \textit{all} AGB winds observed exhibit distinct aspherical geometries (see \textbf{Figure~\ref{Fig:ATOMIUM}}), which have morphological counterparts in the PNe. This led to the inference that the same physics is key in shaping both AGB winds and PNe. The {\sc atomium} data catch the wind-shaping mechanism in the act and constrain the moment in time when AGB morphologies are being transformed into nonspherical geometries.
A strong statistical correlation emerges between the AGB mass-loss rate and the prevailing geometry \citep{DecinScience}: a dynamically complex EDE is often observed for oxygen-rich AGB stars with low mass-loss rates (\Mdot$\,\la$\,10$^{-7}$\,\Msun\,yr$^{-1}$; defined there as ‘Class~1’), a bipolar structure tends to be dominant for stars with medium mass-loss rates (‘Class~2’; 10$^{-7}$\,\Msun\,yr$^{-1}$$\,\la\,$\Mdot$\,\la$\,10$^{-6}$\,\Msun\,yr$^{-1}$), while the wind of high mass-loss rate stars preferentially exhibit a spiral-like structure (‘Class~3’; \Mdot$\,\ga$\,10$^{-6}$\,\Msun\,yr$^{-1}$). This correlation and the observed characteristics are readily explained by binary interaction. The results allowed \citet{DecinScience} to propose an evolutionary scenario for AGB wind morphologies in which early-type AGB stars will often be characterized by an EDE, with complex flow patterns (Class~1), and the wind of late-type high mass-loss rate AGB stars will mainly be shaped by spiral structures (Class~3). 
The proposed evolutionary scheme for AGB wind morphologies can explain various AGB, post-AGB, and PNe phenomena, including for instance why post-AGB star binaries can have nonzero eccentricities \citep{DecinScience}.
The binary scenario gets support from an analysis of the kinematic wind properties \citep{Gottlieb2020}. 
Moreover, it was shown that early-type AGB stars with a low mass-loss rate are prime candidates for detecting planets. 
This prediction aligns with the detection that Jupiter-sized companions reside in the near vicinity of the low mass-loss rate AGB stars L$_2$~Pup and R~Dor \citep{Kervella2016A&A...596A..92K, Homan2018A&A...614A.113H}.

In any case, regardless of the classification proposed by \citet{DecinScience}, it should be emphasized that the same set of observations may on occasion be interpreted in different ways and  may lead to the recognition of different categories. 
An immediate example might be that instead of the mass-loss rate, the wind acceleration is of greater importance for dictating 
the morphological outcome \citep[see the discussion in Section S3.3 in][]{DecinScience}.
When a system can be categorized in more than one way the question as to which category is better may depend on the particular phenomenon under study. 
The mapping is probably not simple, and the sample size needs to be significantly expanded before we can investigate the statistical aspects 
 in detail.

\subsection{Statistics to probe the invisible}\label{Sec:partner_statistic} 

The following logical proposition
($\models \mathcal{S}\!: A \vee \neg A$) is a clear tautology of propositional logic as introduced by the great twentieth-century philosopher Ludwig \citet{Wittgenstein1921} in his famous \textit{Tractatus}. 
This logical tautology can be paraphrased by saying \textit{`a star has either at least one companion, or a star lives its life alone'}. 
I am of the view that this tautology might vanish in the foreseeable future with only the first part remaining --- not only for massive stars, for which it has been established that binary interaction dominates their evolution \citep{Hugues2012Sci...337..444S}; but also for the low and intermediate-mass stars, which are the focus of this review.
It is quite obvious that this statement is inspired by one of the most significant developments in modern astronomy that we seem to find planets around stars almost everywhere we look.
To assess this hypothesis, we consider another discipline in mathematics: statistics. 

We therefore summarize the current state of knowledge concerning the occurrence rate of stellar and planetary companions around low and intermediate mass stars. 
To do so, I borrow heavily from Section S5 in the Supplementary Material of \citet{DecinScience} and references therein. 
For convenience, I here reproduce \textbf{Table~\ref{Table:binary}} which focusses on stars with initial mass, $M_{\rm{ini}}$, between 0.8\,--\,8\,\Msun. 
It is well established that the binary fraction increases for higher initial masses, in this case for the more massive stars with $M_{\rm{ini}}\!\ga\!8$\,\Msun\ that will evolve to the RSG stage, and that some fractions listed in \textbf{Table~\ref{Table:binary}} might be lower limits owing to limitations in detection techniques.

\begin{table}[htp]
\caption{Summary of main-sequence (sub-)stellar multiplicity fraction. \label{Table:binary}}
\setlength{\tabcolsep}{2.00mm}
\begin{tabular}{ll|cccc}
\hline
\rowcolor{shadecolor}
 & & \textbf{Stellar}           & \textbf{Brown}    &            \multicolumn{2}{c}{\textbf{Planets} $\mathbf{M\!>\!5\,M_{\rm{Jup}}}$}   \\
\rowcolor{shadecolor}
 & & \textbf{companions} & \textbf{dwarfs}    & $\mathbf{a\!=\!2\!-\!10}$\,\textbf{au}                       & $\mathbf{a\!=\!10\!-\!100}$\,\textbf{au} \\
 \hline
 0.8$<\!M_{\rm{ini}}\!<\!1.5$\,\Msun & FGK & $\sim$30--37\% & $\sim$0.8\% & $\sim$7\% & $\sim$9\% \\
 1.5$<\!M_{\rm{ini}}\!<\!8$\,\Msun & AB & $\sim$52--60\% & $\sim$0.8\% & $\sim$7\% & $\sim$40\% \\
 \hline
\end{tabular}
\begin{tabnote}
\noindent Notes: The first and second columns give the main-sequence initial mass and related spectral type on the main-sequence. Columns~3--6 list the (sub)-stellar multiplicity fraction for companions with $\log P_{\rm{orb}} \rm{(days})\!<\!6.5$ and $\log P_{\rm{orb}} \rm{(days)}\!>\!2.7$ (or 3) in the case of planets (respectively, stars).
\end{tabnote}
\end{table}

\textbf{Table~\ref{Table:binary}} is constructed on the arguments that we are looking at those binary systems for which the orbital separation (and hence orbital period $P_{\rm{orb}}$) and the companion mass are such that hydrodynamical simulations indicate a perturbation of the smooth envelope structure. This leads to the conservative estimate of $\log P_{\rm{orb}}  \rm{(days)}\!<\!6.5$ and $\log P_{\rm{orb}} \rm{(days)}\!>\!2.7$ (or 3) in the case of planets (or respectively, stars).  Using these boundaries, we assess the main-sequence stellar and sub-stellar multiplicity factor (see \textbf{Table~\ref{Table:binary}}; \citet{Moe2017ApJS..230...15M, Nielsen2019AJ....158...13N, Fulton2018AJ....156..264F, Fulton2019talk}). In a final step, one should account for the evolution of the (sub-)stellar binary fraction from the main-sequence to the AGB (and RSG) phase. A rough estimate is that the binary fraction will decrease by $\sim$10\,--\,20\% as stars evolve toward and on the AGB/RSG phase \citep{DecinScience}.

A last realisation is adopted from population statistics. Based on stellar evolution models, it can be argued that the majority of AGB (and RSG)  stars with a mass-loss rate above $10^{-7}$\,\Msun\,yr$^{-1}$ are of mass above 1.5\,\Msun\ \citep{DecinScience}. Stars of lower mass will only have a very short period in their AGB phase during which the mass-loss rate is greater than $10^{-7}$\,\Msun\,yr$^{-1}$ before they transit into the PN phase, limiting their detection probability. Combining this outcome with \textbf{Table~\ref{Table:binary}} leads to the specific conjecture that most AGB stars with mass-loss rate above $10^{-7}$\,\Msun\,yr$^{-1}$ have masses above $\sim$1.5\,\Msun\ and hence have, on average, $\ga$1 companion(s) with masses above $\sim$5 Jupiter masses. Hence (sub-)stellar binary interaction is the prime wind shaping agent of the majority of cool ageing stars for which the mass-loss rate exceeds the nuclear burning rate (\Mdot$\ga\!10^{-7}$\,\Msun\,yr$^{-1}$), in which case mass loss rules stellar evolution \citep{DecinScience}. 

In conclusion, it took us some time to realize that \textit{roses} flourish during the end stages of stellar evolution (see \textbf{Figure~\ref{Fig:3D}}),  but resorting to one of the best-selling and most translated books ever published  `C'est le temps que tu as perdu pour ta rose, qui fait ta rose si importante' \citep[`It's the time you spent on your rose that makes your rose so important';][]{Saint1943}. 
The question then is how that rose impacts on our \textit{deterministic} insight in the mass-loss rate of AGB and RSG stars.

\subsection{How can a companion change a stellar life}\label{Sec:partner_impact}

Indeed, the stellar and planetary companions not only serve to beautify the stellar winds during the end stages a star's life, they can also have a direct impact on the star's evolution. For close binary systems, a stellar companion or a massive planet can enhance the mass-loss rate by depositing angular momentum into the envelope and by reducing the effective gravity of the mass-losing star. 
Stars that are born single or binary stars isolated from angular momentum deposition hence might suffer from a lower mass-loss rate than stars prone to angular momentum deposition \citep{Sabach2018MNRAS.479.2249S}; for the example shown in \textbf{Figure~\ref{Fig:binary}} this effect is a factor of $\sim$6. This conclusion has a serious repercussion on any empirically \textit{retrieved} mass-loss rate relation discussed in Section~\ref{Sec:Mdots_theor_emp}.  Building on the discussion in Section~\ref{Sec:partner_statistic}, we conjecture that most empirically \textit{retrieved}  mass-loss rates yield mass-loss rate measures that are too high for application in single-star evolution models, since samples of stars will be flawed by a large fraction of stars that experience binary interaction \citep{Gottlieb2020}. 
This brings us back to one of the pitfalls of \textit{retrieval} approaches outlined in Section~\ref{Sec:Setting_theory}, and the caution 
expressed there about unrecognised bias effects in sample selections and the elemental difference between correlation and causal effects.

There is a second no less fundamental problem with empirically \textit{retrieved}  mass-loss rates in the case of unrecognised binary interaction. 
For a mass-losing AGB/RSG star in a binary system, the material will have a directional preference towards the orbital plane, and an equatorial density enhancement (EDE) will form. 
The density contrast between the equator and the pole increases for smaller orbital distance, lower wind velocity and higher companion masses \citep{ElMellah2020arXiv200104482E}, with density contrasts up to an order of magnitude. 
However, dust mass-loss rates --- which in a next step are converted to gas mass-loss rates --- are most often derived from the analysis of near to mid-infrared spectral energy distributions (SEDs) that mainly trace warm dust residing close to the (primary) star, hence in the EDE \citep{Decin2019NatAs...3..408D}. 
Therefore, the analysis of dust spectral features with a simplified 1D approach reflects the higher density in the EDE created by the binary interaction, but not the actual mass-loss rate which will be lower by up to an order of magnitude \citep{Wiegert2020A&A...642A.142W}.
This deduction amplifies the resolution of the previous paragraph, and leads to the inference that scientists modelling single-star evolution by applying empirically \textit{retrieved}  mass-loss rate formulae should be very cautious because the rates might be seriously overestimated. 
It is readily seen that this conclusion directly impacts any estimate of the chemical enrichment of the ISM by cool ageing stars.
In what follows, I will make some suggestions and predictions that flow directly from the above conclusions.

The only prescriptions currently devoid of a binary-induced bias are the \textit{theoretical} and --- to a large extent --- the \textit{semi-empirical} mass-loss rate formulae (see \textbf{Figure~\ref{Fig:Mdots}}). 
However, these mass-loss rate prescriptions for AGB stars are only meant to describe the properties of Mira variables, 
while the only RSG \textit{theoretical} mass-loss prescription from \citet{Kee2020} hinges on knowledge of the turbulent velocity which is barely constraint from observations and which does not yet account for any radial or tangential change.
Supported by almost a billion CPU hours at HPC facilities for theoretical modelling purposes and the provision of specialised computing facilities for data analysis, several groups are now collaborating with the aim 
to provide the community with improved mass-loss rate prescriptions accounting for this 3D binary perspective.

\begin{marginnote}[15cm]
	\entry{Type~IIP supernovae}{Type~II supernovae can be further sub-classified  into  II-P  (plateau  light  curves),  II-L  (linear  decline light curves), IIn (narrow emission lines) and some peculiar events, generically labelled II-pec}
\end{marginnote} 

\
\subsubsection{The red supergiant problem} \label{Sec:RSG_problem}

The preceding discussion will assist us in addressing the  `red supergiant problem', which concerns the question 
as to why the observed upper limit of $\sim$16\,\Msun\ on the masses of Type IIP SN progenitors appears to be significantly lower than the 
maximum mass of $\sim$30\,\Msun\ for stars expected to explode while they are red supergiants \citep{Smartt2009MNRAS.395.1409S}.
Numerous explanations have been considered including: a steeper initial mass function \citep{Smartt2009MNRAS.395.1409S};
the loss of the loosely bound hydrogen envelope of the most massive RSG stars prior to core collapse \citep{Yoon2010ApJ...717L..62Y};
the proposition that enhanced mass loss might limit the RSG mass to $\sim$20\,\Msun\ \citep{Groh2013A&A...558A.131G};
or the fact that massive stars collapse to black holes with optically dark or faint `failed' SNe \citep{Woosley2012ApJ...752...32W}. 
Another direction was to argue that for stars whose mass is greater than $\sim$15\,\Msun, 
the mass-loss rate should be higher than that currently considered in stellar evolution codes \citep{Ekstrom2012A&A...537A.146E};
as cause for this increase in mass-loss rate, it was suggested that some of the most external layers of the stellar envelope might exceed the Eddington luminosity. 
Each of these explanations is grounded on uncertainties in the theoretical models, with the latter proposal of \citet{Ekstrom2012A&A...537A.146E} seeming mutually incompatible with the conjecture formulated at the start of Section~\ref{Sec:partner_impact}.
\begin{marginnote}[]
\entry{Eddington luminosity}{maximum luminosity a star can achieve for there to be a balance between the radiation force acting outward and the gravitational force acting inward}
\end{marginnote}

However, the puzzle that red supergiants with masses $\sim$16\,--\,30\,\Msun\ have not been identified as progenitors of Type~IIP supernovae can also be examined from the perspective of retrieval analyses.  An insufficient circumstellar dust correction for the pre-SN mass loss could bias the progenitor mass estimates to lower values \citep{Walmswell2012MNRAS.419.2054W}. 
Although the  SED analysis of \citet{Walmswell2012MNRAS.419.2054W} was criticized by \citet{Kochanek2012ApJ...759...20K}, the underlying rationale can be reviewed in the light of the scientific discourse embarked upon at the start of Section~\ref{Sec:partner_impact}. Admittedly, the following discussion is narrowly confined to this one characteristic of retrieval modelling, and other potentially influential characteristics are not considered. 

A first point of consideration deals with the binary hypothesis, and the well-established issue that the binary fraction increases with stellar mass. 
The larger the stellar mass of the primary star, the larger the fraction of companions that will inspiral owing to the larger gravitational attraction by the primary star \citep{DecinScience}. 
For a binary system with a decreasing orbital distance, the EDE will get more pronounced. 
 Higher densities in the EDE promote the formation of (crystalline) dust grains \citep{Decin2019NatAs...3..408D}. Hence, it is to be expected that stars of higher initial mass can have a higher dust mass in their circumstellar envelope, so  the circumstellar dust correction is weighted higher in the final estimate of the pre-SN mass loss.
A second manner of reasoning touches upon the issue of 3D clumps. Following the argument of \citet{Schwarzschild1975ApJ...195..137S}, the size of the convection bubbles for stars of higher mass (and larger radius) will be larger. Using the same diameter-to-depth ratio of 3 as \citet{Schwarzschild1975ApJ...195..137S}, this implies that the depth --- and corresponding dust optical depth ---  might be larger for more massive stars so the impact of a proper circumstellar dust correction is larger. 
These assertions permit the extreme hypothesis that the red supergiant problem is potentially not so severe as currently considered, and  the real solution can only be formulated under the condition of a detailed 3D analysis of the pre-SN mass loss.  
Of course this is not a Solomon-like resolution, but  only one step in a holistic approach of tackling this dilemma. Recent studies even argue that the upper limit cut-off is likely to be higher and has large uncertainties ($M$\,=\,19$^{+4}_{-2}$\,\Msun), implying that the statistical significance of the RSG problem is less than 2$\sigma$ \citep{Davies2020MNRAS.493..468D, Davies2020MNRAS.496L.142D}.
\begin{marginnote}[]
\entry{Solomon-like resolution}{named after King Solomon of Israel (Hebrew Bible), a famous practitioner of dispute resolution}
\end{marginnote}

\subsubsection{Survival rate of (exo)planets}\label{Sec:planets} 

The future of our own planet, the Earth, seems far from smooth sailing. 
Cogent modelling by \citet{Schroder2008MNRAS.386..155S} suggests that the Earth will be swallowed by the time that the Sun is near its tip of the red giant branch (RGB) evolution, $\sim$7.6\,Gyr from now. At that moment, the solar radius will reach $\sim$1.2\,au, while the Earth's orbit will hardly ever exceed 1\,au by a significant amount. 
This is caused by a competition between orbital widening induced by the solar RGB wind reducing the Sun's mass, and orbital tightening owing to dynamical drag with the lower chromosphere and to orbital angular momentum loss due to tidal interaction with the giant Sun. 
Tidal interaction is dominant so the planet Earth cannot escape engulfment. 
Although not all parameters are invariably rigorously defined ---  in particular the tidal friction coefficient --- it seems that doomsday is unavoidable.  
There is a future beyond the Sun's RGB evolution only for planets whose current orbital separation is greater than $\sim$1.15\,au.
\begin{marginnote}[]
	\entry{RGB phase}{follows the main-sequence phase; stars have an inert helium core surrounded by a shell of hydrogen fusion}
	\end{marginnote}

A larger RGB mass-loss rate would delay the engulfment and lead to smaller orbital separations for planets still able to survive.
In their modelling approach,  \citet{Schroder2008MNRAS.386..155S} used an adapted form of Reimers' law (\citeyear{Reimers1975MSRSL...8..369R}) to describe the  cool wind --- not driven by dust but presumably by (magneto)acoustic processes --- which was physically motivated by a consideration of global chromospheric properties and wind energy requirements
\begin{equation}
	\Mdot = 4 \times 10^{-13} \eta_{\rm{SC}} \frac{\Lstar \Rstar}{\Mstar} \left( \frac{T_{\rm{eff}}}{4000\,{\rm{K}}}\right)^{3.5} \left(1 + \frac{g_\odot}{4300\,g_\star}  \right)\,,
	\label{Eq:Mdot_Reimers_adapted}
\end{equation}
with $\eta_{\rm{SC}}\!=0.2$ and $g_\odot$ the solar surface gravitational acceleration. 
Following Eq.~\eqref{Eq:Mdot_Reimers_adapted}, the mass-loss rate of the giant Sun would be around $4\!\times\!10^{-8}$\,\Msun\,yr$^{-1}$ at the tip of the RGB.
The feeble RGB mass loss is difficult to constrain from observations.
The first detection of rotational CO line emission arising from  an RGB wind was announced in \citeyear{Groenewegen2014A&A...561L..11G}; the derived mass-loss rate is around a few $10^{-9}$\,\Msun\,yr$^{-1}$ \citep{Groenewegen2014A&A...561L..11G}.
A promising  way to measure the RGB mass loss is to obtain a difference in stellar mass between two points in its evolution, and use this to determine a scaling parameter such as $\eta$ in Eq.~\eqref{Eq:Mdot_Reimers} or $\eta_{\rm{SC}}$ in Eq.~\eqref{Eq:Mdot_Reimers_adapted}  \citep{McDonald2015MNRAS.448..502M}.
This can be achieved in globular clusters, where one can probe the individual stars with accurately known distances, metallicities, abundances and ages. 
Using 56 well-studied global clusters, \citet{McDonald2015MNRAS.448..502M} derived as median values $\eta\!=\!0.477 \pm 0.070$ and $\eta_{\rm{SC}}\!=\!0.172 \pm 0.024$, with very little metallicity dependence.
This provides support to a Reimers-like law being a good RGB mass-loss model and to the model predictions of \citet{Schroder2008MNRAS.386..155S}.

Given the conjecture that AGB mass-loss rates for single stars or stars isolated from angular momentum deposition might be overestimated, the follow-up question  then deals with the impact of that conjecture on the survival rate of planets. At first sight, this question might seem quite semantic, given the current lack of \textit{directly} imaged planets in the close vicinity of cool ageing stars and remembering the critical remark on that matter (see Section~\ref{Sec:partner_observations}). But to quote the old maxim: `Absence of evidence is not evidence of absence'. Given the large amount of exoplanets currently detected and expected to be present in the Universe, and our current knowledge of stellar and planetary evolution, this inquiry on the survival potential of (exo)planets gets more substantial. 
For a reduced AGB/RSG mass-loss rate the predictions are, however, not at all auspicious since the change of orbital separation, ${\rm{d}}a/{\rm{d}}t$, scales with the mass-loss rate. The lower mass-loss rate implies a lower rate of orbital widening, if occurring, and hence more (exo)planets experiencing a catastrophic encounter with their giant mother star due to tidal interaction. Therefore escaping the doomsday scenario is more challenging than previously suggested.

\section{The 3D life of molecules and dust grains}\label{Sec:molgrains}

To comprehend the essence of mass loss  requires an understanding of the driving forces that dictate the nature of the phenomenon. Delving into the \textit{why} question will provide feedback to the \textit{how} question, and vice-versa. 
Our view of the essential processes has gradually been enriched thanks to the enormous capacities of modern telescopes and space observatories designed for optical, IR, and microwave  investigations of cool astronomical objects. 
Moreover, a new window for exploring dust formed in stellar outflows commenced  in 1987 with the detection of small particles in the matrix material of certain meteorites which originated from stars whose lives ended before the formation of our Solar system: the  presolar grains \citep{Nittler2016ARA&A..54...53N}. Long before this detection, \citet{Cameron1973IAUS...52..545C} speculated about `Interstellar grains in museums?' and concluded that primitive carbonaceous chondrites may harbour presolar grains, but it was \citet{Lewis1987Natur.326..160L} who
stated `Interstellar dust contains diamonds' in the final phrase  in the abstract of their  paper in \textit{Nature}. 
Before you consider any space exploration, the diamonds were  tiny --- i.e., only about $\sim$10\,\AA.

This is where we reach the other spectrum of astrophysics, the regime where the small pieces make up the big picture. 
These small pieces seldom bear any resemblance with smooth spherical particles, not even for the micrometer-sized dust grains (see \textbf{Figure~\ref{Fig:3D}}). For chemists, micrometer-sized dust grains are `macroscopic'. Even when staying within the realm of solid-state physics, it is well established that further size reduction from the macroscopic to length scales of only a few nanometers can lead to dramatic changes in a material's properties, including its atomic ordering, quantum, and surface effects \citep[e.g.,][]{LamielGarcia2017, Gobrecht2017ApJ...840..117G}. 
Wondering then how solid-state species form, we enter the world of atoms and molecules which undergo a sequence of collisional association and recombination reactions to form 
polyatomic molecules of ever-increasing size, the gas-phase clusters (see \textbf{Figure~\ref{Fig:3D}} and \textbf{Supplemental Text}). 
Under favourable conditions  the clusters condense,  and form the first dust seeds consisting of 10 to several hundred atoms. 
The seed nuclei  become the substrate to which molecules are added, resulting in further growth of the condensed phase which eventually 
can reach the macroscopic regime. By the end of this process, the macroscopic dust species involve the accumulation of typically $10^6$\,--$10^9$ atoms  in a single grain.

\subsection{Molecules and dust grains identified in stellar outflows} \label{Sec:molgrains_observations} 

\subsubsection{Molecules} Observatories that operate in the micrometer and millimeter-wave  bands such as the 
{\sl Infrared Space Observatory} (ISO), the {\sl James Clerk Maxwell Telescope} (JCMT), the {\sl IRAM-30m} telescope, 
the {\sl Plateau de Bure Interferometer} (PdBI), and ALMA have boosted the detection of cosmic molecules via their rotational and 
vibration-rotation spectroscopic fingerprints. 
The wavelength resolution of most telescopes in the mid and near-IR
is moderate, $\Delta \lambda / \lambda\!\sim\!10^{-4}$, compared with gas-phase spectra whose
frequencies can be determined to an accuracy of about 1 part in $10^{7}$ or better in
narrow line sources with heterodyne receivers on radio telescopes.
In his excellent review, \citet{Olofsson2005IAUS..231..499O} provided a table listing the detection  of 63 molecules in AGB winds. \textbf{Table~\ref{Table:molecules}} gives an update fifteen years later: currently 105 molecules have been detected in stellar outflows. This is roughly 50\% of all molecules currently discovered in outer space: the extensive review by \citet{McGuire2018ApJS..239...17M} lists  204 molecules, to which 6 other ones can be included \citep[MgCCH, MgC$_3$N, MgC$_4$H, CaNC and potential detections of NCCP and FeO;][]{Agundez2014A&A...570A..45A, Cernicharo2019A&A...630L...2C, Cernicharo2019A&A...627L...4C, Decin2018ApJ...855..113D}. 
I opted to include in \textbf{Table~\ref{Table:molecules}}  the polycyclic aromatic hydrocarbon molecules (PAH) since their characteristic C-H and C-C stretching and bending modes have been detected in AGB stars \citep{Boersma2006A&A...447..213B, Smolders2010A&A...514L...1S}, although the  precise carriers have not been determined.

\begin{table}[htp]
\caption{Molecules identified in the winds of cool evolved stars \label{Table:molecules}}
\begin{tabular}{a|llllllll}
\hline
\textbf{2-atoms} &  AlCl &  AlF &  AlO &  C$_2$ &  CN &  CO \\ 
 &  CP &  CS &  FeO\,(?) &  HCl &  HF &  KCl \\ 
 &  NO &  NS &  NaCl &  OH &  PN &  PO \\ 
 &  SO &  SiC &  SiN &  SiO &  SiS &  TiO \\ 
 &  VO &   &   &   &   &   \\ 
\textbf{3-atoms} &  AlNC &  AlOH &  C$_2$H &  C$_2$S &  C$_3$ &  CCN \\ 
 &  CCP &  CO$_2$ &  CaNC &  FeCN &  H$_2$O &  H$_2$S \\ 
 &  HCN &  HCP &  HNC &  KCN &  MgCN &  MgNC \\ 
 &  NaCN &  SO$_2$ &  SiC$_2$ &  SiCN &  SiCSi &  SiNC \\ 
 &  TiO$_2$ &   &   &   &   &   \\ 
\textbf{4-atoms} &  $c$-C$_3$H &  $l$-C$_3$H &  C$_2$H$_2$ &  C$_3$N &  C$_3$O &  C$_3$S \\ 
 &  H$_2$CO &  H$_2$CS &  HC$_2$N &  HMgNC &  MgCCH &  NCCP\,(?) \\ 
 &  NH$_3$ &  PH$_3$ &  SiC$_3$ &   &   &   \\ 
\textbf{5-atoms} &  $c$-C$_3$H$_2$ &  C$_4$H &  C$_5$ &  CH$_2$CN &  CH$_2$NH &  CH$_4$ \\ 
 &  H$_2$CCC &  HCCCN &  HCCNC &  HNC$_3$ &  MgC$_3$N &  SiC$_4$ \\ 
 &  SiH$_4$ &   &   &   &   &   \\ 
\textbf{6-atoms} &  C$_2$H$_4$ &  C$_5$H &  C$_5$N &  C$_5$S &  CH$_3$CN &  H$_2$CCCC \\ 
 &  HC$_4$N &  MgC$_4$H &  SiH$_3$CN &   &   &   \\ 
$\mathbf{\ge}$\textbf{7-atoms} &  C$_6$H &  C$_7$H &  C$_8$H &  CH$_2$CHCN &  CH$_3$CCH &  CH$_3$SiH$_3$ \\ 
 &  H$_2$C$_6$ &  HC$_5$N &  HC$_7$N &  HC$_9$N &  PAH &   \\ 
\textbf{Ions} &  C$_3$N$^-$ &  C$_4$H$^-$ &  C$_5$N$^-$ &  C$_6$H$^-$ &  C$_8$H$^-$ &  CN$^-$ \\ 
 &  HCO$^+$ &   &   &   &   &   \\ 
 \hline
\end{tabular}
\begin{tabnote}
\noindent Notes: (?) indicates a tentative identification
\end{tabnote}
\end{table}

\begin{marginnote}[12cm]
	\entry{Reaction kinetics}{or chemical kinetics, dealing with understanding the rates of chemical reactions}
\end{marginnote}
In addition, there is the obvious difference between \textit{being} and \textit{seeing}. 
Given the current detections, our knowledge of quantum-chemical selection rules, reaction kinetics, and the thermodynamical wind properties, we have no doubt about the presence of some molecular species, but they remain hidden from our telescopes. 
The most obvious and important example is molecular hydrogen, H$_2$, the most abundant molecule in stellar winds. 
This non-polar light molecule has no permanent electric dipole moment and is not easily observed under the conditions prevalent in AGB CSEs. However, H$_2$ is detected in 2\,$\mu$m spectra of AGB variables \citep{Hinkle2000A&A...363.1065H},  but it is not always straightforward to determine whether an infrared line has a circumstellar origin or whether it emerges from the (extended) atmosphere. The same argument holds for other molecules such as CH \citep{Matsuura2007ASPC..378..450M}.  

Of the 105 molecules listed in \textbf{Table~\ref{Table:molecules}}, only 27 do not contain carbon and most  are diatomic species. 
This is due to the reactive nature and unique bonding properties of the carbon atom, but also because the nearest ($D\!\sim\!150$\,pc) and best known carbon star, CW~Leo, has a high mass-loss rate 
\citep[\Mdot$\sim$1.5$\times$10$^{-5}$\,\Msun\,yr$^{-1}$,][]{DeBeck2012A&A...539A.108D, Cernicharo2015A&A...575A..91C}. 
Just for comparison, the nearest O-rich AGB star R~Dor resides at a distance of $\sim$59\,pc and has a mass-loss rate of only $\sim$1.5$\times$10$^{-7}$\,\Msun\,yr$^{-1}$ \citep{Danilovich2016A&A...588A.119D}, while the O-rich supergiant VY CMa has an enormous mass-loss rate of $\sim$2$\times$10$^{-4}$\,\Msun\,yr$^{-1}$, but resides at a distance of $\sim$1150\,pc \citep{Decin2006A&A...456..549D, Zhang2012ApJ...744...23Z}.

In all CSEs the dominant molecule apart from H$_2$ is CO, which forms at local thermodynamic equilibrium (LTE) in the stellar photosphere 
and consumes all of the remaining atomic C or O --- i.e., whichever is the least abundant of the two. (See the sidebar titled Carbon- and oxygen-rich cool stars.)
For a long time, it was therefore thought that O-rich winds are very deficient in other carbon-bearing molecules, while C-rich ones are deficient 
in other oxygen-bearing molecules, but this picture has gradually been amended. 
First of all, there was the detection of HCN, HNC, and CS in O-rich CSEs which were explained  on the assumption of an active photochemistry in the outer CSE layers owing to the penetration of harsh interstellar UV photons \citep{Bujarrabal1994A&A...285..247B}. 
While this process can explain the formation of `atypical' molecules in the outer envelope, it cannot resolve the unanticipated detection of CO$_2$ in the inner wind of O-rich AGB stars and H$_2$O, OH, H$_2$CO, and SiO in carbon-rich winds \citep{Justtanont1998A&A...330L..17J, Ryde1998Ap&SS.255..301R, Melnick2001Natur.412..160M, Ford2003ApJ...589..430F, Ford2004ApJ...614..990F, Schoier2006A&A...454..247S, Cherchneff2008A&A...480..431D, Schoier2013A&A...550A..78S, Velilla2015ApJ...805L..13V}. 
Little by little, it became clear that we should abandon the convenient idea of LTE,  
and need to acknowledge that the chemistry is more complex than inferred by the assumption of thermodynamic equilibrium 
and instead is best described as `non-equilibrium chemistry'.
I say `little by little', since that change in mindset didn't happen silently and overnight. It was not that there was a disregard of unwelcome evidence, but non-equilibrium calculations imply solving the full set of rate equations for the detailed reaction kinetics  in the rapidly expanding and cooling gas. These calculations are not only CPU expensive, the challenge here is also to know the rate coefficients, which commonly depend on the temperature and often also on the gas density or on external sources such as the ISM radiation field or cosmic rays.
Not only can chemical reactions relevant for astrophysics rarely be extracted from industrial resources,  in addition  the gas number densities in CSEs are generally lower than in the laboratory by 10 decades. One should also realize that only $\sim$15\% of the reaction rates listed in astrochemical databases are experimentally obtained at room temperature ($\sim$300\,K), only $\sim$2\% have rate constants at temperatures below 200\,K and less than 0.5\% at temperatures below 100\,K. For some reaction rates, the results of theoretical quantum-chemical calculations are available. However, the complexity of the electronic structures often complicates a reliable theoretical prediction of the rate coefficients. Accurate rates are in particular critical for the rate-limiting steps (see Section~\ref{Sec:dust_nucl_short}).

Photo-induced processes in the outer CSE are not the only cause for out-of-equilibrium chemistry. In general, the low particle densities in CSEs imply that reaction conditions can change considerably before a reaction has run to completion, a particular example is the extended atmosphere which is subject to pulsation-induced shocks \citep{Cherchneff2011A&A...526L..11C,  Gobrecht2016A&A...585A...6G}. 
In addition, the material is subject to chromospheric UV fields (notably for RSG stars) and ionizing radiation from decaying radioactive nuclei. 
We now think that photo-induced processes are not only relevant in dictating the outer-envelope chemistry: given the fact that contemporary observations reveal the envelopes to be non-homogeneous (see Section~\ref{Sec:clumps}), recent theoretical models incorporating the emergence of 3D clumps indicate an enhanced penetration depth of interstellar photons resulting in a vast difference in predicted molecular abundance structures  compared  with the smooth envelope case \citep{Agundez2010ApJ...724L.133A, VandeSande2018A&A...616A.106V, VandeSande2019ApJ...873...36V}. 

In addition, we can get an extra glimpse of the molecular content of the winds of AGB stars, by studying the circumstellar environment of their immediate successors, the young post-AGB stars, where the increased UV flux and the presence of shocks can affect the chemistry. 
Good examples include the envelopes of the C-rich objects AFGL\,618 and AFGL\,2688 and the O-rich object OH\,231.8+4.2, in which molecules such as H$_2$, OCS, HNCO, HNCS, CH$_3$OH, H$_3$O$^+$, SO$^+$, CO$^+$, CH$^+$,  N$_2$H$^+$, H$_2$CO, HC$_4$H, HC$_6$H, CH$_3$C$_2$H, CH$_3$C$_4$H, and C$_6$H$_6$ have been detected. 

These molecules and atoms form the building blocks of the solid-state species which can form if density and temperature are of the right order for dust formation to take place. 
The process of solidification is modelled under the common interpretive framework of \textit{nucleation}, a process during which the nano-to-microscale transition of nuclei form; see also Section~\ref{Sec:grains}. (See the sidebar titled Nucleation.)

\begin{textbox}[htp]\section{Nucleation}Gas does not spontaneously go over to the solid bulk phase because there exists an energy barrier between the metastable and stable phases that makes a global phase change highly improbable. An energetically more favourable pathway is the formation of nano-to-microscale density transition nuclei, called the cluster model of the nucleation process. During this process, atoms and molecules undergo a sequence of collisional association reactions to form polyatomic molecules of ever increasing size. Clusters are generally represented by the symbol $M_N$, which is an assemblage of a (small) number $N$ of atoms or molecules, denoted here as monomer $M$.
Obviously, the term small is ambiguous, and distinguishing between a small gas-phase cluster and a larger particle can be difficult. 
A common practice is to define a small cluster as a system that still has a discrete electronic structure (as in an atom), and therefore differs from a larger system which has electronic bands. The properties of clusters change drastically with their size including optical, electronic, magnetic and chemical properties. Generally, nucleation is a very inefficient process and only a small fraction of the collisions between the gas phase atom or molecule and a small cluster will lead to cluster growth. 
\end{textbox}

\subsubsection{Condensates}
Condensates are characterised by their mineralogical structure (monocrystalline, polycrystalline, or amorphous), their chemical composition (homogeneous, but most often heterogeneous and layered), and their geometric shape. In contrast to molecules, 
 the analysis of IR dust spectral features does not always provide sufficient insight into the mineralogical, chemical, and geometric properties. This stems from the fact that dust spectral features often originate in radiatively excited functional groups of atoms constituting the grain, such as the stretching vibration of the Si-O bond (around 9.5\,$\mu$m) and the bending vibrations of the O-Si-O group (around 18.5\,$\mu$m) within the SiO$_4$ tetrahedron of silicates. As such, it is often impossible to unequivocally determine the exact grain properties from the analysis of IR spectra since, for example, two different amorphous silicate-type grains can each display the broad emission pattern in their mass absorption coefficient around 9.5 and 18.5\,$\mu$m \citep{Jaeger1998A&A...339..904J, Molster2002A&A...382..184M}. This is also the reason why the PAH molecules are sometimes analysed in a similar way as dust species, since also for these macroscopic PAHs we witness the behaviour of the functional C-H and C-C groups.

In particular the {\sl Infrared Space Observatory} was ground-breaking for detecting new dust species \citep{Tielens1998Ap&SS.255..415T, Waters2000ASPC..196....3W, Henning2010ARA&A..48...21H}; an overview of the dust grains identified in the winds of AGB and RSG stars can be found in \textbf{Table~\ref{Table:condensates}}. Instruments such as the ESO VLTI-MIDI and VLT-SPHERE interferometers were crucial for determining the location of the grains  \citep{Ohnaka2006A&A...445.1015O, Norris2012Natur.484..220N, Karovicova2013A&A...560A..75K}. In this context, a very interesting yet unexpected discovery was the spectroscopic identification of pure crystalline silicate grains \citep[pure forsterite, Mg$_2$SiO$_4$, and enstatite, MgSiO$_3$, the Mg-rich end members of olivine and pyroxene;][]{Molster2002A&A...382..184M}. Given an expanding and cooling envelope, grains are predicted to be amorphous and additional processes were sought for a further crystallisation of the grains (see also Section~\ref{Sec:challenges}). Some low-temperature crystallization process of unknown origin acting in circumstellar disks has been proposed \citep{Molster1999Natur.401..563M}. 
However, our recent finding that the majority of cool ageing stars with a measurable mass-loss rate might not live their lives alone, but have a companion offers an alternative explanation (Section~\ref{Sec:partner}): in O-rich winds with low wind acceleration properties the companion can quite easily induce the formation of an equatorial density enhancement \citep{ElMellah2020arXiv200104482E}, which offers the perfect conditions for amorphous grains to be annealed into crystalline minerals owing to shock formation  with an associated increase in temperature in the orbital plane \citep{Decin2019NatAs...3..408D}.

What is really fascinating is that physical samples of these dust grains can be assembled here on Earth: these microscopic stardust grains are found in meteorites (see \textbf{Figure~\ref{Fig:3D}}). After their journey through the ISM, these grains survived the formation of the Solar System, where they were trapped inside asteroids, the parent bodies of the meteorites. These presolar grains have isotopic compositions indicating an AGB or RSG origin \citep{Lodders2005ChEG...65...93L, Nittler2008ApJ...682.1450N, Davis2011PNAS..10819142D, Nittler2016ARA&A..54...53N}. A common property is that they are chemically stable and for this reason have survived their journey to Earth. The grain's morphology, mineralogy, and composition reflect the formation conditions in the CSE, and hence complement the studies of the IR spectral bands. The cosmochemical studies of these extraterrestrial rocks even allowed the identification of AGB and RSG dust grains that escape identification via infrared spectral features (see bottom part of \textbf{Table~\ref{Table:condensates}}). 
Typical grain sizes range between a few nanometer up to 1\,$\mu$m \citep{Goderis2016}. 
One should realize, however, that the laboratory techniques induce a bias in species detected due to the fact that some species might be dissolved out during the preparation process. In addition, smaller grains are more difficult to detect, resulting in a grain size bias.

\begin{table}[htp]
\caption{Condensates identified in the winds of cool evolved stars}
 \label{Table:condensates}
\begin{tabular}{@{}l|l||l|l@{}}
\hline
\rowcolor{shadecolor}
\multicolumn{1}{c|}{\textbf{Species}} &  \multicolumn{1}{c||}{\textbf{Formula}} & \multicolumn{1}{c|}{\textbf{Species}} & \multicolumn{1}{c}{\textbf{Formula}}\\
\hline
Alumina & Al$_2$O$_3$	&  Diopside & MgCaSi$_2$O$_6$ \\
Olivine & Mg$_{(2-2x)}$Fe$_{2x}$SiO$_4$ & Water ice & H$_2$O \\
Pyroxene &  Mg$_{(1-x)}$Fe$_{x}$SiO$_3$ & Metallic Fe & Fe \\
Melilite & Ca$_2$Al$_2$SiO$_7$ & Carbon & C \\
Magnesiow\"ustite & Mg$_{x}$Fe$_{(1-x)}$O & Silicon Carbide & SiC \\
Spinel & MgAl$_2$O$_4$ & Magnesium sulfide & MgS \\
\hline
Silica & SiO$_2$ & Hibonite & CaAl$_{12}$O$_{19}$ \\
Titanium oxide & TiO$_2$ & Titanium Carbide & TiC \\
\hline
\end{tabular}
\begin{tabnote}
\noindent Notes: 
Minerals listed in the second part of the table are only identified in presolar grains formed in AGB/RSG winds.
More detailed information can be found in the \textbf{Supplemental Tables~\ref{Table:condensates_IR_Suppl}\,--\,\ref{Table:condensates_presolar_Suppl}}.
\end{tabnote}
\end{table}

\subsection{Challenging the {\normalfont\bfseries\itshape{why}} question}\label{Sec:challenges}

The wealth of observations has guided the theoretical models and conveys the principles to which the models must adhere. 
The ISO observations of circumstellar dust around AGB stars have been interpreted within the context of thermodynamic equilibrium condensation sequences \citep{Onaka1989AA...218..169O, Tielens1998Ap&SS.255..415T}. Gearing the discussion toward oxygen-rich environments, the silicate thermodynamic condensation sequence predicts two branches to occur \citep{Grossman1974RvGSP..12...71G, Tielens1990fmpn.coll..186T}. The condensation sequence starts with that of Al$_2$O$_3$ (around 1\,700\,K). Gas-solid reactions lead to the formation of the first silicates, melilite (Ca$_2$Al$_2$SiO$_7$) and diopside (CaMgSi$_2$O$_6$) around 1\,500\,K. Most of the silicon will condense out in the form of pure Mg-rich silicates, first forsterite (Mg$_2$SiO$_4$, around 1\,440\,K) and later enstatite (MgSiO$_3$). Finally, reactions with gaseous iron will convert some enstatite into fayalite (Fe$_2$SiO$_4$) at about 1\,100\,K.

Observations indicate a very low Fe/Mg content of crystalline silicates \citep[Fe$<$10\%;][]{Molster1999A&A...350..163M}. This can be explained in a scenario proposed by \citet{Tielens1998Ap&SS.255..415T} in which crystalline Mg-rich silicates form close to the photosphere. These grains are thermally coupled to the gas via collisions and attain temperatures well above the glass temperature (of $\sim$1\,050\,K), resulting in a crystalline structure. 
Upon the expansion and cooling of the gas, it becomes  possible for Fe to be gradually incorporated in the silicate grains. 
The inclusion of Fe results in a considerable increase in the near-IR absorption coefficient so that just enough Fe can be incorporated for the radiative temperature of the grain to remain below the sublimation temperature of around 800\,--\,1100\,K with an Fe-inclusion of around 20\%. 
\begin{marginnote}[]
\entry{Glass temperature}{above the glass temperature atoms in an amorphous solid are very mobile (i.e.\ low viscosity) and can rearrange themselves in a crystalline lattice}
\end{marginnote} 
\hspace*{-.34cm} However, at these low temperatures the Fe-rich lattice cannot reach its energetically most favourable structure leading to the formation of amorphous Fe-rich silicates. The Fe-rich silicates efficiently absorb the stellar light in the near-IR so that the radiation pressure on the grains becomes large enough to initiate a stellar wind. The scenario proposed by \citet{Tielens1998Ap&SS.255..415T} rests on an important conjecture, namely that efficient dust nucleation and growth actually take place close to the stellar photosphere. As discussed by \citet{Tielens1990fmpn.coll..186T} this is far from obvious not only since the densities have to be high ($n \ga 10^{13}$\,cm$^{-3}$) for efficient nucleation, but also since there seems to be a lack of stable monomers.

Given the observations, thermodynamics seemed to be well obeyed, if one accounts for the `freezing-out' of chemical reactions which occurs when the density and temperature are too low and chemical reactions cease \citep{Tielens1998Ap&SS.255..415T}. In this way, the `macroscopic' world of dust grains was handled under the convenient assumption of thermodynamic equilibrium. If so, one should not care about the exact pathway of dust condensation since the most stable condensates will form under the general temperature-pressure conditions and relevant element mixtures. One can compare it with the analogy of a ball rolling down  a --- rather smooth --- hill: the ball will come at rest in the valley independent of the path taken. 
So, the thermodynamic equilibrium principles provided an answer to the \textit{why} question.

A striking disruption of this comfortable interpretation occurred in 2006, when Woitke applied an elegant deduction --- by essentially solving Eq.~\eqref{Eq:Gamma} and Eq.~\eqref{Eq:Tdust} --- to prove that the qualitative scenario proposed by \citet{Tielens1998Ap&SS.255..415T} for the wind driving in O-rich AGB stars actually does not hold  \citep{Woitke2006A&A...460L...9W}. In addition, through the adoption of detailed RHD models, Woitke showed that, in fact,  two dust layers are formed, an almost pure glassy Al$_2$O$_3$ layer close to the star, and a more opaque Fe-poor Mg-Fe-silicate layer further out. However, since only a small fraction of Fe can be incorporated in the silicate grains, because otherwise they become too hot, almost no mass loss occurred (\Mdot$\la$10$^{-10}$\,\Msun\,yr$^{-1}$). This paper not only re-opened the quest for the driving mechanism of O-rich AGB winds, but also touches on the fundamental \textit{why} question.

Various solutions have been offered to solve this dilemma. \citet{Hofner2007A&A...465L..39H} provided an unorthodox suggestion --- their words --- involving the formation of both carbon and silicate grains. This scenario implies departures from chemical equilibrium (CE), since a fraction of the carbon is not bound in gaseous CO, but available for grain formation. One year later, \citet{Hofner2008A&A...491L...1H} provided an alternative mechanism in which the Fe-free grains can grow to sizes large enough ($\sim$200\,nm -- 1\,$\mu$m) close to the star for photon scattering to compensate for their low near-infrared absorption cross-sections and to trigger the onset of a stellar wind. Recent VLT-NACO and SPHERE data support the presence of large transparent grains ($\sim$0.3\,$\mu$m) at a distance of $\sim$0.5\,\Rstar\ in some AGB stars \citep{Norris2012Natur.484..220N, Khouri2016A&A...591A..70K}, but these data cannot pinpoint the chemical composition of the grains. 
The current line of reasoning stipulates that Al$_2$O$_3$ grains can grow around $\sim$0.5\,\Rstar\ from the star and reside there in a gravitationally bound dust shell \citep{Khouri2015A&A...577A.114K, Hofner2016A&A...594A.108H, Hofner2019A&A...623A.158H}. Once the temperature gets below $\sim$1\,200\,K (around 4\,\Rstar), silicates can form as a mantle around the aluminium-oxide cores. For grain sizes above $\sim$100\,nm, a dust-driven outflow can be expected owing to the operation of scattering.

Each of the proposed solutions is, however, a `macroscopic' solution and bypasses the question of the molecular formation routes towards the predicted grain types. 
Although \citet{Hofner2007A&A...465L..39H} suggested a non-equilibrium formation route for carbon grains, they did not present a detailed 
non-CE study, and the abundance of available carbon was a parameter in the models. 
In the same vein, most modern radiation hydrodynamic wind models for O-rich AGB winds assume the first dust seeds are present, after which further growth to micrometer-sized dust grains can take place \citep{Hofner2018A&ARv..26....1H}. 
Hence, the conjecture posed by \citet{Tielens1990fmpn.coll..186T} is still open some thirty years later: we still assume in the models that efficient dust nucleation occurs close to the star, although we now acknowledge the shortcomings of thermodynamic equilibrium to understand dust nucleation.

\medskip

Leaving behind the comfortable landscape of thermodynamic equilibrium, one enters a new gigantic playground in which the rules of the game are not always very clear, and sometimes even undefined.
Given the earlier discussion on the challenges of gas-phase chemical kinetics (Section~\ref{Sec:molgrains_observations}), it is to be acknowledged that a proper description of dust nucleation is a fundamentally unattainable endeavour.  The reaction rate coefficients entering the description of  cluster growth are often unknown; the knowledge of the geometric configurations of the energetically low-lying structures is often lacking \citep{Bromley2016PCCP...1826913B, Gobrecht2017ApJ...840..117G, Gobrecht2018CPL...711..138G, Boulangier2019MNRAS.489.4890B}. The same applies to the description of grain growth and destruction in which case the adsorption and desorption coefficients are often only approximately known since the atomistic details of this process are unknown \citep{Cuppen2017SSRv..212....1C}. 
In aiming to answer the \textit{why} question, we have to enter this kinetics playground. Unlike the situation of thermodynamic equilibrium, for which the path is unimportant, an understanding of the path(s) taken toward dust nucleation will be central in our quest. In fact, this addition of specific paths implies an important change in the topology of the landscape. In TE the topology is convergent toward the most stable solutions, the thermodynamic sink. 
Turned around, this convergent TE solution in the forward direction becomes divergent in the backward direction prohibiting one from going 
back in time and addressing the \textit{why} question. Complicating the situation in a non-CE sense implies that we move away from this forward convergent/backward divergent topology, and that we can seek the real roots of mass loss.

\subsection{Dust nucleation: top-down versus bottom-up}\label{Sec:grains}

The realistic description of dust formation and growth  requires a treatment based on reaction kinetics, which is practically impossible since it would require the solution of a set of order $10^9$ rate equations for which most of the reaction rate coefficients are unknown and potentially not even critical. This task is considerably simplified by applying a two-step process in which first suitable seed particles have to form from the gas phase (the nucleation process;  see the sidebar titled Nucleation), and in a second step the seed nuclei form the substrate to which molecules are added \citep{Gail2013pccd.book.....G}. 
During the last decade, we have witnessed considerable progress in our understanding of the formation of this first seed nuclei using first principles, i.e.\ following a bottom-up approach. After delving through this section, however, the reader will realise that the situation is still far from satisfactory.

\subsubsection{Some theoretical aspects} \label{Sec:dust_nucl_short}

Various levels of theory exist for the description of nucleation. One can broadly classify them as the top-down and bottom-up approaches, with some variants in between; a schematized representation is shown in Figure~\ref{Fig:levels_of_theory}. In order to understand the challenges in this field, I have summarized the main principles behind the various approaches in the \textbf{Supplemental Text}. The key points for consideration are summarized here. Before doing so, a general note is appropriate. The term \textit{non-equilibrium chemistry}, so frequently used in the literature, may convey some information but is devoid of specific information. Only when delving through the detailed mathematical derivations, does it often become clear which physical quantity is not adhering to an equilibrium condition.

\begin{figure}[htp]
\includegraphics[width=\textwidth]{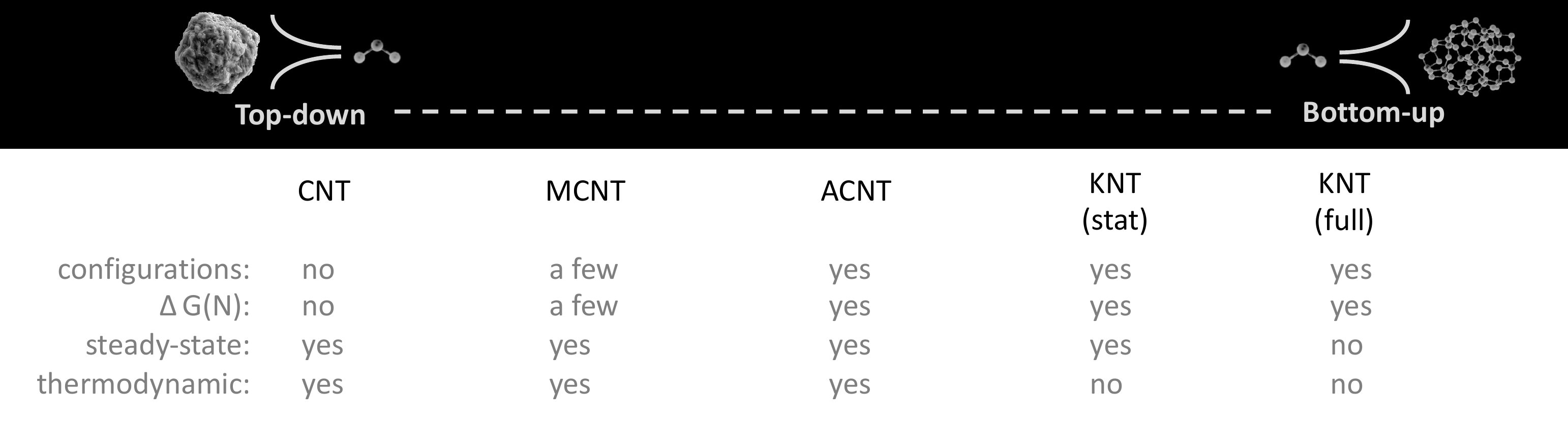}
\caption{Schematic representation of the different levels of theory used for the calculation of the nucleation rate. The different approaches include the classical nucleation theory (CNT), the modified classical nucleation theory (MCNT), the atomistic classical nucleation theory (ACNT), and the kinetic nucleation theory (KNT). 
Within KNT, the full time-resolved kinetic model can be solved (`full') or a stationary approach  can be pursued (${\rm{d}}(n,t)/{\rm{d}}t\!=\!=0$; `stat'). 
The different rows indicate if ground-state configurations are calculated using density functional theory (DFT), including the calculation of $\Delta G(N)$, 
if the nucleation rate is calculated under the assumption of steady-state, and
if thermodynamic quantities are needed --- in which case `no' indicates that the nucleation rate can in principle be calculated without recourse to these quantities, although rarely all rate efficiencies are available for this situation to occur.}
\label{Fig:levels_of_theory}
\end{figure}

The challenge of the bottom-up approach is that one needs the geometric configurations with the lowest potential energy (i.e.\ the global minima) and their respective binding energies for each cluster of size $N$ forming along the pathway toward the macroscopic solid compound. 
In a first step, one determines a good approximation of the geometric configuration of the molecular clusters by searching the potential energy surfaces  for the increasingly complex space of possible arrangements. This first step can already be quite CPU demanding. The candidate configurations are then used as the initial configurations in density functional theory (DFT) calculations to find the lowest energy structures. (See the sidebar titled Density functional theory.) For these structures, the thermodynamic properties (entropy, enthalpy of formation $\Delta_f H$, and Gibbs free energy of formation $\Delta_f G$) are then computed. The Gibbs free energies of formation are then used to calculate $\Delta G(N)$, which is the free energy change associated with the formation of a cluster of size $N$ from the saturated vapour. Using the most modern quantum chemical codes, the DFT method currently allows  systems containing some 50 atoms to be calculated \citep{Gobrecht2018CPL...711..138G}.
Owing to the computational challenges, only a few systems of relevance for astrophysical research have been computed at this level of theory, notable examples include (SiO)$_N$ \citep[$N\!=\!1-20$;][]{Bromley2016PCCP...1826913B}, (TiO$_2$)$_N$ \citep[$N\!=\!1-38$;][]{Jeong2000JPhB...33.3417J, Lee2015A&A...575A..11L, LamielGarcia2017, Boulangier2019MNRAS.489.4890B}, (C)$_N$ \citep[$N\!=\!1-99$;][]{Mauney2015ApJ...800...30M}, (Al$_2$O$_3$)$_N$ \citep[$N\!=\!1-8$;][]{Li2012, Gobrecht2018CPL...711..138G, Boulangier2019MNRAS.489.4890B}, and (MgO)$_N$ \citep[$N\!=\!1-10$;][]{Chen2014JPCA..118.3136C, Boulangier2019MNRAS.489.4890B}; each of them being proposed as dust precursor candidates in stellar winds.
\begin{marginnote}[]
	\entry{Binding energy}{energy required to separate constituent atoms/molecules from a system of particles}
	\entry{Gibbs free energy}{$\Delta G$, part of the heat of a reaction that can be converted to useful work; if $\Delta G\!<\!0$ the reaction is spontaneous, otherwise the reaction is non-spontaneous}
	\entry{Thermodynamic relations}{at constant temperature $\Delta G\!=\! \Delta H - T \Delta S$, with $H$ the enthalpy, $T$ the temperature and $S$ the entropy of the system}
	\end{marginnote}

\begin{textbox}[htp]\section{Density functional theory}To derive the cluster configurations, one needs to solve the Hamiltonian of a system of $N_e$ interacting electrons and fixed nuclei for the many-body eigenstate $\Psi_k$, $H | \Psi_k \rangle = E_k | \Psi_k \rangle$. Here $H$ denotes the Hamiltonian, $E_k$ the energy of the system, and $\Psi_k$ the wave function. The basis for density functional theory (DFT) is the Hohenberg-Kohn theorem \citep{Hohenberg1964PhRv..136..864H}, which states that the ground-state energy for an electronic system is a functional of the electron density.
Hence, in DFT calculations, the electron density rather than the wave function is used to describe the system. 
This reduces the dimensions of the system drastically, from $3N_e$ variables  to only 3, the three spatial coordinates.
As a result of the Hohenberg-Kohn theorems, the ground-state energy functional \(E[n]\) is stationary with respect to the number density $n$.
Determination of the ground-state number density \(n_0 (\mathbf{r})\) is then a solution to the
constrained variational equation
\begin{equation} \label{eq:dft_numberdens}
\frac{\delta}{\delta n(\mathbf{r})} \left[ E[n(\mathbf{r})] - \Lambda \left(\int {\rm{d}}\mathbf{r} \, n(\mathbf{r}) - N_e \right) \right] = 0\,, 
\nonumber
\end{equation}
with $\mathbf{r}$ the space coordinate and $\Lambda$ the mathematical variable Lagrange multiplier \citep{Kristyan2013JTAP....7...61K}.
\end{textbox}

Given the general lack of quantum chemical data on the cluster properties, various simplifications have been pursued along the path (see \textbf{Figure~\ref{Fig:levels_of_theory}}). In classical nucleation theory (CNT), one uses the capillary approximation to represent the potentially unknown $\Delta G(N)$ via the surface tension of the bulk material, $\sigma_\infty$ (see Eq.~\eqref{Eq:delta_G_CNT}). This approximation is only valid for $N\!\gg\!1$ and is also based on the assumption of thermal and chemical equilibrium. It also bypasses the well-known fact that at the nanoscale quantum effects become important, causing the nano-sized particles to behave very differently from their bulk counterpart.
Modified classical nucleation theory (MCNT) aims to improve CNT by linking the macroscopic solid bulk and microscopic cluster properties. Using DFT, the cluster properties of a few of the smallest clusters are calculated. The derived $\Delta G(N)/N$ values are then fitted using an analytical formula (see Eq.~\eqref{Eq:MCNT}) to allow extrapolation to larger values of $N$. That way, one accounts for the curvature of the small clusters. For values of $N\!\to\!1$, the atomistic version (ACNT) is recommended, in which the binding energies as calculated from DFT are explicitly used (Eq.~\ref{Eq:Delta_G_ACNT}). 

\medskip

In the full time-resolved kinetic model, one solves the full time-dependent set of coupled ordinary differential equations describing the rates of change in numbers densities of each cluster of size $N$, ${\rm{d}}n(N,t)/{\rm{d}}t$; see Eq.~\ref{Eq:conserv_N}. If the reaction efficiencies are known, the growth and evaporation time scale can be calculated directly, without additional need of the thermodynamic quantities. However, the growth efficiencies are often uncertain and the evaporation efficiencies are mostly unknown. Therefore, methods have been developed based on kinetic nucleation theory (KNT) and thermodynamic arguments (see \textbf{Supplemental Text}). 
Within KNT, CNT, MCNT, and ACNT it is possible to formulate the stationary nucleation rate, $J_\star(t)$, assuming that the cluster densities remain constant over time, i.e.\ ${\rm{d}}n(N,t)/{\rm{d}}t\!=\!0$. Assuming that the growth of clusters only occurs via monomer addition within the context of homogeneous nucleation, $J_\star(t)$ can be written as 
\begin{equation}
J_\star(t) \simeq z(N_\star) f(N_\star,t) \stackrel{\circ}{n}\!(N_\star)\,,
\end{equation}
where $\stackrel{\circ}{n}\!(N)$ is the equilibrium density of a cluster of size $N$, and $f(N,t)$ is the monomer attachment rate (see Eq.~\eqref{Eq:J_KNT} and Eq.~\eqref{Eq:attachment_rate}). In the framework of CNT, MCNT, and ACNT $z$ is the  kinetic Zel'dovich factor which takes a different value for each level of theory (see Eq.~\eqref{Eq:z_CNT}, Eq.~\ref{Eq:z_MCNT}, and Eq.~\ref{Eq:z_ACNT}), but $z\!=\!1$ within KNT. $N_\star$ is the critical cluster size where $\stackrel{\circ}{n}$ reaches its absolute minimum. The growth reaction of this cluster of size $N_\star$ is the bottleneck of the chemical network and hence is the rate limiting step. 

For each level of theory, in addition one needs reaction rate coefficients for the calculation of attachment rate $f(N,t)$. One of the most detailed methods (used in astrophysical research) is based on the Rice-Ramsperger-Kassel-Markus (RRKM) theory of chemical activity, employing a solution of the master equation based on the inverse Laplace transform method \citep[][in the case of (SiO)$_N$]{Bromley2016PCCP...1826913B}. 
In the absence of accurate reaction rates, the rate can be approximated by using the geometrical cross-section and the Maxwell-Boltzmann distribution, so that 
\begin{equation}
k = \pi (r_a +r_b)^2 \sqrt{\frac{8 k T}{\pi \mu}} \left(1 +\frac{E_a}{k T}\right) \exp\left(-\frac{E_a}{kT}\right)\,,
\label{Eq:k_rate}
\end{equation}
with $r_a$ and $r_b$ the radii of the colliding species, $\mu$ the reduced mass of the system, and $E_a$ the activation energy which is assumed to be zero if unknown \citep{Boulangier2019MNRAS.489.4890B}. 
When dealing with large clusters $N\gg1$, this reaction rate is often denoted as the sticking coefficient $\gamma$ --- i.e., the probability that an incoming monomer binds to the target cluster. While in principle it can differ from unity, it is a customary assumption in classical nucleation theory 
(CNT, MCNT, and ACNT) to assume $\gamma\!=\!1$.

\subsubsection{Recent theoretical outcomes} \label{Sec:theoretical_outcomes_nucleation}
Given these various levels of theory, one wonders if taking the bottom-up directive of full kinetic modelling is like taking a sledgehammer to crack a nut. 
The nucleation process can, \textit{mutatis mutandis}, be modelled using CNT, the theory which is currently most often used in astrophysical problems. 
Only a few authors address a comparison between the different approaches \citep{Jeong2003A&A...407..191J, Goumans2012MNRAS.420.3344G, Mauney2015ApJ...800...30M, Bromley2016PCCP...1826913B, Lee2018A&A...614A.126L}. These outcomes are summarized, in a somewhat encyclopedic way, in the \textbf{Supplemental Text} (Section~\ref{Sec:comparison_NT_theory}).
Before jumping to the bottom line of these comparisons, it is important to realize that each answer given should be interpreted within the context of the particular thermodynamical properties of the medium modelled.

Given the discussion in Section~\ref{Sec:comparison_NT_theory}, it seems that resorting to ACNT is sufficient to study dust nucleation in stellar outflows. On the condition that large enough $N$ are reached in the DFT calculations, ACNT yields a fast method to compute the stationary nucleation rate. 
The  only concern then is whether steady state is reached in the pulsation-dominated region where dust nucleation is thought to occur. 
This point has been addressed by \citet{Boulangier2019MNRAS.489.4890B} who developed a full kinetic model to study the nucleation of (TiO$_2$)$_N$, (SiO)$_N$, (MgO)$_N$, and (Al$_2$O$_3$)$_N$ in O-rich AGB CSEs. Moreover, they extended the model by including not only monomer interaction, but also polymer interaction. 
Their results show that cluster growth is time dependent, but for some species the largest clusters are formed in only a few hours, while others  
take more than $\sim$50 days  --- i.e., a significant fraction of the AGB/RSG pulsation period during which the thermodynamic properties can change drastically. 

Assuming the monomers are present, the full kinetic models of \citet{Boulangier2019MNRAS.489.4890B} favour Al$_2$O$_3$ as the primary dust precursor candidate owing to the high temperature at which nucleation can occur ($\sim$1\,800\,K). However, if one starts from an atomic mixture, an acute problem arises because the concentration of the Al$_2$O$_3$ monomer is almost negligible in the relevant temperature range, and most of the aluminium is in the atomic form. These results suggest that among the possible candidates, TiO$_2$ clusters are the only possible dust precursors. However, the kinetic temperature corresponding to the spatial regions where dust is found is roughly 1\,500\,--\,2\,000\,K \citep{Norris2012Natur.484..220N, Khouri2016A&A...591A..70K}, well above the formation temperature of (TiO$_2$)$_N\!=\!10$ which is around 1\,000\,--\,1\,200\,K. 
Moreover, a significantly larger fraction of presolar grains of AGB origin contain Al$_2$O$_3$  rather than TiO$_2$ grains \citep{Choi1998Sci...282.1284C, Stroud2004Sci...305.1455S, Bose2010ApJ...714.1624B}, although  the expected smaller size of TiO$_2$ grains makes them more difficult to detect. 
So,  alternative routes are now being explored  to determine whether larger (Al$_2$O$_3)_n$ clusters can form  bypassing the monomer, and  using for example the (Al$_2$O$_3)_2$ dimer as the main building block, where the dimer is formed from oxidation reactions of the abundant Al$_2$O$_2$ molecule with AlO, AlOH, OH, and H$_2$O (Gobrecht et al.\ \textit{in prep.}). 
Another possibility might be that large clusters can exist at higher kinetic temperatures than currently suggested by the model calculations, because of  thermo-ionic electron emission \citep{Demyk2004A&A...420..547D}, or possibly radiative cooling  (potentially via recurrent fluorescence or vibrational radiative cooling) rather than dissociating. 
The Nobel-prize winning physicist Richard \citeauthor{Feynman} famously used as the title of his talk presented to the American Physical Society in Pasadena on December 29, 1959 \textit{`There's Plenty of Room at the Bottom'}. This statement remains a stellar  example of physics prognostication; much of that room has yet to be explored.

\subsection{Discover the path by tracing the route}

So this is where we are. 
In 2020, we still remain in the realm of the argument formulated by Tielens in 1990, and this despite considerable progress in observations, theoretical models, quantum chemical calculations, and laboratory experiments. The non-CE path followed during dust formation in O-rich environments has still not been identified.
The path leading from bottom to up is not merely conceptual --- i.e., a \textit{gedanken} experiment --- but rather it is a genuine path that  
the chemical system adheres to. It is the topology invoked by that path that allows us to reverse the time axis. And actually, we do not need to discover dozens of paths, but one is sufficient since \textit{the winner takes it all}: if for a given chemical mixture several reaction chains lead to different condensation products, then the reaction chain with the highest nucleation rate wins. Once seeds of one kind start to form, other thermally stable materials will condense on the surface of these seeds, rather than forming seeds themselves. 
One might question if the quest for this path is too ambitious a target. 
My answer to this question is `no'. 
Thanks to the impressive increase in HPC facilities, it appears we have the beginnings of a solution to this tantalising problem. And although nature is allowed the prerogative of thinking out of the box, even the conventional methods have only been vaguely explored. 
We have a few indications on the complex quantum chemical route map, but \textit{divergent thinking} might be needed to identify the correct direction. 
I use the words `divergent thinking' because of the huge clumps of dust that are detected in the close vicinity of RSG stars \citep{Cannon2020}: circumstances are far from ideal for dust to form under conditions of chromospheric and pulsation activity in a low density region, but far from optimal is clearly not a synonym for impossible. Rather like extremofiles an Earth indicating that life is not vulnerable but persistent, the same seems to hold for dust nucleation.

Not only will quantum chemical calculations help astronomers to discover the path, a new take on understanding circumstellar chemistry was recently launched with the ambitious {\sl Stardust} project (https://nanocosmos.iff.csic.es/). By building a novel ultra-high vacuum machine combining atomic gas aggregation with in situ characterization techniques, the aim is to reproduce and characterize the bottom-up dust formation process under conditions similar to those of evolved stars. The prime focus of the project is currently on carbon chemistry. By feeding the reactions with individual C atoms and H$_2$ molecules, it recently has been shown that aromatic species and fullerenes do not form efficiently under these conditions \citep{Martinez2020NatAs...4...97M}. Furthermore, surface reactions can be studied with {\sl Stardust}. This reveals that reactions on surfaces can process the deposited material and lead to the formation of aromatic systems (in this case benzene, C$_6$H$_6$, and naphthalene, C$_{10}$H$_6$, one of the smallest PAHs), similar to what might be happening in CSEs. However, it should also be noted that the results of \citet{Martinez2020NatAs...4...97M} do not rule out gas-phase pathways for forming aromatic molecules. Although one can conclude from experimental results subject to restricted conditions which reactions are possible, it is logically unsound to conclude which reactions are not possible, particularly over a time span of millions of years within an environment where temperature and density change by orders of magnitude. The phrase by Charles \citet{Dickens} in his first novel \textit{The Pickwick Papers} is particularly apt: `never say never'.

Moreover, current observational techniques allow one to tackle the problem from an empirical perspective. We are at an important point of history and are now able to spatially resolve the surfaces of stars other than the Sun and to follow the change of the 3D structures through time, and hence gain access to an incredible amount of detail in a 4-dimensional spacetime. 
Using for example ALMA, we can follow the change in the radial abundance structure of molecules thought to be the building blocks of dust grains 
\citep[TiO, TiO$_2$, SiO, AlO, AlOH;][]{Kaminski2017A&A...599A..59K, Decin2017AA...608A..55D, Danilovich2020} to constrain the fraction of molecules left after dust formation. 
The challenges here are situated in (1)~the lack of collisional rates for the species under study, (2)~the need of radiative excitation rates (Einstein A-coefficients) for highly excited states, and (3)~the non-spherical and time-varying geometry of the stellar winds \citep{Danilovich2020}. 
But again, these challenges are within reach thanks to the modern HPC facilities. 
In collaboration with software engineers, we are on the verge of developing highly efficient algorithms for analysing these data \citep{DeMijolla2019A&A...630A.117D, DeCeuster2020MNRAS.492.1812D, DeCeuster2021}. 
The ALMA observatory is not the only one to offer some resolution on this issue. 
The strongest constraints are obtained from a multi-wavelength multi-instrument setup allowing for contemporaneous observations, a strategy increasingly endorsed by intergovernmental organisations in astronomy.
 How else would we be able to explain the \textit{great dimming} of 
Betelgeuse, the famous red supergiant in the constellation of Orion that is visible with the naked eye and that started fading  in December 2019? Not only astronomers, but also the broad public\footnote{The ESO press-release --- https://www.eso.org/public/news/eso2003/ --- was by far ESO's most popular press release of 2020, the study being covered in media outlets with a combined reach of over 2\,300 million.}, were wondering if Betelgeuse’s dimming meant it was about to explode. Like all red supergiants, Betelgeuse will one day go supernova. One of the currently most favoured working hypothesis for this dimming is the ejection or instant formation of a huge dust clump \citep[][Montarg\`es et al., \textit{in prep.}]{Cotton2020RNAAS...4...39C, Safonov2020arXiv200505215S}; again the seemingly impossible  becomes possible, although this might not be exactly what the public has hoped for. 
The principle by which dust clouds arise turns out to be extraordinarily effective but still obscure. It is clear that a fascinating scientific journey awaits us.

\section{Epilogue} \label{Sec:Epilogue}

The mathematical logic of living in a world existing of {\sl 4} (classical) spacetime dimensions has guided me through the writing of this review: {\sl 1} section was devoted to introducing the reader to the 1D world of cool ageing stars, {\sl 3} sections to three different levels of 3D complexities; $1\!+\!3\!=\!4$. These 3D spatial complexities steer the 4D path through spacetime of any star.

This discourse brings us naturally to reflections on the next events that might occur on the 4D path, not only the future of science in this field, but also the future of the stars themselves. 
A 4-dimensional future in which the time scale of evolution of the 3D complexities cannot be considered as being `secular' and phenomena are not stationary; $\partial/\partial t \ne 0$.
There are two prominent changes which I foresee occurring, the first one within reach of the next few years, the other which might take substantially longer. 
First, there is the inclusion of binarity in both the \textit{forward} theoretical and the \textit{retrieval} modelling of the mass-loss rates of cool ageing stars. 
This reorientation from a single-star perspective to the inclusion of (sub-)stellar companions will address the \textit{how} question in a profound way.  
Second, the exploration of the 3D reality of gas-phase clusters for our understanding of extraterrestrial dust formation is ripe with promise.
 This is an area within the field of astrochemistry where the \textit{why} and \textit{how} question meet each other in an intimate way. For these two advances to occur,  both the bottom-up and top-down approach should be pursued in a collaborative effort between astrophysicists, laboratory experimentalists, and quantum chemists.
 In particular the bottom-up methodology might induce considerable progress in the field given the expected revolution in HPC facilities. I therefore express here the explicit desire for a more intense collaboration with computational experts and HPC facilities. Often we are too addicted to software codes inherited from the past, which are not adapted to current computer performances. 
That attitude limits the rate of progress in the field, and the rate of improving our understanding of the \textit{how} and 
\textit{why} questions. 

\medskip

This address to the \textit{how} and \textit{why} questions is only one small piece in a highly intricate puzzle. And just as you have the medium-sized \textit{Golden Russets} and the large \textit{Haralsons} apples, you have \textit{why} and \textit{Why} questions. 
I reserve the capital \textit{Why} questions for these deeply grounded origin's questions of which the (known) sample size is just one: 
`Why did life emerge?' and `Why did the Universe come about?'. 
The \textit{why} question addressed here enters a statistically very different domain with billion of stars exhibiting similar characteristics 
of `old and simple', and at the same time of lively and creative as extraterrestrial laboratories. The link between the \textit{whys} and the \textit{Whys} can be expressed through the metaphor with the Russian dolls, where each doll gives birth to a smaller doll, logically connected, seemingly separate, but never independent. 
This brings me to the very first sentences of this review: 
the \textit{Why?}, \textit{why?}, and \textit{how?} questions driving the curiosity of the human race.

\section*{DISCLOSURE STATEMENT}
The author is not aware of any affiliations, memberships, funding, or
financial holdings that might be perceived as affecting the objectivity of this
review.

\section*{ACKNOWLEDGMENTS}

The author acknowledges the work sessions with the postdoctoral researchers and PhD students during the past few years in Leuven, in particular within the context of the team L.E.E.N. (`Low mass Evolved stars and their ENvironments') meetings, and the various discussions with the colleagues in the ATOMIUM consortium (https://fys.kuleuven.be/ster/research-projects/aerosol/atomium/atomium).
The author is grateful for constructive feedback on drafts of the manuscript from Carl Gottlieb, Tom Millar, David Gobrecht, Rens Waters, and the editor Ewine van Dishoeck.
C.\ Gielen is thanked for having produced Figure~\ref{Fig:Jels}, H.-P.\ Gail for Figure~\ref{Fig:model}, and J.\ Bolte and F.\ De Ceuster for Figure~\ref{Fig:binary}.
The author received funding from the European Research Council (ERC) under the
European Union's Horizon 2020 research and innovation programme (grant
agreements No.\ 646758: AEROSOL with PI L.\ Decin) and the KU Leuven C1 excellence grant MAESTRO C16/17/007 (PI L.\ Decin).
The author acknowledges the UK Science and Technology Facilities Council (SFTC) IRIS for provision of high-performance computing facilities.
This work was partly performed using the Cambridge Service for Data Driven Discovery (CSD3), part of which is operated by the University of Cambridge Research Computing on behalf of the STFC DiRAC HPC Facility. The DiRAC component of CSD3 was funded by BEIS capital funding via STFC capital grants ST/P002307/1 and ST/R002452/1 and STFC operations grant ST/R00689X/1. 


\bibliographystyle{ar-style2_LD.bst}
\bibliography{Decin_ARAA}

\newpage


\section{SUPPLEMENTAL ONLINE MATERIAL}

\subsection{Mass-loss rate prescriptions} \label{Sec:Mdots}

The first \textit{empirical} mass-loss rate prescription for cool ageing stars was derived in 1975 by \citet{Reimers1975MSRSL...8..369R}
\begin{equation}
\Mdot_{\rm{R}} = 4\times10^{-13} \eta L/g R\,,
\end{equation}
with \Mdot\ the mass-loss rate in units of \Msun\,yr$^{-1}$, $\eta$ a unitless parameter of the order of unity, and the stellar luminosity $L$, gravity $g$, and radius $R$ in solar units; see also Eq.~\eqref{Eq:Mdot_Reimers} This relation was derived for RSG stars, but later was often applied to AGB stars by adapting the $\eta$-parameter.
\citet{Kudritzki1978A&A....70..227K} calibrated the $\eta$-parameter for RSG stars and arrived at $\eta\!\simeq\!1.375$; \citet{Renzini1988ARA&A..26..199R} derived $\eta\!\simeq\!0.4$ for AGB stars.
Since then, various other \textit{empirical}, \textit{semi-empirical}, and \textit{theoretical} mass-loss rate prescriptions have been derived, such as the ones mentioned in Section~\ref{Sec:limitations}. In this section, I review in chronological order some of these parametric relations often used in the AGB and RSG community, in particular for stellar evolution and population synthesis modelling. 
 With the aim of comparing various mass-loss prescriptions for their explicit dependence on the fundamental parameters \Lstar\ and \Teff, Reimers' relation is rewritten as
\begin{equation}
\log \Mdot_{\rm{R}} =  - 4.876 +\log \eta + 1.5 \log \Lstar - 2 \log \Teff -2 \log \Mstar\,,
\end{equation}
where $\log$ refers to the base-10 logarithm, and \Mstar\ is in solar units. 


\citet{deJager1988A&AS...72..259D} collected mass-loss rates determined for 271 Galactic stars of spectral types O through M. The mass-loss rates could be reproduced by a single \textit{empirical} formula dependent on \Teff\ and \Lstar
\begin{equation}
\log \Mdot_{\rm{J88}} = -8.158 + 1.769 \log \Lstar - 1.676 \log \Teff\,.
\end{equation}
The original de Jager formulation did not account for the stellar mass. An improved relation was therefore proposed by \citet{Nieuwenhuijzen1990A&A...231..134N}
\begin{equation}
\log \Mdot_{\rm{NJ90}} = -7.93 + 1.64 \log \Lstar -1.61 \log \Teff + 0.16 \log \Mstar\,,
\end{equation}
but it can be seen that the dependence on the stellar mass is weak.

\citet{Vassiliadis1993ApJ...413..641V} combined an \textit{empirical} relation between the mass-loss rate and the pulsation period ($P$, in days) for Mira-type variables
\begin{eqnarray}
    \log \Mdot_{\rm{VW93}} & = &  -11.4 + 0.0123 P \hspace{3.25truecm} 
     {\text{(for \Mstar $\le$ 2.5\,\Msun)}} \label{Eq:Mdot_VW93a}\\
   \log \Mdot_{\rm{VW93}} &= & -11.4 + 0.0125 [P - 100 (\Mstar-2.5)] \qquad
    {\text{(for \Mstar $>$ 2.5\,\Msun)}} \label{Eq:Mdot_VW93b}
\end{eqnarray}
with a theoretical period-mass-radius relation (or period-mass-luminosity-effective temperature relation)
\begin{equation}
\log P = 12.52 +0.97 \log \Lstar -3.88 \log \Teff -0.9 \log \Mstar
\label{Eq:period_VW93}
\end{equation}
to obtain a \textit{semi-empirical} mass-loss rate formula for stars with periods between $\sim$300-800 days. For larger periods, the mass-loss rate is given by the single-scattering radiation pressure limit $\Mdot\,=\,\Lstar / c v_e$,
where the expansion velocity, $v_e$ (in km\,s$^{-1}$), was calculated using
\begin{equation}
v_e = -13.5 + 0.056 P\,.
\end{equation}

\citet{Blocker1995A&A...297..727B} used a modification of the Reimers' law for his theoretical AGB evolutionary calculations, by multiplying the Reimers' rate by $L^{2.7}$ and adding a dependence on the mass at the zero-age-main-sequence (ZAMS) as $M_{\rm{ZAMS}}^{-2.1}$. The modifications were guided by theoretical mass-loss rate calculations for fundamental mode Mira-type pulsators for which the period-mass-radius relation was taken as
\begin{equation}
\log P = -1.92 +0.93 \log \Lstar -3.72 \log \Teff -0.73 \log \Mstar\,.
\end{equation}
By setting, somewhat arbitrarily, $M_{\rm{ZAMS}}$ to the actual total mass, the following \textit{theoretical} relation was obtained
\begin{equation}
\log \Mdot_{\rm{B95}} = -13.19 + \log \eta +3.7 \log \Lstar -2 \log \Teff -3.1 \log \Mstar\,.
\end{equation}

Using the period-luminosity relation as derived by \citet{Feast1992iesh.conf...18F} for 15 RSG stars in the Large Magellanic Cloud (LMC) and mass-loss rates derived by \citet{Reid1990ApJ...348...98R},  \citet{Salasnich1999A&A...342..131S} derived an \textit{empirical} relation
\begin{equation}
	\log \Mdot_{\rm{S99}} = -14.5 + 2.1 \log \Lstar\,.
\end{equation}

Using the same methodology discussed in Section~\ref{Sec:limitations}, \citet{Wachter2002A&A...384..452W} presented an improved mass-loss rate description for carbon-rich AGB stare. The dependence of the mass-loss rate on the pulsation period in the model has been accounted for by applying an observed period-luminosity relation. The resulting \textit{semi-empirical} relation is 
\begin{equation}
\log \Mdot = 8.86 +2.47 \log \Lstar - 6.81 \log \Teff- 1.95 \log \Mstar\,.
\end{equation}

In 2005, \citet{vanLoon2005A&A...438..273V} analysed the optical spectra and infrared photometric data of a sample of dust-enshrouded RSG and O-rich AGB stars in the LMC to derive an \textit{empirical} mass-loss rate formula, under the assumption of a gas-to-dust ratio, $\psi = \rho_{\rm{gas}}/\rho_{\rm{dust}}$,  of 500: 
\begin{equation}
\log \Mdot_{\rm{vL05}} =12.478 + 1.05 \log \Lstar -6.3 \log \Teff\,.
\end{equation}
In the same year, \citet{Bergeat2005A&A...429..235B} derived an \textit{empirical} mass-loss rate formula for carbon-rich AGB stars by extracting the mass-loss rate, expansion velocities, and dust-to-gas density ratios from millimeter observations reported in the literature. Distances and luminosities previously estimated from {\sl Hipparcos} data, masses from pulsations, C/O abundance ratios from spectroscopy, and effective temperatures from a new homogeneous scale were used. The derived \textit{empirical} relation is dependent on the effective temperature:
\begin{eqnarray}
\log \Mdot_{\rm{B05}} & = & 24.93 -9.20 \log \Teff \qquad (\Teff  < 2400\,{\rm{K}}) \label{Eq:Mdot_B05a}\\
\log \Mdot_{\rm{B05}} & = & -3.6 -0.8 \log \Teff \qquad (2400\,{\rm{K}} < \Teff < 2900\,{\rm{K}}) \label{Eq:Mdot_B05b}\\
\log \Mdot_{\rm{B05}} & = & 20.65 -7.83 \log \Teff \qquad (2900\,{\rm{K}} < \Teff) \label{Eq:Mdot_B05c}.
\end{eqnarray}

 \citet{Srinivasan2009AJ....137.4810S} used Spitzer Space Telescope data to relate the excess emission at 8 and 24\,$\mu$m from evolved AGB stars in the LMC to their mass-loss rate. The derived \textit{empirical} relations are
  \begin{eqnarray}
  \log \Mdot_{\rm{S09}} & = & -14.2 +1.7 \log \Lstar \qquad (\text{C/O $<$ 1})\\
  \log \Mdot_{\rm{S09}} & = & -14.96 +1.87 \log \Lstar \qquad (\text{C/O $>$ 1})\,.
  \end{eqnarray}

\citet{Goldman2017MNRAS.465..403G} used a survey of 1612\,MHz OH maser observations of AGB and RSG stars in the LMC. This sample of sub-solar metallicity OH/IR stars was supplemented with data from the galactic centre and galactic bulge to derive a new \textit{empirical} mass-loss rate prescription which depends on luminosity, pulsation period, and gas-to-dust ratio $\psi$:
\begin{equation}
\log \Mdot_{\rm{G17}} = -10.6 +0.9 \log \Lstar + 0.75 \log P -0.03 \log (\psi/200)\,.
\label{Eq:Mdot_G17}
\end{equation}
By neglecting the weak dependence on the gas-to-dust ratio and adopting the same period-radius-mass relation as \citet{Vassiliadis1993ApJ...413..641V}, Eq.~\eqref{Eq:Mdot_G17} turns into
\begin{equation}
\log \Mdot_{\rm{G17}} = -1.28 +1.62 \log \Lstar -2.91 \log \Teff -0.675 \log \Mstar\,.
\end{equation}

Lately, \citet{Beasor2020MNRAS.492.5994B} measured the mass-loss rate and luminosities for RSG stars that reside in two clusters where the 
age and initial mass, $M_{\rm{ini}}$, are known. By combining the results with those of clusters with a range of ages, mass-loss rate prescriptions 
are derived that show an explicit dependence on the initial mass
\begin{equation}
\log \Mdot_{\rm{B20}} = (-26.4 -0.23 M_{\rm{ini}}) + 4.8 \log \Lstar\,.
\end{equation}

Recently, \citet{Kee2020} have studied the role of vigorous atmospheric turbulence in initiating and determining the mass-loss rate of RSG stars, following both an analytical and numerical approach. They provide the first \textit{theoretical} and fully analytic mass-loss rate prescription for RSG stars. They derived that
\begin{equation}
	\Mdot_{\rm{K20}} = \Mdot_{\rm{an}} \left(\frac{v_{\rm{turb}}/(17\,{\rm{km\,s^{-1}}})}{v_{\rm{esc}}(\Mstar,\Rstar)/(60\,{\rm{km\,s^{-1}}})} \right)^{1.3}\,,
\end{equation}
where $v_{\rm{turb}}$ and $v_{\rm{esc}}$ denote the turbulent and escape velocity, respectively.  \Mdot$_{\rm{an}}$ is given by
\begin{equation}
	\Mdot_{\rm{an}} = 4 \pi \rho(R_{\rm{p,mod}}) \sqrt{c_s^2+v_{\rm{turb}}^2} \,	R_{\rm{p,mod}}^2\,,
\end{equation}
with $c_s$ the sound speed and $R_{\rm{p,mod}}$ the modified Parker radius
\begin{equation}
	R_{\rm{p,mod}} = \frac{G \Mstar (1-\Gamma_{\rm{Edd}})} {2 (c_s^2+v_{\rm{turb}}^2)}\,,
\end{equation}
with $\Gamma_{\rm{Edd}}$ the Eddington factor, $\Gamma_{\rm{Edd}}\!=\!\kappa \Lstar/(4\pi G \Mstar c)$, and $\kappa$ the flux weighted mean opacity taken to be 0.01\,cm$^2$\,g$^{-1}$. 
The density at the modified Parker radius, $\rho(R_{\rm p,mod})$, is computed from hydrostatic stratification up to $R_{\rm p,mod}$ so that
\begin{equation}
	\rho(R_{\rm p,mod}) = \frac{4}{3} \frac{R_{\rm p,mod}}{\kappa R_\star^2} \frac{\exp\left[-\frac{2 R_{\rm p,mod}}{R_\star} +\frac{3}{2}\right]}{1-\exp\left[-\frac{2 R_{\rm p,mod}}{R_\star}\right]}.
\end{equation}
We here use a turbulent velocity of 18.2\,km\,s$^{-1}$, which is the mean turbulent velocity required for their theoretical model to reproduce observationally inferred gas mass-loss rates.

\bigskip

Each of the above mass-loss rate prescriptions shows an explicit dependence on the luminosity,  $\Mdot \propto \Lstar^{\alpha_L}$, with regression coefficient $\alpha_L$ ranging between 0.9\,--\,4.8. For AGB stars, the \textit{semi-empirical} formulae of \citet{Blocker1995A&A...297..727B} and \citet{Vassiliadis1993ApJ...413..641V}  have the steepest dependence on \Lstar; in the first case $\alpha_L$\,=\,3.7 and in the latter case the mass-loss rate is exponentially dependent on \Lstar. The recently derived RSG mass-loss rate prescription  by \citet{Beasor2020MNRAS.492.5994B} displays a very steep dependence on the luminosity of $\alpha_L$\,=\,4.8 when compared with those of \citet{deJager1988A&AS...72..259D} and \citet{Nieuwenhuijzen1990A&A...231..134N} which are also valid for RSG stars.

To illustrate the range of AGB mass-loss rates and their dependence on the luminosity, a set of fundamental stellar parameters was calculated for values of the hydrogen-depleted core mass, $M_c$, ranging between 0.56\,--\,1.4\,\Msun. The luminosity was calculated using the Paczysnki core-mass-luminosity (CML) relation \citep{Paczynski1970AcA....20...47P}:
\begin{equation}
\Lstar = 59250 (M_c - 0.522)\,
\end{equation}
with both quantities in solar units. Paczynski's relation is valid for $M_c\!>\!0.57$\,\Msun. 
Over the years, continual upgrades in the input physics and more detailed stellar evolution calculations resulted in various other CML relations being proposed \citep[see, e.g.][]{Groenewegen2004agbs.book..105G}. 
For values of $M_c\!<\!0.68$\,\Msun, we have opted to rely an the CML relation derived by \citet{Wagenhuber1998A&A...340..183W}, where the core mass limit is chosen to guarantee a smooth transition between both CML relations.
For a fixed effective temperature or stellar mass, the other fundamental parameter can be determined from the \citet{Iben1984ApJ...277..333I} radius-luminosity-mass relation for evolving AGB stars
\begin{equation}
\log \Rstar = \log(312) +0.68  \log \left(\frac{\Lstar}{10^4}\right) -0.31 \cdot \mathcal{S}   \log\left(\frac{\Mstar}{1.175}\right) + 0.088 \log\left(\frac{Z}{0.001}\right) -0.52 \log\left(\frac{l}{H_p}\right)\,,
\end{equation}
with $Z$ the initial mass abundance of  elements heavier than helium, $l/H_p$ the ratio of mixing length to pressure scale height, and the dimensionless parameter $\mathcal{S}\!=\!0$ for \Mstar\,$\le$\,1.175\,\Msun\ and $\mathcal{S}\!=\!1$ otherwise. 
We  here take $Z$\,=\,0.02 and assume $l/H_p$\,=\,1.
For a stellar mass of 2\,\Msun\ or an effective temperature of 2\,800\,K, the various mass-loss rate relations discussed here are shown in the upper panels of \textbf{Figure~\ref{Fig:Mdots}}. 

For the RSG stars, we follow the procedure of \citet{Kee2020} for an effective temperature of 3\,500\,K. The radius is computed from the luminosity-radius-temperature relation $\Lstar\!=\!4\pi \Rstar^2 T_{\rm{eff}}^4$. 
For the bottom right panel of \textbf{Figure~\ref{Fig:Mdots}}, in which only the effective temperature is fixed, the mass is computed from the mass-luminosity relation $\Mstar\!=\!(L/15.5)^{1/3}$, following \citet{Kee2020}.

\bigskip
The impact of a particular choice of mass-loss rate prescription on stellar evolution calculations is shown in \textbf{Figure~\ref{Fig:evol_tracks}}. A set of simplified evolutionary tracks has been constructed for stars with an AGB mass at the first thermal pulse of
(0.8, 1.0, 1.2, 1.4, 1.6, 2.0, 2.5, 3)\,\Msun. 
The corresponding core mass at the first thermal pulse was calculated using Eq.~13 of \citet{Wagenhuber1998A&A...340..183W}, 
and the luminosity was derived from the core-mass-luminosity (CML) relation described above. 
The stars were assumed to be Mira-type variables with fundamental mode period given by Eq.~\eqref{Eq:period_VW93}. 
The core mass increases owing to nuclear burning where the rate is given by \Mdot$_c$\,=\,1.02$\times$10$^{-11}$\Lstar. 
The evolution of mass and luminosity for solar composition stars is shown in \textbf{Figure~\ref{Fig:evol_tracks}}. 
The CML relation implies that ${\rm{d}}\Lstar/{\rm{d}}t \propto {\rm{d}}M_c/{\rm{d}}t$, and since $\Lstar \propto$ the hydrogen fusion rate 
$\propto {\rm{d}}M_c/{\rm{d}}t$, the rate of change of the abscissa is ${\rm{d}}\log\Lstar/{\rm{d}}t \sim 59250 \cdot 1.02 \times 10^{-11}$\,yr$^{-1}$ or 0.605\,Myr$^{-1}$ (as indicated by the black arrow in the lower right panel of \textbf{Figure~\ref{Fig:evol_tracks}}). 
The remnant mass for the planetary nebula nucleus or white dwarf is indicated by the intersection of the evolutionary track with the CML relation.

\subsection{Condensates identified in the winds of cool evolved stars} \label{Sec:condensates}

\textbf{Table~\ref{Table:condensates_IR_Suppl}} and \textbf{Table~\ref{Table:condensates_presolar_Suppl}} give more detailed information about the dust condensates identified in the winds of cool ageing stars, including references to the optical constants and to identifications from observations or presolar grains. 

\begin{table}[htp]
\caption{Condensates identified in the winds of cool evolved stars using infrared data}
 \label{Table:condensates_IR_Suppl}
\setlength{\tabcolsep}{0.5mm}
\begin{tabular}{@{}l|l|l|l@{}}
\hline
\rowcolor{shadecolor}
\textbf{Name} &  \multicolumn{1}{c|}{\textbf{Lattice}} & \multicolumn{1}{c}{} & \multicolumn{1}{|c}{}\\
\rowcolor{shadecolor}
 \textbf{\hspace*{0.5cm}Composition} &  \multicolumn{1}{c|}{\textbf{structure}} & \multicolumn{1}{c}{\multirow{-2}{*}{\textbf{Optical constants}$^{\rm a}$}}&  \multicolumn{1}{|c}{\multirow{-2}{*}{\textbf{References}$^{\rm b}$}}\\
\hline
Alumina$^\dagger$ & & & \\
\hspace*{0.5cm}Al$_2$O$_3$	&  AM 	& \citet{Koike1995Icar..114..203K}	 & \citet{Onaka1989AA...218..169O}\\
		&			 CR 			& \citet{Begemann1997ApJ...476..199B}	&  \citet{Glaccum1995ASPC...73..395G}\\[1.4truemm]

Olivine$^\dagger$ & & & \\
\hspace*{0.5cm}Mg$_{(2-2x)}$Fe$_{2x}$SiO$_4$\,$^{\rm c}$ & & & \\
\hspace*{1cm}($x=0.6$) & 	AM	& \citet{Dorschner1995AA...300..503D} & \citet{Woolf1969ApJ...155L.181W} \\
			&								&	 & \citet{Verhoelst2009AA...498..127V} \\
\hspace*{1cm}($x=0$)$^{\rm c}$ 			&  CR & \citet{Koike1993MNRAS.264..654K} & \citet{Waters1996AA...315L.361W}\\
			&											&	 \citet{Jaeger1994AA...292..641J} & \citet{Molster2002AA...382..184M}\\
Pyroxene$^\dagger$ & & & \\
\hspace*{0.5cm}Mg$_{(1-x)}$Fe$_{x}$SiO$_3$\,$^{\rm d}$ & & & \\
\hspace*{1cm}($x=0$)$^{\rm d}$			&   CR & \citet{Koike1993MNRAS.264..654K} & \citet{Waters1996AA...315L.361W}\\
			&											&	 \citet{Jaeger1994AA...292..641J} & \citet{Molster2002AA...382..184M}\\[1.4truemm]

Melilite	& & &  \\
\hspace*{0.5cm}Ca$_2$Al$_2$SiO$_7$\,$^{\rm e}$&  AM & \citet{Mutschke1998AA...333..188M} & \citet{Heras2005AA...439..171H} \\
			&												& \citet{Jaeger1994AA...292..641J} & \citet{Verhoelst2009AA...498..127V}\\[0.5truemm]

Magnesiow\"ustite 	& & &  \\
\hspace*{0.5cm}Mg$_{x}$Fe$_{(1-x)}$O\,& & & \\
\hspace*{1cm}($x=0.1$)&   AM & \citet{Henning1995AAS..112..143H} & \citet{Posch2002AA...393L...7P} \\[1.4truemm]

Spinel$^\dagger$ 		& & &  \\
\hspace*{0.5cm}MgAl$_2$O$_4$&   AM & \citet{Tropf1991} & \citet{Posch1999AA...352..609P} \\[1.4truemm]

Diopside 	& & &  \\
\hspace*{0.5cm}MgCaSi$_2$O$_6$&   AM & \citet{Koike2000AA...363.1115K} & \citet{Hony2009AA...501..609H}\\[1.4truemm]

Water ice	& &  & \\
\hspace*{0.5cm}H$_2$O &  CR & \citet{Bertie1969JChPh..50.4501B} & \citet{Justtanont1992PhDT.......104J}\\
			&												& \citet{Warren1984ApOpt..23.1206W} & \citet{Kemper2002AA...384..585K}\\[1.4truemm]

Metallic Fe & & &  \\
\hspace*{0.5cm}Fe&   CR & \citet{Henning1996AA...311..291H} & \citet{Kemper2002AA...384..585K} \\[1.4truemm]

Amorphous carbon 		& & &  \\
\hspace*{0.5cm}C& 	 AM & \citet{Preibisch1993AA...279..577P} & \citet{Martin1987ApJ...322..374M} \\[1.4truemm]

Silicon carbide$^\dagger$	& &  & \\
\hspace*{0.5cm}SiC&   AM & \citet{Pitman2008AA...483..661P} & \citet{Gilra1973IAUS...52..517G} \\ [1.4truemm]

Magnesium sulfide& &  & \\
\hspace*{0.5cm}MgS	&   AM & \citet{Begemann1994ApJ...423L..71B} & \citet{Goebel1985ApJ...290L..35G} \\
\hline
\end{tabular}
\begin{tabnote}
Abbreviations: AM, amorphous; CR, crystalline; U, crystal structure is unknown.\\
$^\dagger$Identified both from astronomical (infrared) observations and in presolar grains. 
$^{\rm a}$Reference to source for optical constants; 
$^{\rm b}$Reference to identification from observations; 
$^{\rm c}$Olivine, with the Mg-rich end member being forsterite; 
$^{\rm d}$Pyroxene, with the Mg-rich end member being enstatite; 
$^{\rm e}$Gehlenite; 
\citep{Takigawa2014LPICo1800.5389T}; in this table only TiO$_2$ has been listed for convenience.
\end{tabnote}
\end{table}

\begin{table}[htp]
\caption{Condensates only identified in presolar grains with AGB/RSG origin}
 \label{Table:condensates_presolar_Suppl}
\setlength{\tabcolsep}{0.5mm}
\begin{tabular}{@{}l|l|l|l@{}}
\hline
\rowcolor{shadecolor}
\textbf{Name} &  \multicolumn{1}{c|}{\textbf{Lattice}} & \multicolumn{1}{c}{} & \multicolumn{1}{|c}{}\\
\rowcolor{shadecolor}
 \textbf{\hspace*{0.5cm}Composition} &  \multicolumn{1}{c|}{\textbf{structure}} & \multicolumn{1}{c}{\multirow{-2}{*}{\textbf{Optical constants}$^{\rm a}$}}&  \multicolumn{1}{|c}{\multirow{-2}{*}{\textbf{References}$^{\rm b}$}}\\
 \hline
Magnesiow\"ustite & & & \\
\hspace*{0.5cm}Mg$_x$Fe$_{(1-x)}$O &  &  &  \\
\hspace*{1cm}($x=0$)\,$^{\rm c}$ & U & \citet{Tikhonov2018AdSpR..62.2692T} & \citet{Floss2008ApJ...672.1266F} \\
\hspace*{1cm}($x=1$)\,$^{\rm c}$ & U & \citet{Schrettle2012} & \citet{Bose2012GeCoA..93...77B} \\[1.4truemm]
Silica & & & \\
\hspace*{0.5cm}SiO$_2$ & U & \citet{Kitamura2007ApOpt..46.8118K} & \citet{Bose2010LPI....41.1812B} \\[1.4truemm]

Titanium oxide & & & \\
\hspace*{0.5cm}TiO$_2$\,$^{\rm d}$ & CR & \citet{Posch2003ApJS..149..437P} & \citet{Nittler2008ApJ...682.1450N}\\[1.4truemm]

Hibonite & &  & \\
\hspace*{0.5cm}CaAl$_{12}$O$_{19}$ & CR & \citet{Mutschke2002AA...392.1047M} & \citet{Choi1999ApJ...522L.133C}\\[1.4truemm]

Diamond \& Graphite & &  & \\
\hspace*{0.5cm}C & CR & \citet{Jager2003JQSRT..79..765J} & \citet{Amari1990Natur.345..238A}\\[1.4truemm]

Titanium carbide & & & \\
\hspace*{0.5cm}TiC & CR & \citet{Henning2001AcSpe..57..815H} & \citet{Bernatowicz1991ApJ...373L..73B}\\
\hline
\end{tabular}
\begin{tabnote}
Abbreviations: AM, amorphous; CR, crystalline; U, crystal structure is unknown.\\
$^{\rm a}$Reference to source for optical constants; 
$^{\rm b}$Reference to identification from observations; 
$^{\rm c}$ Likely originated in an AGB star, FeO (w\"ustite) and MgO (periclase) are the end members of magnesiow\"ustite; 
$^{\rm d}$ For the Ti-rich grains, possible phases of titanium oxides are TiO$_2$ (anatase), Ti$_4$O$_7$, Ti$_3$O$_5$, Ti$_2$O$_3$, and TiO, although rutile-structured TiO$_2$ was ruled out \citep{Takigawa2014LPICo1800.5389T}; in this table only TiO$_2$ has been listed for convenience.
\end{tabnote}
\end{table}

\subsection{Theory of dust nucleation} \label{Sec:nucleation_theory}

Following the cluster model of nucleation, atoms and molecules undergo a sequence of association and recombination reactions to form 
polyatomic molecules of ever-increasing size, the gas-phase clusters (see \textbf{Figure~\ref{Fig:3D}}). 
In general, association reactions can be divived in two types --- radiative association (which is difficult to detect in the laboratory) and collisional association (often ‘association’ is used as a shorthand) --- that describe what happens to the collisional intermediate, AB$^\star$:
\begin{itemize}
\item[(1)] A + B $\to$  AB$^\star$ $\to$  AB + photon --- the intermediate either stabilises to AB by emitting a photon, or returns to reactants A and B
 
\item[(2)] A + B $\to$  AB$^\star$ + M $\to$ AB + M --- the intermediate stabilises through collision with a third body M or returns to reactants.
\end{itemize} 
 
In order for reaction~(2) to proceed to AB, the density of M must be high, around 10$^{13}$\,cm$^{-3}$, typically, so this process is usually negligible in the ISM.  
At densities lower than $\sim\!10^{13}$\,cm$^{-3}$, reaction~(1) can occur but whether it does efficiently depends on whether the lifetime of the complex AB$^\star$ is longer than the radiative time scale. If so, then the complex can emit a photon and stabilise, if not it falls back to reactants. Generally, the larger the complex and the lower the temperature, the longer it lives and the more chance it has to emit a photon. 
 
Dust formation occurs at high density and temperature under conditions under which one would expect collisional association reactions to be feasible. They may dominate over normal two-body reactions at high density. In this regard, collisional association reactions are acting essentially like three-body reactions.

\bigskip

In what follows, we consider both the kinetic theory and the classical thermodynamic theory of nucleation to describe the nucleation process. This section is mainly based on \citet{Patzer1998A&A...337..847P, Gail2013pccd.book.....G}, and \citet{Mauney2015ApJ...800...30M}.

\subsubsection{Kinetic nucleation theory} \label{Sec:theory_kinetic}
The master equation for the particle number density $n(N,t)$ of a cluster of size $N$ is
\begin{equation}
\frac{{\rm{d}} n(N,t)}{{\rm{d}}t} = \sum_{i=1}^{I} J_i^c(N,t) - \sum_{i=1}^{I} J_i^c(N+i,t)\,,
\label{Eq:conserv_N}
\end{equation}
where $i$ denotes the $i$th geometrical configuration of a molecule (henceforth called $i$-mer),  $I$ the maximum molecular $i$-mer contributing, and $J_i^c(N,t)$ the effective flux (or transition rate) for the growth of the particle of size $N-i$ to size $N$. The effective transition rate is given by
\begin{equation}
J_i^c(N,t) = \sum_{r_i-1}^{R_i} \left(\frac{n(N-i,t)}{\tau_{\rm{gr}}(r_i, N-i, t)} - \frac{n(N,t)}{\tau_{\rm{ev}}(r_i, N, t)} \right)\,,
\label{Eq:rate}
\end{equation}
with $r_i$ denotes a chemical reaction in which the $i$-mer is involved. The time scale of the growth reaction from the cluster of size $N-i$ to size $N$, $\tau_{\rm{gr}}(r_i, N-i, t)$ is given by
\begin{equation}
\tau_{\rm{gr}}^{-1}(r_i, N-i, t) = A(N-i) \, \alpha(r_i, N-i) \, v_{\rm{rel}}(n_f(r_i), N-i) \,n_f(r_i,t)
\label{Eq:tau_gr}
\end{equation}
and the evaporation time scale leading from size $N$ to size $N-i$ by
\begin{equation}
\tau_{\rm{ev}}^{-1}(r_i, N, t) = A(N) \, \beta(r, N) \, v_{\rm{rel}}(n_r(r_i), N) \,n_r(r_i,t)\,,
\label{Eq:tau_ev}
\end{equation}
where $A(N)$ is the surface area of the cluster of size $N$, $\alpha$ and $\beta$ the reaction coefficients for the forward and backward process for reaction $r_i$ respectively, $n_f(r_i)$ and $n_r(r_i)$ the number density of the molecule of the growth (forward) and the evaporation (reverse) process for reaction $r_i$, and $v_{\rm{rel}}\!=\!\sqrt{k T/2 \pi \mu} \!\approx\!\sqrt{k T/2 \pi m_1} $ the average relative velocity between the molecule of mass $m_1$ and the cluster of mass $m_c$ with $\mu\!=\!1/(1/m_1 + 1/m_c)$. The reaction rates are determined from chemical kinetic theory or laboratory measurements. 

Homogeneous nucleation is the process in which a single kind of gas-phase species is involved, while heterogeneous nucleation implies the action of several gas-phase species. They are principally the same process, but from a theoretical point of view, homogeneous nucleation allows a simpler mathematical treatment of the process. In that case, Eq.~\eqref{Eq:rate} is described by one of the $r_i$ reactions.

If the reaction efficiencies were all known, the growth and evaporation time scales could be calculated directly from Eqs.~\ref{Eq:tau_gr}--\ref{Eq:tau_ev} and the transition rate could be derived without recourse to thermodynamic arguments. However, the growth efficiencies are often uncertain and the  evaporation coefficients $\beta$ are mostly unknown. Therefore, in order to describe the nucleation process under the conditions of chemical non-equilibrium in the gas-phase and of thermal non-equilibrium between the gas and the clusters or dust particle, methods have been develop based on kinetic nucleation theory and thermodynamic arguments.

\medskip
One therefore needs to define a reference equilibrium state which is characterized by thermal equilibrium with equal temperatures of the solid, the clusters and the gas, by phase equilibrium between the monomers as well as the $i$-mers and the bulk solid, and by simultaneous chemical equilibrium in the gas phase, i.e.\ it is a state of local thermodynamic equilibrium (LTE). All quantities referring to the LTE-state will be denoted with an `$^{\circ}$'.  For the LTE-state, the internal temperature of the $N$-clusters, $T_d(N)$, is equal to  the equilibrium temperature, $\stackrel{\circ}{T}=\!T_d(N)$. Local thermodynamic equilibrium between the gas phase and clusters implies that the principle of detailed balance holds for the growth processes and their respective reverse reactions, and thus that the effective transition rate equals zero for all cluster size, $J_i^c(N,t)\!=\!0$. It follows that in LTE
\begin{equation}
\frac{\stackrel{\circ}{n}\!(N-i)}{\stackrel{\circ}{\tau}_{\rm{gr}}(r_i, N-i)} = \frac{\stackrel{\circ}{n}\!(N)}{\stackrel{\circ}{\tau}_{\rm{ev}}(r_i, N)}\,.
\end{equation}
The equilibrium particle densities can be determined from the law of mass action
\begin{equation}
\left(\frac{\stackrel{\circ}{n}\!(N-i) \stackrel{\circ}{n}_f(r_i)} {\stackrel{\circ}{n}\!(N) \stackrel{\circ}{n}_r(r_i)}  \right) = \exp \left(\frac{\Delta_r G^{\minuso}(r_i, N, T_d(N))}{R \, T_d(N)} \,, \right)
\end{equation}
with $R$ the gas constant and where the Gibbs free energy of the reaction $\Delta_r G^{\minuso}(r_i, N, T_d(N))$ is given by the standard molar Gibbs free energy of formation of all reaction participants at temperature $T_d(N)$ in the reference state `$\minuso$'; see Eq.~(7) in \citet{Patzer1998A&A...337..847P}. These Gibbs free energies of formation are determined from (complex) chemical modelling.

\medskip

In what follows, we will derive the stationary nucleation rate for a situation of a homogeneous, homomolecular process in thermal equilibrium. For a stationary flow, ${\rm{d}}n(N,t)/{\rm{d}}t\!=\!0$. From Eq.~\ref{Eq:conserv_N}, it follows that the transition rate for clusters with $N \ge 2$ are equal in size space at time $t$, $J^c_1(N,t)=J^c_1(N+1,t) \equiv J_\star (t)$. $J_\star (t)$ is the size-independent rate of cluster formation.
It then can be shown
\begin{equation}
J_\star(t) = \left[\sum_{N=1}^{N_{\rm{max}}} \left( \frac{\tau_{\rm{gr}}(r_1, N, t)}{\stackrel{\circ}{n}\!(N)} \right) \right]^{-1}\,,
\label{Eq:rate_kin_stationary}
\end{equation}
with 
\begin{equation}
\tau_{\rm{gr}}^{-1}(r_1, N, t) = A(N)\, \alpha(N)\, v_{\rm{rel}}(n(1),N) \,n(1,t)\,.
\end{equation}
Consequently, the task of calculation the rate of cluster formation implies the determination of the growth time scales, $\tau_{\rm{gr}}$, and the determination of the thermodynamic properties of the clusters from equilibrium thermodynamics. 

For a process in thermal equilibrium with the gas and further assuming that the temperature of the cluster does not depend on size, $T_d(N) = T_g = T$,
the equilibrium cluster size distribution can be expressed by a Boltzmann-like distribution
\begin{equation}
\stackrel{\circ}{n}\!(N) = \stackrel{\circ}{n}\!(1) \exp\left(-\frac{\Delta G(N)}{R T} \right)\,,
\label{Eq:n_equil}
\end{equation}
with $\stackrel{\circ}{n}\!(1)$ the equilibrium distribution of the monomer corresponding to the vapour saturation pressure $p_{\rm{sat}}(T)$.
$\Delta G(N)$ denotes the free energy change associated with the formation of a cluster of size $N$ from the saturated vapour, which is related to the standard molar Gibbs free energy of formation of the $N$-cluster $\Delta_f G^{\minuso}(N)$ by
\begin{equation}
\Delta G(N) = \Delta_f G^{\minuso}(N) + R T \ln \left(\frac{p_{\rm{sat}}(T)}{p^{\minuso}} \right) - N \Delta_f G^{\minuso}_1(s)\,,
\end{equation}
where $\Delta_f G^{\minuso}_1(s)$ is the standard molar Gibbs free energy of formation of the solid phase and $p^{\minuso}$ is the pressure of the standard state.


The time scale $\tau_{gr}(r_{1}, N, t)$ on the right hand size of Eq.~\eqref{Eq:rate_kin_stationary} grows with increasing cluster size $N$ less than exponentially; for spherical grains it is $\propto N^{2/3}$. This implies that for sufficiently large $N$, the terms of the sum become negligible. The size $N_{\rm{max}}$ then denotes the maximum cluster size for which all $N$ above $N_{\rm{max}}$ do not contribute to the sum. Eq.~\eqref{Eq:rate_kin_stationary} can be further simplified by realising that there exists some critical size $N_\star$ where $\stackrel{\circ}{n}$
 attains its absolute minimum. 
 The growth reaction of this cluster of size $N_\star$ is the bottleneck of the chemical network. It is energetically favourable for clusters of size $N=N_\star+1$ to continue to form larger clusters that eventually form a `macroscopic' dust grain. We hence may approximate the whole sum by the largest terms, so that
\begin{equation}
J_\star(t) \simeq \frac{\stackrel{\circ}{n}\!(N_\star)}{\tau_{\rm{gr}}(r_1,N_\star,t)} = \stackrel{\circ}{n}\!(N_\star)\, n(1,t) \,A(N_\star) \,\alpha(N_\star) \,v_{\rm{rel}} \equiv\, \stackrel{\circ}{n}\!(N_\star) \, f(N_\star,t)\,,
\label{Eq:J_KNT}
\end{equation}
where $f(N_\star,t)$ is the monomer attachment rate and
the reaction $r_{1}$ is the  rate determined by the reaction
\begin{equation}
M_{N_\star} + M_1 \to M_{N_\star+1}\,.
\label{Eq:M_star}
\end{equation}
In general, truncating the summation in Eq.~\eqref{Eq:rate_kin_stationary} at $N_{\rm{max}} = 2 N_\star$ has negligible effect on the result.
Note that Eq.~\eqref{Eq:M_star} inherently assumes that the addition of a monomer, $A_1$, is the rate limiting step. This implies that the ratio of association with the least abundant cluster along the chain of reactions determines the rate at which macroscopic particles form. For that reason, $J_\star$ is also called the \textit{rate of seed formation} or the \textit{nucleation rate}. For the example of (SiO)$_n$, the critical cluster size is $N_\star\!=\!4-6$ for temperatures slightly below 630\,K \citep{Gail2013pccd.book.....G, Bromley2016PCCP...1826913B}\,. 

\subsubsection{Classical nucleation theory} \label{Sec:theory_CNT}
For species for which no thermochemical data are available, one often resorts to \textit{classical nucleation theory} (CNT), an effective theoretical tool originally developed in the early 20th century and based on the concepts of classical continuum theory. It refers to a state of constrained equilibrium between the clusters and the supersaturated vapor at the same supersaturation ratio $S$ and temperature $T$. Although widely used in the astronomical community, it can only be applied if (1)~$\Delta G(N)$ can be approximated by macroscopic parameters such as the surface tension of the bulk solid, (2)~the cluster are in thermal equilibrium,  (3)~small clusters are present in equilibrium concentrations, and (4)~chemical equilibrium prevails in the gas phase. 

CNT uses the capillary approximation to represent the potentially unknown $\Delta G(N)$ via the surface tension of the bulk material, $\sigma_\infty$, as measured in the laboratory:
\begin{equation}
\frac{\Delta G(N)}{R T} = - N \ln(S) + \sigma_\infty A(N)\,.
\label{Eq:delta_G_CNT}
\end{equation}
Assuming the monomers to be small spheres of radius $a_1$, $A(N) = 4 \pi a_1^2$ and we can write that 
\begin{equation}
\frac{4}{3} \pi a_1^3 = N V_1
\end{equation}
with $V_1$ the volume corresponding to one monomer in the bulk solid. Under this assumption
\begin{equation}
\frac{\Delta G(N)}{R T} = - N \ln(S) + \sigma_\infty (36 \pi)^{1/3} N^{2/3} V_1^{2/3} 
\end{equation}
or
\begin{equation}
\frac{\Delta G(N)}{RT} = - N \ln(S) + \theta_\infty N_\star^{2/3}  \text{\ \ with\ \ } \theta_\infty = \frac{4 \pi a_1^2 \sigma_\infty}{R T}\,.
\end{equation}

Clearly, the capillary approximation is inadequate for small values of $N$ and where a spherical approximation of the cluster geometry is poor.
Moreover, it does not yield the correct limit $\Delta G(N) \to 0$ for $N \to 1$. Therefore, a modification has been proposed to allow for a consistent formulation
\begin{equation}
\frac{\Delta G(N)}{R T} = - (N-1) \ln(S) + \theta_\infty (N_\star^{2/3} -1)\,.
\end{equation}

Since CNT refers to a constrained equilibrium, the equilibrium particle density of the monomer in this state is equal  to the monomer particle density in the actual situation, 
$\stackrel{\circ}{n}\!(1) = n(1,t)$.
For a homogeneous and homomolecular process, the CNT stationary nucleation rate for the critical cluster of size $N_\star$ is  given by 
\begin{equation}
J_\star(t) = z(N_\star) \, \gamma \, A(N_\star) \, \left(n(1,t)\right)^2 \, v_{\rm{rel}} \, \exp({-\Delta G_\star/kT})\,,
\label{Eq:rate_CNT}
\end{equation}
where $z$ is the Zel'dovich factor (see below), $\gamma$ the sticking coefficient (the probability that an incoming monomer binds to the target cluster, often assumed to be unity in CNT) and $\Delta G_\star = \Delta G(N_\star)$. 
Using Eq.~\eqref{Eq:n_equil} and writing the monomer attachment rate for a cluster of size $N$ as
\begin{equation}
f(N,t) = \gamma \, A(N) \,  n(1,t)\, v_{\rm{rel}}\,,
\label{Eq:attachment_rate}
\end{equation}
Eq.~\eqref{Eq:rate_CNT} can be written in a more concise way 
\begin{equation}
J_\star(t) = z(N_\star) f(N_\star,t) \stackrel{\circ}{n}\!(N_\star)\,,
\label{Eq:rate_CNT2}
\end{equation}
hence resembling Eq.~\eqref{Eq:J_KNT}, except for the correction factor $z$. 
The kinetic Zel'dovich factor $z$ accounts for the loss of nuclei during their Brownian motion, and in the classical theory it takes the form of 
\begin{equation}
z(N_\star) = \left(\frac{1}{2 \pi R T} \left| \frac{\partial^2 \Delta G(N)}{\partial N^2} \right|_{N_\star} \right)^{1/2}\,.
\label{Eq:Zeldovich}
\end{equation}
It is straightforward to derive that 
\begin{equation}
z(N_\star) = \left( \left(\frac{2 V_1}{9 \pi} \right)^{2/3} \frac{\sigma_\infty}{R T} \frac{1}{N_\star^{4/3}} \right)^{1/2}\,.
\label{Eq:z_CNT}
\end{equation}
The minimum value of $\stackrel{\circ}{n}\!(N)$ is encountered where $\partial \Delta G/\partial N\,=\,0$; the corresponding value of $N$ (i.e., the critical cluster size $N_\star$) is then
\begin{equation}
N_\star = \left( \frac{8\pi}{3} \left(\frac{3V_1}{4 \pi}\right)^{2/3} \frac{\sigma_\infty}{\ln S} \right)^3\,.
\end{equation} 

The critical size of refractory grains is usually small owing to the large temperatures at which nucleation occurs. 
This implies that the capillary approximation can destroy the accuracy of the results since the nucleation rate depends exponentially on the energy of cluster formation, which in this case, is only poorly approximated via the surface tension of the bulk. 

\subsubsection{Modified classical nucleation theory} \label{Sec:theory_MCNT} 
Aiming to improve CNT, modified classical nucleation theory (MCNT) links the macroscopic solid bulk and microscopic cluster properties by accounting for the curvature on the surface for small clusters. 
As a first step, one obtains the thermochemical properties of some small clusters, for example via density functional theory (DFT) and the calculation of the Gibbs free energy. This input is then used to approximate the bulk surface tension, $\sigma_\infty$, by fitting the following equation
\begin{equation}
\frac{\Delta G(N)}{k T} = \theta_\infty \frac{N-1}{(N-1)^{1/3} + N_f^{1/3}}.
\label{Eq:MCNT}
\end{equation}
to the available data. The parameter $N_f$ is a fitting factor defined as the (integer) monomer cluster size where the surface energy is reduced 
to half of the bulk value, or is left as a free parameter in the fitting procedure \citep[see, for example, Fig.~2 in][]{Lee2015A&A...575A..11L}. Eq.~\eqref{Eq:MCNT} can then be used to calculate the Gibbs free energies of formation and partial pressures for arbitrary values of $N$.
The Zel'dovich factor then reduces to
\begin{equation}
z(N_\star) = \left( \frac{\theta_\infty}{9 \pi (N_\star-1)^{4/3}} \frac{1+2\left(\frac{N_f}{N_\star-1}\right)^{1/3}}{\left(1+\left(\frac{N_f}{N_\star-1}\right)^{1/3}\right)^3} \right)^{1/2}\,.
\label{Eq:z_MCNT}
\end{equation} 
where the critical cluster size $N_\star$ is
\begin{equation}
 N_{*} - 1 = \frac{N_{*,\infty}}{8}\left( 1 + \sqrt{1 + 2\left(\frac{N_{f}}{N_{*,\infty}}\right)^{1/3}} - 2\left(\frac{N_{f}}{N_{*,\infty}}\right)^{1/3}\right)^{3} \,,
\end{equation}
and
\begin{equation}
 N_{*,\infty} = \left(\frac{\frac{2}{3}\theta_{\infty}}{\ln S(T)}\right)^{3}.
\end{equation}

\subsubsection{Atomistic classical nucleation theory}
The capillary approximation in CNT offers a reasonable description of nucleation when $n\!\gg\!1$. In the case of $n\!\to\!1$, the more exact atomistic model is required.
One therefore constructs the free energy of a cluster of size $N$ as
\begin{equation}
G(N) = G_V(N) + G_S(N)\,,
\end{equation} 
where $G_V(N$) is referred to as the volume term, $G_V(N) = - N k T \ln(S)$, and $G_S(N)$ as the surface term. In the atomistic case, one uses
\begin{equation}
G_{\rm{S}}(N) = \lambda N - E_b(N)\,,
\label{Eq:Delta_G_ACNT}
\end{equation}
where
$E_b(N)$ is the binding energy of a cluster of size $N$ and $\lambda N$ represents the binding energy of $N$ monomers in the bulk solid phase with bulk cohesive energy $\lambda$.
The binding energy is given by
\begin{equation}
E_b(N) = N E_1 - E_t\,,
\end{equation}
with $E_t$ the total energy and $E_1$ the single monomer total energy. 
The difficulty lies in determining $E_b(N)$ which --- similar  to the kinetic description  -- is the result of complex chemical and quantum mechanical interactions within the molecule and requires DFT calculations. If DFT calculations are available, the number densities are calculated as
\begin{equation}
n(N) = n(1) \exp\left(-\frac{G(N)}{kT}\right)
\end{equation}
 and Eq.~\eqref{Eq:rate_CNT2} yields the stationary nucleation rate. Recourse to the kinetic formulation, shows that the Zel'dovich factor then can be given in terms of $n(N)$:
\begin{equation}
z(N_\star) = \left( \sum_{N=1}^{N_{\rm{max}}} \frac{n(N_\star)}{n(N)} \right)^{-1}\,,
\label{Eq:z_ACNT}
\end{equation}
where $N_{\rm{max}} \!>\! N_\star$ is a reasonable cut-off of the summation.
The atomistic formulation of CNT (ACNT) hence also accounts for clusters with irregular shapes.

\subsubsection{Comparing the various levels of theory}\label{Sec:comparison_NT_theory}

Starting at the lowest level of complexity, it first should be noted that publications are not always clear as to whether CNT or MCNT was used as the modelling approach and hence if the bulk or a fictitious surface tension was used to represent the Gibbs free energies. Aiming to unravel the role of various nucleation candidates in O-rich AGB winds, \citet{Jeong2003A&A...407..191J} have pursued a mixed approach with TiO$_2$ studied at the level of MCNT based on \citet{Jeong2000JPhB...33.3417J}, and Fe, SiO$_2$, SiO, TiO, and Al$_2$O$_3$ at the level of CNT. 
Of the candidate materials considered, only TiO and TiO$_2$ reach nucleation rates greater than $\sim$10$^{-24}$\,s$^{-1}$ for the typical pressures in the dust condensation region of $10^{-3}$\,--\,$10^{-1}$\,dyn cm$^{-2}$ \citep{Hofner2016A&A...594A.108H}. (See the sidebar titled Nucleation rates from models.) A problem, however, is that the kinetic temperature should be lower than $\sim$1\,200\,K for the TiO/TiO$_2$ nucleation to start, which is below the typical kinetic temperatures of $\sim$1\,500\,--\,1\,800\,K in the dust formation region.

\begin{textbox}[htp]\section{Nucleation rates from models}
	The latest radiation hydrodynamical models for O-rich winds of \citet{Hofner2016A&A...594A.108H} indicate that the fractional seed particle abundance, $n_d/n_{\rm{H}}$, should be above $\sim$10$^{-16}$ for a dust-driven wind of mass-loss rate 1$\times$10$^{-7}$\,\Msun\,yr$^{-1}$ to be initiated. Winds of mass-loss rate 7.8$\times$10$^{-5}$\,\Msun\,yr$^{-1}$ are generated for $n_d/n_{\rm{H}}\!=\!10^{-15}$ \citep{Bladh2019A&A...626A.100B}. After further grain growth, the final $n_d/n_{\rm{H}}$ ratio is around $10^{-12}$\,--\,$10^{-13}$ for C-rich winds (see Figure~\ref{Fig:model}).
	With a typical AGB pulsation period ranging between $10^7$\,--\,$10^8$\,sec, this translates into a stationary nucleation rate $J_\star/n_{\rm{H}}\!\approx\!n_d/(n_{\rm{H}} \Delta t)$ of $\sim$10$^{-19}$\,--\,$10^{-24}$\,s$^{-1}$.
\end{textbox}

\cite{Lee2018A&A...614A.126L} used the small cluster thermochemical data of (TiO$_2$)$_N$ ($N$\,=\,1--\,10)  to study the stationary nucleation rate following both a MCNT and KNT approach. For regions where the supersaturation ratio $S\!\gg\!1$ both results agree within one order of magnitude. However, as the values of the gas temperature and pressure reach the $S\!=\!1$ equilibrium zone, larger differences between both methods occur, reaching four orders of magnitude. 
This difference is caused by  limitations in both MCNT and KNT.
For $S \to 1$, KNT reaches its limits since the critical cluster size $N_\star$ increases rapidly in this regime. 
By truncating Eq.~\eqref{Eq:rate_kin_stationary} at $N=10$, an unphysical KNT nucleation rate can be produced in regions of $S\!\la\!1$. 
This argues for further computational chemistry and experimentally based investigations into the small cluster properties for sufficiently large $N$.
\begin{marginnote}[]
	\entry{Supersaturation ratio}{defined as the ratio $p_1/p_{\rm{vap}}$ with $p_1$ the partial pressure of the monomers and $p_{\rm{vap}}$ the total vapour pressure. This definition is only useful for homogeneous nucleation.}
\end{marginnote}

\citet{Mauney2015ApJ...800...30M} computed the thermochemical properties of (C)$_N$ ($N$\,=\,1\,--\,99) and used that input to compare CNT, ACNT, and KNT stationary nucleation rates. By considering large cluster geometries, they were able to capture $N_\star$ across a wide range of temperatures and pressures for carbon nucleation. The ACNT and KNT rates agree to very high order. This is a particular useful result for numerical modelling since the prescription of ACNT is much faster than KNT. However, compared to KNT (and hence ACNT), the CNT rates are much faster at temperatures relevant for dust nucleation in stellar winds. For low values of the supersaturation ratio, a difference of 60 orders of magnitude is deduced due to the much larger critical size in CNT. For values of the supersaturation rate above $\sim$7 the CNT and KNT approach tend to agree.

Trying to understand under which conditions (SiO)$_N$ nucleation occurs, \citet{Bromley2016PCCP...1826913B} followed a bottom-up kinetic modelling approach and have compared the ACNT and KNT stationary nucleation rates and performed a full time-resolved kinetic model. For pressure values below 0.1\,dyn cm$^{-2}$ (relevant for stellar winds) ACNT agrees reasonably well with KNT, but overestimates $J_\star$ by up to a factor 1\,000 for higher pressure regions. The KNT stationary and full kinetic model agree well between 300\,--\,620\,K for pressures above 1\,dyn\,cm$^{-2}$, but for higher temperatures and high pressure values the KNT stationary approach predicts higher nucleation rates. This difference arises because the KNT stationary approach assumes chemical steady-state which is less valid at higher temperatures where the dissociation rates of some (SiO)$_N$ clusters become very fast. Independent of the model approach, the kinetics of SiO are much too slow to explain dust formation above 1\,000\,K under the conditions of stellar outflows. \citet{Goumans2012MNRAS.420.3344G} followed a KNT approach to study the heteromolecular nucleation of Mg, SiO, and H$_2$O to form silicate clusters with an enstatite stoichiometry (MgSiO$_3$)$_2$, but found the nucleation to be only efficient at 1\,000\,K for pressure exceeding 0.01\,dyn cm$^{-2}$; at higher temperatures the small clusters become thermodynamically unstable.

\end{document}